\documentclass[aps,prb,reprint,superscriptaddress,twocolumn]{revtex4-2}

\usepackage{bm,mathrsfs,dcolumn,graphicx,color} 
\usepackage{amsmath}
\usepackage{graphicx}
\usepackage[T1]{fontenc}
\usepackage{hyphenat}
\usepackage{setspace}
\usepackage{url}
\usepackage{xcolor}
\usepackage[utf8]{inputenc}
\usepackage{amssymb}
\usepackage{epsfig}
\usepackage[english]{babel}
\usepackage{empheq}
\usepackage[bbgreekl]{mathbbol}

\definecolor{myblue}{rgb}{.8, .8, 1} 

\newcommand{\bra}[1]{\left< #1 \right|}
\newcommand{\ket}[1]{\left| #1 \right>}

\newcommand{\br}{\mathbf{r}}
\newcommand{\bk}{\mathbf{k}}

\newcommand{\bq}{\mathbf{q}}
\newcommand{\bl}{\mathbf{l}}
\newcommand{\bp}{\mathbf{p}}
\newcommand{\bG}{\mathbf{G}}
\newcommand{\bR}{\mathbf{R}}

\newcommand{\bn}{\mathbf{0}}

\newcommand{\ev}[1]{\big< #1 \big>}

\newcommand{\en}[2]{\varepsilon^{#1}_{#2}}
\newcommand{\ent}[2]{\tilde{\varepsilon}^{#1}_{#2}}

\newcommand{\wfc}[5]{\phi^{#1,#2}_{#3,#4}(#5)}

\usepackage[toc,page]{appendix}
\newcommand{\nocontentsline}[3]{}
\newcommand{\tocless}[2]{\bgroup\let\addcontentsline=\nocontentsline#1{#2}\egroup}

\newcommand{\pap}[5]{\Xi^{#1,#2}_{#3,#4,#5}}

\begin{document}

\title{Wannier-Function-Based Approach to Coupled Exciton-Phonon-Photon Dynamics in Two-Dimensional Semiconductors}

\author{Alexander Steinhoff}
\email{asteinhoff@itp.uni-bremen.de}
\affiliation{Institut für Theoretische Physik, Universität Bremen, 28334 Bremen, Germany}
\affiliation{Bremen Center for Computational Materials Science, Universit\"at Bremen, 28334 Bremen, Germany}
\author{Frank Jahnke}
\affiliation{Institut für Theoretische Physik, Universität Bremen, 28334 Bremen, Germany}
\affiliation{Bremen Center for Computational Materials Science, Universit\"at Bremen, 28334 Bremen, Germany}
\author{Matthias Florian}
\affiliation{University of Michigan, Department of Electrical Engineering and Computer Science, Ann Arbor, Michigan 48109, USA}

\begin{abstract}

Marrying the predictive power of \textit{ab initio} calculations with many-body effects 
remains a challenging task
in two-dimensional (2d) materials, where efficient carrier-carrier interaction challenges established approximation schemes.
In particular, understanding exciton-phonon interaction from first principles is a field of growing interest.
Here, we present a many-body theory for coupled free-carrier, exciton, phonon and photon dynamics based on carrier-carrier and carrier-phonon interaction matrix elements obtained via projection on Wannier orbitals. The framework is applied to study the impact of carrier-phonon correlations on optical spectra and coupled nonequilibrium carrier-phonon kinetics in monolayer MoSe$_2$.
We find that 
non-Markovian effects and dynamical buildup of quasi-particles are only properly described if correlations at least on the two-phonon level are included.
Our studies open a perspective to advance the material-realistic description of nonequilibrium physics in 2d nanostructures to new many-body realms.

\end{abstract}

\maketitle

\section{Introduction}

Understanding the nonequilibrium dynamics of charge carriers and lattice vibrations in low-dimensional systems in the presence of efficient many-body interaction is an ongoing endeavor for computational materials science. The challenge is to take into account correlations between different particles at sufficiently high order, while adequately considering the rich landscape of electronic and vibrational basis states. In the case of atomically thin semiconductors, extensive research activities have been dedicated to exciton-phonon coupling as the key interaction mechanism to understand optical \cite{moody_intrinsic_2015,dey_optical_2016,cadiz_excitonic_2017,christiansen_phonon_2017,shree_observation_2018,jakubczyk_coherence_2019, li_excitonphonon_2021,purz_imaging_2022} and transport \cite{zipfel_exciton_2020,rosati_non-equilibrium_2021} properties. 
On the one hand, excitons have been identified as fundamental inter-band excitations due to their prominent signatures in optical spectra and large exciton binding energies, which are a consequence of strong carrier-carrier Coulomb interaction in van der Waals materials \cite{qiu_optical_2013, chernikov_exciton_2014,wang_colloquium:_2018}. On the other hand, phonons are present even in the limit of low excitation density and exhibit a characteristic temperature dependence that can be compared between theory and experiment \cite{moody_intrinsic_2015,dey_optical_2016,selig_excitonic_2016,jakubczyk_radiatively_2016,cadiz_excitonic_2017,jakubczyk_coherence_2019,rosati_temporal_2020,martin_encapsulation_2020}. Besides the above-mentioned demands on the predictive power of a theory, a major experimental challenge is to disentangle intrinsic (homogeneous) phonon-induced effects from extrinsic inhomogeneities \cite{jakubczyk_radiatively_2016, martin_encapsulation_2020}.
\\
\\While model-based approaches to exciton-phonon coupling in TMDs have been put forward for several years \cite{selig_excitonic_2016, christiansen_phonon_2017, selig_dark_2018,lengers_theory_2020}, \textit{ab initio} descriptions of these processes moved into the focus more recently. The latter are usually based on vibrational properties obtained from first principles via the well-established density-functional perturbation theory (DFPT) \cite{giannozzi_ab_1991,baroni_phonons_2001}. Combining DFPT with many-body perturbation theory (MBPT), the exciton-phonon scattering efficiency has been quantified for WSe$_2$ \cite{zhang_ab_2022,zhang_phonon-mediated_2024} and hexagonal boron nitride (hBN) \cite{chen_exciton-phonon_2020}, and exciton thermalization has been studied for MoS$_2$ \cite{chan_exciton_2024}.
MBPT has also been used to describe phonon-induced exciton energy renormalizations in monolayer TMDs \cite{shi_dark-exciton_2023,yu_anomalous_2024}.
Recent conceptional developments include 
off-diagonal exciton-phonon self-energy contributions to the MoS$_2$ exciton lineshape and relaxation \cite{chan_exciton_2023} as well as coupling of excitons to coherent phonons \cite{perfetto_theory_2024} and phononic screening of excitons \cite{lee_phonon_2024}.
On the other hand, the concept of taking those excitons visible in optical spectra as real quasi-particles participating in scattering processes has been questioned \cite{paleari_exciton-phonon_2022}. 
In Ref.~\cite{amit_ultrafast_2023}, a different first-principle perspective on exciton-phonon scattering is taken by computing the phonon-driven dynamics of the composition of excitons in terms of electron-hole pairs.
As an alternative to DFPT-based methods, it has been proposed to combine real-time GW-BSE calculations with nonadiabatic molecular dynamics to describe exciton spin and valley dynamics \cite{jiang_real-time_2021}.
Besides exciton-phonon scattering, phonon-induced effects in TMDs have been discussed on a single-particle level, from the limitation of carrier mobility due to electron-phonon scattering \cite{kaasbjerg_phonon-limited_2012} to spin-flip valley depolarization \cite{molina-sanchez_ab_2017} and excitation-induced nonequilibrium lattice dynamics \cite{caruso_nonequilibrium_2021}.
\\
\\So far, \textit{ab initio} approaches have either been formulated in terms of unbound carriers, or in a representation of bound exciton states, which are interpreted as well-defined quasi-particle eigenstates. There are several directions in which this picture needs to be extended. First of all, it has been pointed out that the fermionic substructure of excitons can significantly influence the phonon-assisted thermalization of excitons and inhibit bosonic behavior \cite{katzer_exciton-phonon_2023,katzer_fermionic_2024}.
Moreover, depending on the excitation conditions and lattice temperature, it is not a priori clear that photoexcited electron-hole pairs form a purely bosonic quantum gas. At low and high excitation densities, a significant fraction of excitons might be ionized into unbound electrons and holes \cite{semkat_ionization_2009,steinhoff_exciton_2017}. It is therefore desirable to have an overarching theory that can describe both, correlated and uncorrelated electron-hole pairs, on the same footing. 
While the previous statement refers to quasi-equilibrium situations reached at long timescales after excitation, quantum dynamics at very short times can exhibit non-Markovian behavior that modifies the often invoked adiabatic Boltzmann-type carrier scattering. This holds in particular in the presence of efficient many-body interaction such as in TMDs \cite{lengers_theory_2020}.   
It has been concluded in \cite{paleari_exciton-phonon_2022} that 
a promising route lies in the development of \textit{ab initio} theories that consistently describe the interplay of coherent and incoherent excitonic regimes including dynamical effects such as electronic screening of carrier-phonon interaction. 
Also, a non-Markovian first-principle treatment of exciton-phonon coupling is envisioned in \cite{amit_ultrafast_2023}. 
A possible approach to arrive at these long-term goals, i.e., describing the dynamical buildup of nonequilibrium correlations between carriers and phonons on the basis of first-principle electronic and vibrational input, is introduced and demonstrated in the following.
\\
\\In this paper, we present a nonequilibrium many-body theory for the coupled carrier-phonon-photon dynamics based on \textit{ab initio} band structures as well as matrix elements for carrier-photon, carrier-carrier, and carrier-phonon interaction in two-dimensional semiconductors. To this end, we apply density functional theory (DFT) and DFPT and project the resulting data to a localized basis of Wannier functions \cite{steinhoff_influence_2014, steinhoff_exciton_2017,berges_phonon_2023}. Subsequently, matrix elements are transformed to a set of Bloch states that capture the low-energy physics around the fundamental bandgap. The distinct advantages of this procedure are that (i) matrix elements and band structures in a minimal energy subspace can be obtained on an arbitrary mesh in reciprocal space via Fourier interpolation and (ii) complex phase factors of the Bloch states are consistently included in all types of many-body interaction processes. 
The Wannier-function based approach also allows us to include screening of Coulomb interaction within the 2d semiconductor due to a dielectric environment via the Wannier functon continuum electrostatics (WFCE) method \cite{rosner_wannier_2015}. 
Due to the downfolding of \textit{ab initio} electronic state information to a low-energy subspace with a minimal number of Bloch states, we are able to treat many-body effects of photoexcited carriers on an unprecedented level of complexity.
\\
\\To develop a many-body hierarchy that can be systematically truncated at a desired order, we use the so-called cluster expansion method \cite{fricke_transport_1996, kira_many-body_2006}, which formulates dynamical equations in terms of correlation functions, or cumulants, instead of expectation values. In contrast to the nonequilibrium Green function method, the cluster expansion is based on single-time quantities. Rather than resumming interaction processes in terms of self-energies, higher-order correlations are explicitly propagated. This fact leads to a less intricate structure of the EOM when treating nonequilibrium situations in the time domain, with a much easier inclusion of vertex corrections. 
Another advantage is the more direct access to many-body correlations and their temporal buildup, which allows the physics of correlated and uncorrelated electrons and holes to be disentangled to some extent. On the other hand, the graphical representation of interaction processes is not as simple as the well-known Feynman diagrams, and spectral properties of the interacting many-body system are not as readily accessible as in the case of retarded Green functions and self-energies.
We note that the cluster expansion is closely related to the coupled-cluster method frequently used in quantum chemistry \cite{purvis_full_1982}. Also, Cudazzo recently proposed an approach to augment GW+BSE calculations by cumulants, which enables a more accurate descripton of phonon sidebands, or satellites, than standard Green function methods \cite{cudazzo_first-principles_2020}.
Starting from a many-particle Hamiltonian that describes carriers, phonons, and photons as well as ther mutual interaction, the cluster expansion unifies various phenomena discussed before for atomically thin semiconductors. These include the above-mentioned coupling between carriers and coherent or incoherent phonons, nonequilibrium lattice dynamics, photoluminescence \cite{steinhoff_efficient_2015,selig_dark_2018} and carrier-carrier scattering \cite{steinhoff_nonequilibrium_2016,hader_importance_2022}.
\\
\\As a prototypical material system for our studies, we choose monolayer MoSe$_2$, which features an optically bright exciton ground state~\cite{gillen_light-matter_2017,druppel_electronic_2018}. To simulate a realistic experimental situation, we assume that the monolayer is encapsulated in hBN and placed on top of a SiO$_2$/Si substrate.
We study the impact of carrier-phonon coupling in two distinct regimes of photoexcitation: In the linear regime, optical absorption is extracted as a linear response to the external field, while decidedly nonequilibrium carrier-phonon dynamics is studied for nonlinear, resonant excitation with ultrashort laser pulses.
Carrier-carrier scattering can be systematically included on the level of approximation that we discuss in this work. However, for the presented applications at low excitation densities, it has no significant impact.
Moreover, we treat light-matter coupling in semiclassical approximation, solving the wave equation for external electromagnetic fields self-consistently with the quantum dynamics inside the TMD semiconductor. Therefore, spontaneous emission as a purely quantum-electrodynamical phenomenon is not included. 
To reduce numerical complexity, we also neglect electron-hole exchange interaction and thereby the exciton finestructure \cite{qiu_nonanalyticity_2015}, which we do not expect to significantly influence our results. Finally, we assume that spin-flip processes are not relevant on the ultra-short time scales discussed here.
\begin{figure}
\centering
\includegraphics[width=\columnwidth]{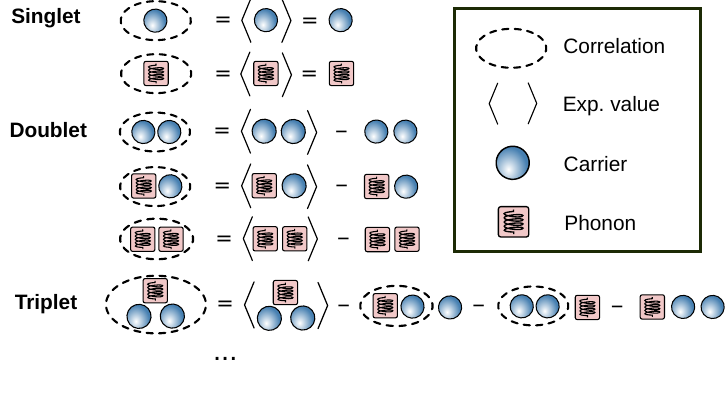}
\caption{Scheme of the \textit{cluster expansion} method: Expectation values are systematically replaced by correlations according to Eq.~(\ref{eq:cluster_hierarchy}).
Blue bubbles symbolize the smallest unit of carrier operators, which is given by polarization-type operators $a^{\dagger}a$. Springs symbolize the corresponding unit for phonon operators, which is given by single operators $D^{\dagger}$ or $D$. Correlations, or clusters, are classified as \textit{singlets}, \textit{doublets}, \textit{triplets}, etc. Singlets are correlations by construction, where we drop the correlation symbols for notational simplicity.
}
\label{fig:sketch_hierarchy}
\end{figure}

\begin{figure*}
\centering
\includegraphics[width=\textwidth]{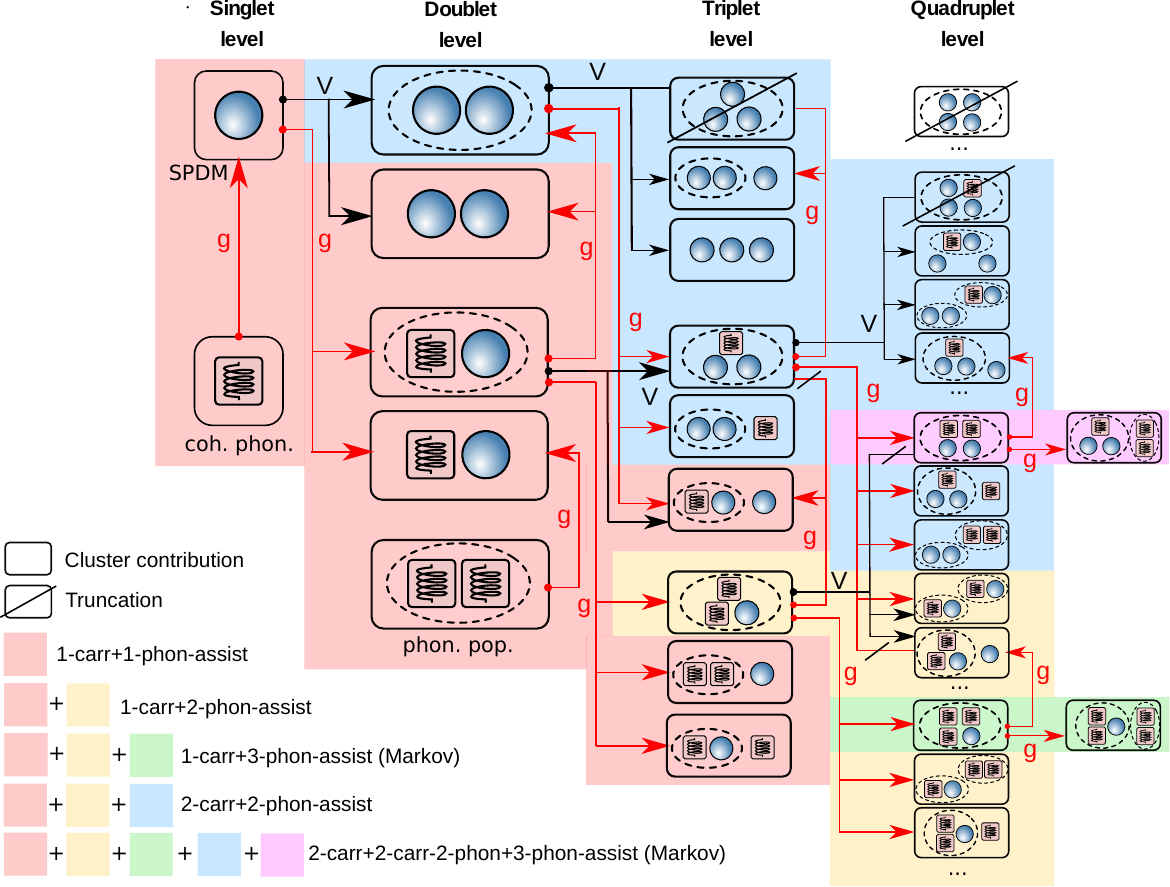}
\caption{Scheme of cluster hierarchy used in this work after replacing expectation values by correlations according to Eq.~(\ref{eq:cluster_hierarchy}) and Fig.~\ref{fig:sketch_hierarchy}. Coupling of correlations to higher-order correlations as well as to factorizations into smaller clusters is shown. Black arrows depict coupling via Coulomb interaction matrix elements $V$, and red arrows symbolize coupling via carrier-phonon interaction $g$. Coupling to external (semiclassical) fields is not shown.
Truncation of the hierarchy in a certain direction is symbolized by crossed out correlations. Some couplings are neglected to reduce numerical complexity, which is symbolized by crossed out interaction lines. The single-particle density matrix (SPDM), coherent phonons (coh. phon.) and phonon populations (phon. pop.) are marked explicitly. 
Physically, n-th order correlations induce scattering and dephasing on the (n-1)-th hierarchy level.
Colors symbolize subsets of clusters that are taken into account on different levels of approximation used throughout this work. \textit{1(2)-carr} stands for cluster contributions containing correlations with at most one (two) carriers, \textit{n-phon-assist} denotes polarizations assisted by $n$ phonons, \textit{2-carr-2-phon} denotes correlations of two carriers and two phonons, and \textit{(Markov)} means that quadruplet correlations are taken into account in Markov approximation.}
\label{fig:sketch_cluster}
\end{figure*}
Our key finding is that two-phonon processes significantly influence the optical and kinetic properties of monolayer MoSe$_2$ due to efficient exciton-phonon coupling. 
A cluster theory on the level of 
first-order carrier-phonon correlations 
underestimates the linewidth in optical spectra at elevated temperatures and 
introduces artificial heating of the lattice during nonequilibrium carrier-phonon dynamics. The reason for the latter is that
microscopic dephasing due to coupling to higher-order correlations is replaced by a phenomenological damping.
A simpler approach yielding physical results is given by first-order perturbation theory, which describes exciton-phonon scattering with exact energy conservation using delta functions for the scattering integrals. However, in the presence of efficient coupling, 
its quantitative reliability is questionable
and quantum-kinetic effects as well as the interplay of bound and unbound carriers are not accessible at all.
Our approach offers a way to go beyond perturbation theory by systematically including many-body correlations,
but one has to climb up high enough in the many-body hierarchy to obtain meaningful results.

\section{Theory}

Our starting point is the Hamiltonian for interacting Bloch electrons, phonons and photons,
\begin{equation}
 \begin{split}
H&=H_{\textrm{carr}}+H_{\textrm{phon}}+H_{\textrm{field}}+H_{\textrm{Coul}}+H_{\textrm{c-ph}}+H_{\textrm{LM}}\,,
\end{split}
\label{eq:Hamiltonian_start}
\end{equation}
with the individual terms being defined in the Appendix. 
The Hamiltonians of free carriers, phonons and photons are given by Eqs.~(\ref{eq:H_Bloch}), (\ref{eq:H_phon}) and (\ref{eq:H_field}), respectively.
The mutual interaction of carriers is given by the Coulomb-type Hamiltonian $H_{\textrm{Coul}}$ (\ref{eq:carr_carr_Hamiltonian}), the carrier-phonon interaction is given by $H_{\textrm{c-ph}}$ (\ref{eq:carr_phon_Hamiltonian}) and the light-matter interaction is given by $H_{\textrm{LM}}$ (\ref{eq:H_LM}).
To obtain information about the excited-state properties of the system, we compute the temporal dynamics of observables after pulsed optical excitation. Observables, such as optically induced polarizations or occupation functions of Bloch electrons and phonons, evolve according to the Heisenberg equations of motions (EOM) for fermionic and bosonic creation and annihilation operators:
\begin{equation}
 \begin{split}
i\hbar\frac{d}{dt} \hat{A}= \left[ \hat{A}, H \right]\,.
\end{split}
\label{eq:heisenberg_start}
\end{equation}
We denote creation and annihilation operators for a carrier with momentum $\bk$ in band $\lambda$ by $a^{\dagger}_{\bk,\lambda}$ and $a^{\phantom\dagger}_{\bk,\lambda}$, respectively. Phonons with momentum $\bq$ in branch $j$ are described by $D^{\dagger}_{\bq,j}$ and $D^{\phantom\dagger}_{\bq,j}$, while photons with (three-dimensional) momentum $\bq$ and polarization $\sigma$ are created and annihilated by $b^{\dagger}_{\bq,\sigma}$ and $b^{\phantom\dagger}_{\bq,\sigma}$, respectively.
The resulting many-body hierarchy is treated by applying the so-called \textit{cluster expansion technique} \cite{fricke_transport_1996, kira_many-body_2006}, which formulates dynamical equations in terms of correlation functions instead of expectation values. While expectation values are moments of the quantum-statistical ensemble, correlation functions correspond to cumulants, or semi-invariants \cite{kubo_generalized_1962,malyshev_cluster_1980}. The fundamental assumption therefore is that at any time, the ensemble is better characterized by cumulants than by moments (the Gaussian distribution being a particularly impressive example).
Correlation functions can be expressed recursively as the difference between an expectation value and all of its possible factorizations into smaller correlation functions. Denoting an $N$-particle cluster by $\Delta\ev{N}$, this procedure can be schematically written as \cite{kubo_generalized_1962,hall_non-equilibrium_1975,malyshev_cluster_1980,schoeller_new_1994,fricke_transport_1996, kira_many-body_2006}:
\begin{equation}
 \begin{split}
\Delta\ev{1}&=\ev{1}\,, \\
\Delta\ev{2}&=\ev{2}-\Delta\ev{2}_{S}\,, \\
\Delta\ev{3}&=\ev{3}-\Delta\ev{3}_{S}-\Delta\ev{1}\Delta\ev{2}\,, \\
&\,... \\
\Delta\ev{N}&=\ev{N}-\Delta\ev{N}_{S}-\Delta\ev{N-2}_{S}\Delta\ev{2}\,,\\
&-\Delta\ev{N-4}_{S}\Delta\ev{2}\Delta\ev{2}-\Delta\ev{N-3}_{S}\Delta\ev{3}-...
\,.
\end{split}
\label{eq:cluster_hierarchy}
\end{equation}
The first line of Eq.~(\ref{eq:cluster_hierarchy}) defines correlations of operators that do not allow for further factorization and therefore equal the corresponding expectation values. These are classified as \textit{singlets}, while higher-order correlation functions are referred to as \textit{doublets}, \textit{triplets} and so on. Factorizations into $N$ singlets are denoted by $\Delta\ev{N}_{S}$.
In general, each term in Eq.~(\ref{eq:cluster_hierarchy}) corresponds to 
a sum over all permutations when distributing the operators among the clusters. The summation is required to obey the indistinguishability of particles, where fermionic operators require an extra sign that takes into account the parity of permutations \cite{kira_many-body_2006}. For a mixed system of carriers and phonons, we visualize the cluster expansion in Fig.~\ref{fig:sketch_hierarchy}. In the presence of many-body interaction processes, clusters in general couple to higher-order clusters, i.e., singlets couple to doublets and so on. 
The correlation hierarchy is truncated at a desired level of many-body complexity by systematically discarding certain classes of correlation functions.
While here we use the cluster expansion to express single-time expectation values at any time in terms of correlations, similar \textit{linked cluster} methods are applied for equilibrium many-body systems to evaluate, e.g., the grand potential in terms of cumulants \cite{mahan_many_2000}. The latter diagrammatically correspond to all different connected graphs. Again, truncating a summation over cumulants at a given order might be more accurate than truncating a corresponding S-matrix expansion, as in a Dyson equation approach.
In a similar spirit, the \textit{coupled-cluster} (CC) method, frequently used as gold standard in
quantum chemistry, approximates the ground-state wave function of molecules as an expansion into $n$-th order clusters instead of $n$-fold excited determinants \cite{purvis_full_1982}.
The CC method classifies the used level of approximation in terms of contributing singles, doubles, triples, etc., similar to our case.
\\
\\For electrons, singlet clusters are the two-operator functions constituting the single-particle density matrix, see the schematic representation of the cluster hierarchy in Fig.~\ref{fig:sketch_cluster}. 
Its matrix elements are the single-particle occupation functions $f_{\bk}^{\lambda}=\ev{a_{\bk,\lambda}^{\dagger}a_{\bk,\lambda}^{\phantom\dagger}}$ and microscopic polarizations $\psi_{\bk}^{\lambda,\lambda'}=\ev{a_{\bk,\lambda}^{\dagger}a_{\bk,\lambda'}^{\phantom\dagger}}$. 
Note that spatial homogeneity of the two-dimensional crystal requires that the net in-plane momentum of any expectation value or correlation function is zero.
While the occupation functions determine the carrier density in every Bloch band $\lambda$,
\begin{equation}
\begin{split}
n^{\lambda}=\frac{1}{\mathcal{A}}\sum_{\bk} f_{\bk}^{\lambda}\,,
\end{split}
\label{eq:carr_dens_band_main}
\end{equation}
with the crystal area $\mathcal{A}$, the off-diagonal terms determine the optical response.
In the case of bosonic particles, single operators have finite expectation values and are therefore classified as singlets, which is consistent with the fact that they appear together with pairs of fermionic operators in the Hamiltonian. One example is the light-matter interaction Hamiltonian, Eq.~(\ref{eq:H_LM}). In the course of cluster expansion, its action leads to correlations between carrier operators and operators of the electric field $\boldsymbol{E}(\bq_{\parallel})$, which contains photon and carrier operators on the singlet level, see Eq.~(\ref{eq:E_field_op}). In this work, we focus on semiclassical light-matter interaction by discarding such correlations, which leads to approximations like:
\begin{equation}
\begin{split}
\ev{a^{\dagger}_{\bk+\bq_{\parallel},\bar{\nu}}\boldsymbol{E}(\bq_{\parallel})a^{\phantom\dagger}_{\bk,\nu}}&\approx
\ev{\boldsymbol{E}(\bq_{\parallel})}
\ev{a^{\dagger}_{\bk+\bq_{\parallel},\bar{\nu}}a^{\phantom\dagger}_{\bk,\nu}}
\\
&=
\ev{\boldsymbol{E}(\bn)}\psi^{\bar{\nu},\nu}_{\bk} \delta_{\bq_{\parallel},\bn}
\,.
\end{split}
\label{eq:light_matter_neglect_main}
\end{equation}
Hence, we do not take into account quantum-electrodynamical effects such as spontaneous emission \cite{kira_quantum_1999, kira_many-body_2006,steinhoff_efficient_2015,selig_dark_2018}.
As shown in the Appendix, the dynamics of the electric-field singlets $\boldsymbol{E}^{\textrm{2d}}=\ev{\boldsymbol{E}(\bn)}$ follows from a semi-classical wave equation
\begin{equation}
\begin{split}
\Big[\frac{\partial^2}{\partial z^2}-\frac{n^2(z)}{c^2} \frac{\partial^2}{\partial t^2}\Big]\boldsymbol{E}(z,t)=\mu_0 \frac{\partial^2}{\partial t^2} \boldsymbol{P}^{\textrm{2d}}(t) |\xi(z)|^2
\end{split}
\label{eq:E_T_wave_FT_class_main}
\end{equation}
by averaging the solution over the two-dimensional layer in the z direction using the confinement functions $\xi(z)$. The two-dimensional polarization $\boldsymbol{P}^{\textrm{2d}}(t)$ is defined in Eq.~(\ref{eq:def_pol_2d}).
The coupled EOM for carrier singlets in the semiclassical limit are known as semiconductor Bloch equations (SBE) \cite{haug_quantum_2004,kira_many-body_2006}:
\begin{equation}
\begin{split}
\frac{d }{dt}f^{\nu}_{\bk}&=\frac{2}{\hbar}\,\textrm{Im}\Big\{ 
\sum_{\bar{\nu}} \big(\Omega_{\bk}^{\bar{\nu},\nu}\big)^*\psi^{\bar{\nu},\nu}_{\bk}
\Big\} \\
&+\frac{d }{dt}f^{\nu}_{\bk}\Big|_{\textrm{D, c-c}} +\frac{d }{dt}f^{\nu}_{\bk}\Big|_{\textrm{D, c-ph}}\, \\
i\hbar\frac{d }{dt}\psi^{v,c}_{\bk}&=(\ent{c}{\bk}-\ent{v}{\bk})\psi^{v,c}_{\bk} 
+\sum_{\bar{v}} \Omega_{\bk}^{v,\bar{v}} \psi^{\bar{v},c}_{\bk}
-\sum_{\bar{c}} \psi^{v,\bar{c}}_{\bk} \Omega_{\bk}^{\bar{c},c} \\
&+\sum_{j,\lambda}2\,\textrm{Re}\big\{\ev{D^{\phantom\dagger}_{\bn,j}}\big\}\Big( 
g^{c,\lambda}_{\bk,\bn,j}\psi^{v,\lambda}_{\bk} -\big(g^{v,\lambda}_{\bk,\bn,j}\psi^{c,\lambda}_{\bk}\big)^* \Big) \\
&+i\hbar\frac{d }{dt}\psi^{v,c}_{\bk}\Big|_{\textrm{D, c-c}}
+i\hbar\frac{d }{dt}\psi^{v,c}_{\bk}\Big|_{\textrm{D, c-ph}}\,,
\end{split}
\label{eq:EOM_SBE}
\end{equation}
where coupling to doublets (D) introduces additional scattering and dephasing processes.
The microscopic polarizations couple to singlets $\ev{D^{\dagger}_{\bn,j}}$, which represent coherent phonons \cite{kira_many-body_2006} with the EOM (\ref{eq:EOM_D}).
Carrier-phonon interaction matrix elements $g$ are defined in Eq.~(\ref{eq:carr_phon_ME_bloch}).
Hartree-Fock renormalizations of electronic states due to photoexcited electrons and holes are included in $\ent{\lambda}{\bk}$, see Eq.~(\ref{eq:def_eps_tilde}), while Coulomb-induced renormalizations of light-matter coupling are contained in the Rabi energy
\begin{equation}
\begin{split}
\Omega_{\bk}^{\nu,\bar{\nu}}=\boldsymbol{d}^{\bar{\nu},\nu}_{\bk}\cdot \boldsymbol{E}^{\textrm{2d}}+\sum_{\bk',\lambda,\lambda'}
V^{\bar{\nu},\lambda,\nu,\lambda'}_{\bk, \bk', \bk, \bk'}\psi_{\bk'}^{\lambda,\lambda'}
\,.
\end{split}
\label{eq:def_Rabi_main}
\end{equation}
Here we assume that only momentum-direct optical transitions are driven due to small photon momenta in the optical spectrum.
Light-matter interaction matrix elements $\boldsymbol{d}$ are defined in Eq.~(\ref{eq:dip_ME}), while Coulomb interaction matrix elements $V$ are given by Eq.~(\ref{eq:Coul_ME}).
The SBE also describe coupling of the single-particle density matrix to doublets, e.g. via
\begin{equation}
\begin{split}
\frac{d }{dt}f^{\nu}_{\bk}\Big|_{\textrm{D, c-c}}&=\frac{2}{\hbar}\,\textrm{Im}\Big\{ \sum_{\substack{\bk',\bq \\ \beta,\lambda,\lambda'}} 
V^{\nu,\lambda,\lambda',\beta}_{\bk, \bk', \bk'+\bq, \bk-\bq} c_{\nu,\lambda,\lambda'\beta}^{\bq,\bk',\bk}  \Big\}\,, \\
\frac{d }{dt}f^{\nu}_{\bk}\Big|_{\textrm{D, c-ph}}&=\frac{2}{\hbar}\,\textrm{Im}\Big\{\sum_{\bq,j,\lambda}g^{\nu,\lambda}_{\bk,\bq,j}\Big(   
\big(\Xi^{\lambda,\nu}_{\bk,\bq,j}\big)^*+\Xi^{\nu,\lambda}_{\bk-\bq,-\bq,j}
\Big) \Big\}\,.
\end{split}
\label{eq:SBE_doublets}
\end{equation}
The so-called phonon-assisted polarization
\begin{equation}
\begin{split}
\Xi^{\nu,\nu'}_{\bk,\bq,j}= 
\Delta\ev{D^{\dagger}_{\bq,j} a_{\bk-\bq,\nu}^{\dagger}a_{\bk,\nu'}^{\phantom\dagger}}
\end{split}
\label{eq:def_Xi_main}
\end{equation}
and the carrier-carrier correlation
\begin{equation}
\begin{split}
c_{\lambda_1,\lambda_2,\lambda_3\lambda_4}^{\bq,\bk',\bk}= 
\Delta\ev{a_{\bk,\lambda_1}^{\dagger}a_{\bk',\lambda_2}^{\dagger}a_{\bk'+\bq,\lambda_3}^{\phantom\dagger}a_{\bk-\bq,\lambda_4}^{\phantom\dagger}  }
\end{split}
\label{eq:def_C_main}
\end{equation}
fulfill their own EOM Eq.~(\ref{eq:EOM_Xi}) and (\ref{eq:EOM_C}), respectively. 
We note that by including the coupling of carrier-carrier doublets to pure singlet factorizations on the triplet level, see Fig.~\ref{fig:sketch_cluster}, we capture carrier-carrier scattering 
on the single-particle level, which becomes significant at elevated carrier densities.
In the limit of low excited-carrier density, the carrier-carrier correlations are connected to 
the correlated densities of two-particle states via the projections \cite{kira_many-body_2006}
\begin{equation}
\begin{split}
n^{\nu}_{\mathrm{corr}}=\frac{1}{\mathcal{A}}\sum_{\bk,\bk',\bq,c,v,c',v'}
\phi^{v'c'}_{\nu,\bq}(\bk')\big(\phi^{vc}_{\nu,\bq}(\bk)\big)^*
c_{v,c',v',c}^{\bq,\bk'-\bq,\bk} \,,
\end{split}
\label{eq:n_corr_proj}
\end{equation}
where the two-particle wave functions $\phi^{vc}_{\nu,\bq}(\bk)$ fulfill a Bethe-Salpeter equation (\ref{eq:BSE}).
Summation over all two-particle states yields the total correlated two-particle density:
\begin{equation}
\begin{split}
n_{\mathrm{corr, tot}}=\sum_{\nu}n^{\nu}_{\mathrm{corr}}=\frac{1}{\mathcal{A}}\sum_{\bk,\bq,c,v} c_{v,c,v,c}^{\bq,\bk-\bq,\bk}\,.
\end{split}
\label{eq:n_corr}
\end{equation}
A distribution function $n_{\mathrm{corr, tot}}(\bq)$ for two-particle correlations can be defined according to
\begin{equation}
\begin{split}
n_{\mathrm{corr, tot}}=\frac{1}{\mathcal{A}}\sum_{\bq} n_{\mathrm{corr, tot}}(\bq)\,.
\end{split}
\label{eq:n_corr_q}
\end{equation}
\\
\\Phonon-assisted polarizations couple to two-phonon-assisted polarizations
\begin{equation}
\begin{split}
S^{(-),1,2}_{\bk,\bq,\bq',j,j'}&=\Delta\ev{D^{\dagger}_{\bq,j}D^{\phantom\dagger}_{\bq',j'}a^{\dagger}_{\bk-\bq+\bq',1}a^{\phantom\dagger}_{\bk,2}}\,, \\
S^{(+),1,2}_{\bk,\bq,\bq',j,j'}&=\Delta\ev{D^{\dagger}_{\bq,j}D^{\dagger}_{-\bq',j'}a^{\dagger}_{\bk-\bq+\bq',1}a^{\phantom\dagger}_{\bk,2}}
\end{split}
\label{eq:def_DDaa_main}
\end{equation}
via carrier-phonon interaction and to carrier-carrier-phonon correlations  
\begin{equation}
T^{\bl,\bq,\bk',\bk}_{1,2,3,4,j}=\Delta\ev{D^{\phantom\dagger}_{\bl,j}a_{\bk,1}^{\dagger}a_{\bk',2}^{\dagger}a_{\bk'+\bq,3}^{\phantom\dagger}a_{\bk-\bq-\bl,4}^{\phantom\dagger}}
\label{eq:def_Daaaa_main}
\end{equation}
via Coulomb interaction. Also, the carrier-carrier correlations couple to carrier-carrier-phonon correlations via carrier-phonon interaction and to pure carrier triplets via Coulomb interaction. Since we focus on the low-density regime in this work, we discard the coupling to pure carrier triplets via carrier-carrier Coulomb interaction and truncate the hierarchy of EOM in this direction.
The coupling terms to triplet correlations cause scattering and dephasing of doublets, which is essential for a proper description of doublet dynamics, as we will show later on. Triplets in turn experience scattering and dephasing contributions from quadruplets. As shown in the Appendix, we treat the coupling to quadruplets in Markov approximation.
\\
\\Via the phonon-assisted polarizations, the single-particle density matrix is coupled to phonon doublets $\Delta\ev{D^{\dagger}_{\bq,j}D^{\phantom\dagger}_{\bq,j'}}$ and $\Delta\ev{D^{\dagger}_{\bq,j}D^{\dagger}_{-\bq,j'}} $, the first one being connected to the phonon occupation $n_{\bq,j}=\ev{D^{\dagger}_{\bq,j}D^{\phantom\dagger}_{\bq,j}}$.
The EOM for the density-like phonon doublets are:
\begin{equation}
\begin{split}
&i\hbar\frac{d }{dt} \Delta\ev{D^{\dagger}_{\bq,j}D^{\phantom\dagger}_{\bq,j'}} = (\hbar\omega_{\bq,j'}-\hbar\omega_{\bq,j})\Delta\ev{D^{\dagger}_{\bq,j}D^{\phantom\dagger}_{\bq,j'}} \\ 
&-i\hbar2\gamma_{\textrm{phon}}(\Delta\ev{D^{\dagger}_{\bq,j}D^{\phantom\dagger}_{\bq,j'}}-n^{\textrm{B}}_{\bq,j}\delta_{j,j'}) \\
&+\sum_{\bk,\lambda,\lambda'} \Big(g^{\lambda,\lambda'}_{\bk,-\bq,j'}\Xi^{\lambda,\lambda'}_{\bk+\bq,\bq,j}
-g^{\lambda,\lambda'}_{\bk,\bq,j}\big(\Xi^{\lambda',\lambda}_{\bk,\bq,j'}\big)^*
\Big)
\,.
\end{split}
\label{eq:EOM_DD_main}
\end{equation}
As shown in the Appendix, we use the Lindblad formalism to introduce a phenomenological phonon damping rate $\gamma_{\textrm{phon}}$, which is due to lattice anharmonicities. As a consequence, phonon occupations $n_{\bq,j}$ relax towards Bose functions at the bath temperature with the relaxation time $\tau=1/(2\gamma_{\textrm{phon}})$. 
The coupling among correlations is schematically summarized in Fig.~\ref{fig:sketch_cluster}.
\\
\\In the following, we solve the cluster equations in two regimes of light-matter interaction to obtain complementary results, assuming perpendicular incidence of the electromagnetic field to the TMD layer. To describe a realistic experimental setup, we embed the TMD in a dielectric heterostructure.
(i) The linear optical response in the absence of electron-hole pairs is probed by exciting the semiconductor with a weak, spectrally broad electric field. The probe field drives inter-band polarizations, which correspond to coherent or ``virtual'' excitons for energies below the bandgap. The optical response is encoded in the Fourier-transformed macroscopic polarization $\boldsymbol{P}^{\textrm{2d}}(t)$, from which the absorption can be computed for a given dielectric heterostructure using transfer matrices, see Appendix~\ref{sec:linear_optics}. Since we are interested in the zero-density regime, no correlations involving carrier operators on the doublet level need to be involved.
(ii) The nonlinear dynamics of photoexcited carriers and phonons is investigated by applying a strong, spectrally narrow pump field. To capture the backaction of the carriers onto the electromagnetic field, we solve the cluster equations and the semiclassical wave equation self-consistently, which also gives access to the propagation of the pump pulse across the heterostructure. The numerical procedures are described in detail in the Appendix~\ref{sec:nonlinear_optics}. In this case, a full description involves carrier doublets and the corresponding phonon-assisted triplets. A certain subset of carrier doublets corresponds to incoherent excitons, as opposed to the coherent excitonic singlets. The interplay between coherent and incoherent regimes will be discussed later on. 
%
\begin{figure*}
\centering
\includegraphics[width=1.\textwidth]{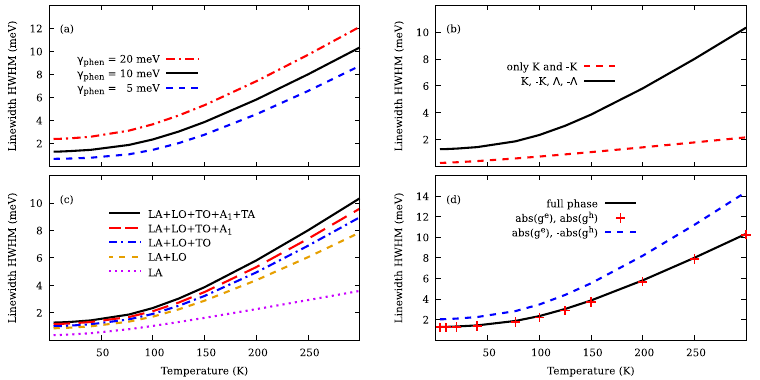}
\caption{Perturbative phonon-induced broadening of A$_{\textrm{1s}}$-excitons in hBN-encapsulated monolayer MoSe$_2$ as a function of temperature. (a): Different quasi-particle broadenings $\gamma_{\textrm{phen}}$ using all phonon branches and $K$,$-K$ as well as $\Lambda$,$-\Lambda$-valleys. (b): Effect of scattering between $K$,$-K$ and $\Lambda$,$-\Lambda$-valleys using all phonon branches and a broadening $\gamma_{\textrm{phen}}=10$ meV. (c): Effect of different phonon branches using all valleys and $\gamma_{\textrm{phen}}=10$ meV. (d): Influence of phase factors in the electron-phonon and hole-phonon scattering matrix elements using all phonon branches, all valleys and $\gamma_{\textrm{phen}}=10$ meV.}
\label{fig:PT}
\end{figure*}
\section{Results}

We study the coupled carrier-phonon dynamics in hBN-encapsulated monolayer MoSe$_2$, which is a direct semiconductor with an optically bright (excitonic) ground state. The used single-particle band structure, phonon dispersion and interaction matrix elements are provided in Appendix~\ref{sec:abinitio} and \ref{sec:abinitio_cp}. To simplify the numerical treatment, we limit our calculations to one conduction and valence band per spin, see Appendix~\ref{sec:numerical} for further details. The dielectric heterostructure, in which the TMD is embedded, is characterized in Fig.~\ref{fig:sketch_HS}.

\subsection{Perturbation theory for exciton broadening}

To quantify the influence of various carrier and phonon degrees of freedom, we study the efficiency of exciton-phonon coupling in first-order perturbation theory. To this end, we transform the EOM for the inter-band polarizations $\psi^{v,c}_{\bk}$ and the phonon-assisted polarizations $\Xi^{\nu,\nu'}_{\bk,\bq,j}$ into a representation of two-particle eigenstates, see Appendix~\ref{sec:perturbation}. 
By adiabatically eliminating the phonon-assisted polarizations, we obtain explicit expressions for the phonon-induced exciton broadening (\ref{eq:EOM_psi_X_Born_Markov_coeff}) similar to those derived in Refs.~ \cite{selig_dark_2018,chan_exciton_2023}.
We emphasize that, unlike the usual notion of perturbation theory, we include a finite phenomenological damping $\gamma_{\textrm{phen}}$ of the energy denominator. Since truncation of the cluster hierarchy necessitates a phenomenological damping of top-level correlations (e.g., of doublets, if triplets are neglected), we thus establish a certain comparability between perturbation theory and cluster expansion.
The results for phonon-induced broadening of A$_{\textrm{1s}}$-excitons are collected in Fig.~\ref{fig:PT}. Panel (a) shows the impact of 
$\gamma_{\textrm{phen}}$.
While the temperature dependence is not influenced by $\gamma_{\textrm{phen}}$, the latter introduces a temperature-independent offset to the exciton linewidth due to an artificial increase of the number of scattering channels.
In the following, we choose $\gamma_{\textrm{phen}}=10$ meV. As Fig.~\ref{fig:PT}(b) shows, the most significant impact on the exciton linewidth (of excitons residing in the $K$- and $-K$-valleys) stems from scattering processes involving Bloch states from the $\Lambda$- and $-\Lambda$-valleys (the latter being halfway between $\Gamma$ and $K$/ $-K$, respectively, as shown in Fig.~\ref{fig:DFT_bands}). As a conclusion, all the valleys need to be considered in the following calculations. We note that the precise numbers depend on the lattice constant used in the DFT band-structure calculations, which determines the energetic difference between the valleys and thereby possible phonon-assisted scattering channels. The impact of inter-valley scattering has been discussed in detail in Refs.~\cite{selig_excitonic_2016,selig_dark_2018}.
The contributions of the different phonon branches \cite{kaasbjerg_phonon-limited_2012}, which are additive in first-order perturbation theory, are displayed in Fig~\ref{fig:PT}(c). The strongest contributions are given by polar longitudinal acoustic (LA) and polar longitudinal optical (LO) phonons, see the phonon dispersion in Fig.~\ref{fig:phonon_dispersion}. Evidence for efficient LA-assisted inter-valley scattering in MoSe$_2$ is given by photoluminescence excitation \cite{shree_observation_2018} and pump-probe spectroscopy \cite{bae_k-point_2022} experiments.
Corrections are given by transverse acoustic (TA), polar transverse optical (TA) and out-of-plane optical ($A_1$) modes. The latter have been identified experimentally to couple efficiently to excitons in MoSe2 \cite{li_excitonphonon_2021}. The remaining branches have no effect at all on the exciton linewidth. To keep the following numerical treatment as simple as possible, we limit ourselves to LA and LO phonons for the solution of the cluster equations.
Exciton-phonon interaction is sensitive to effective matrix elements that involve weighted differences between electron-phonon and hole-phonon interaction matrix elements, see Eq.~(\ref{eq:eff_elph_ME}). This makes the exciton linewidth in principle sensitive to complex phases of the first-principle matrix elements $g^e$ and $g^h$. The impact of these phase factors is shown in Fig~\ref{fig:PT}(d). It turns out that using the modulus of $g^e$ and $g^h$, respectively, is already a very good approximation for monolayer MoSe$_2$. Only an artificial opposite phase between the two contributions would lead to a significant increase of scattering efficiency.
The small impact of relative phases between electron and hole scattering events can be understood from the fact that the dominant scattering processes involve inter-valley transitions in the conduction band only, where $g^h$ and hence its complex phase plays no role.
%
\begin{figure}
\centering
\includegraphics[width=1.\columnwidth]{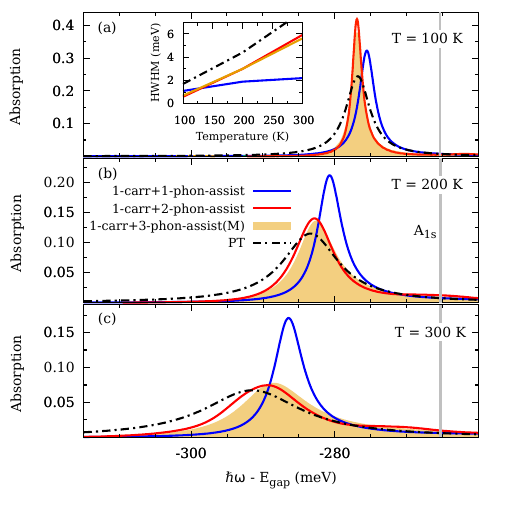}
\caption{A$_{\textrm{1s}}$-exciton absorption of hBN-encapsulated monolayer MoSe$_2$ at different temperatures.  
Different levels of approximation including polarizations assisted by one, two, or three phonons are denoted by \textit{1-carr+n-phon-assist}, see the scheme in Fig.~\ref{fig:sketch_cluster}. Results from the highest level of approximation are emphasized by filled curves. 
Additionally, perturbative results (PT) using a phenomenological broadening $\gamma_{\textrm{phen}}=10$ meV are shown. The vertical line indicates the A$_{\textrm{1s}}$-exciton energy without carrier-phonon interaction. Optical transition energies $\hbar\omega$ are shown relative to the band gap $E_{\textrm{gap}}$. The inset summarizes the HWHM linewidth of A$_{\textrm{1s}}$-excitons as extracted from the linear susceptibility. The perturbative results (PT) correspond to the "LA+LO"-curve in Fig.~\ref{fig:PT}(c).}
\label{fig:absorption}
\end{figure}

\subsection{Linear optical response}

Next we discuss the linear optical response of monolayer MoSe$_2$ under the influence of carrier-phonon interaction. To this end, we solve the cluster equations at different temperatures
using a weak, spectrally broad electric probe field. We assume that the non-excited system is probed. As the weak probe field excites no incoherent carriers, we can discard all correlations involving carrier operators on the two-particle level. This means that we rely on subsets of clusters denoted by \textit{1-carr}, see Fig.~\ref{fig:sketch_cluster}. For each temperature, we compare in Fig.~\ref{fig:absorption} calculations taking into account one-phonon-, two-phonon- and three-phonon-assisted polarizations, denoted by \textit{n-phon-assist}, as well as results from perturbation theory as discussed in the previous section. The level of one-phonon correlations is obtained by solving the SBE, Eq.~(\ref{eq:EOM_SBE}), together with the EOM for the phonon-assisted polarizations (\ref{eq:EOM_Xi}). The cluster hierarchy is truncated by discarding triplet correlations. Instead, dephasing of phonon-assisted polarizations is phenomenologically described by a damping term $\frac{d }{dt}\Xi^{\nu,\nu'}_{\bk,\bq,j}\Big|_{\textrm{phen}}=-\gamma_{\textrm{phen}}\Xi^{\nu,\nu'}_{\bk,\bq,j} $ with $\gamma_{\textrm{phen}}= 10$ meV, similar to the perturbative treatment. The next step is the inclusion of two-phonon-assisted polarizations, which are triplets, as a source of scattering and dephasing in the $\Xi$-equations. The hierarchy is truncated by discarding quadruplet correlations in Eq.~(\ref{eq:EOM_DDaa}), i.e., the two-phonon-assisted polarizations are damped by a phenomenological term. Finally, three-phonon-correlations are taken into account by including quadruplets in Markov approximation, see Eqs.~(\ref{eq:EOM_DDaa_Q}) and (\ref{eq:EOM_DDaa_Q_final}). This means that three-phonon-assisted polarizations are not obtained from their own EOM, but follow the evolution of lower-order correlations adiabatically, with a damping $\gamma_{\textrm{phen}}$. Therefore, scattering with the ``third'' phonon can be treated only in the single-particle picture. Nevertheless, this treatment represents a significant advancement over the phenomenological damping of two-phonon triplets, since scattering and dephasing terms of triplets have a proper momentum dependence.
On the level of perturbation theory, an analytic expression for the optical susceptibility can be derived, see Eq.~(\ref{eq:chi_PT}), for which we also use $\gamma_{\textrm{phen}}= 10$ meV. 
\\
\\For the A$_\textrm{1s}$-excitons, we find significant deviations between cluster expansion calculations on the one-phonon- and two-phonon level. For $T=200$ K and $T=300$ K, the one-phonon calculation underestimates the efficiency of exciton-phonon coupling, with the deviations becoming larger at increasing temperature. In particular, the HWHM of the zero phonon line (ZPL) as well as the polaron shift (peak position vs. grey vertical line) increases and the phonon sideband (PSB) becomes more pronounced. This means that higher-order scattering processes of K- and -K-valley-excitons with two phonons open up new relaxation channels that decrease the exciton lifetime significantly. Considering the efficiency of inter-valley processes extracted from the perturbative calculation, these higher-order processes involve the $\Lambda$- and $-\Lambda$-valleys. At $T=100$ K, we find the opposite effect, with the one-phonon calculation overestimating the linewidth.
The main difference between two- and three-phonon-calculations is that the latter yield a stronger spectral smearing of the PSB.
In the one-phonon calculation, second-order scattering processes are mimicked via the phenomenological damping $\gamma_{\textrm{phen}}$, which corresponds to a spectral broadening of the phonon-assisted polarization. Since we choose a fixed value of $\gamma_{\textrm{phen}}$ for all temperatures, it depends to some degree on this specific value whether the one-phonon calculation over- or underestimates exciton broadening at a given temperature. However, the obvious convergence of results from the doublet to the triplet to the quadruplet level shows that the arbitrary choice of $\gamma_{\textrm{phen}}$ matters less and less for observables, the higher above the hierarchy is truncated.
Results for the B-exciton are provided in Appendix~\ref{sec:abs_B}.
Similar effects of higher-order processes on the excitonic absorption have been found based on a cluster expansion in the exciton representation using model interaction matrix elements \cite{lengers_theory_2020}. There it was shown that two-phonon processes (termed 4th order Born approximation) yield a strong correction, while three-phonon processes (= damped 4th order Born approximation) introduce an additional smoothing of the absorption lineshape. The PSB of MoSe$_2$ has also been studied in Ref.~\cite{christiansen_phonon_2017}, with two-phonon processes treated in Markov approximation. 
\begin{figure*}
\centering
\includegraphics[width=1.\textwidth]{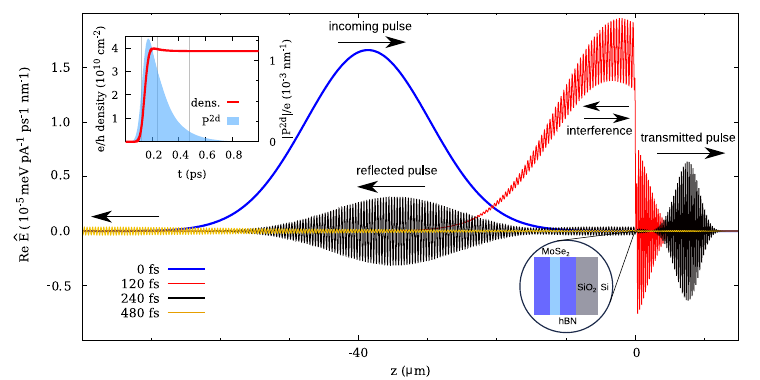}
\caption{Propagation of an external light pulse through a dielectric heterostructure hosting a monolayer of MoSe$_2$. Shown is the spatial dependence of (real part of) electric field $\hat{E}(z)$ in the rotating frame at different times for resonant pump of the A$_{\textrm{1s}}$-exciton with fluence $0.1$ $\mu$J cm$^{-2}$ including correlations involving one carrier and two phonons (\textit{1-carr+2-phon-assist}, as in Fig.\ref{fig:phonon_dyn1}(b)) at $T=300$ K. The heterostructure, with monolayer MoSe$_2$ located at $z=0$, is schematically shown in the zoom-in and specified in detail in Fig.~\ref{fig:sketch_HS}. The inset shows the electron-hole pair density $n^{e/h}$ and the modulus of the macroscopic polarization $P^{\textrm{2d}}$ (normalized by the elementary charge) as functions of time, with the snapshots of the electric field marked by vertical grey lines.}
\label{fig:E_field}
\end{figure*}
\\
\\To translate our calculated relative exciton energies to absolute numbers, we note that the gap energy including GdW corrections due to substrate screening (see Appendix \ref{sec:abinitio}) is $E_{\textrm{gap}}=1890$ meV. Hence the A-exciton and B-exciton energies at $T=300$ K are $1600$ meV and $1800$ meV, respectively. These numbers match well with the experimental observation of excitonic resonances at room temperature at $1.57$ eV and $1.75$ eV \cite{li_excitonphonon_2021}. Further corrections to the exciton energies are expected from temperature-dependent changes of the lattice constant, which are not included in our calculation.
\\
\\We now compare the one-phonon calculation to the perturbative results. The difference is that the former involves the full time evolution of phonon-assisted polarizations, while the latter results from an adiabatic elimination of phonon-assisted polarizations by means of a Markov approximation. This means that the perturbative calculation neglects non-Markovian effects, replacing the frequency-dependent exciton broadening by a frequency-independent (Lorentzian) one. 
Perturbation theory can not explicitly capture PSB effects, but maps them effectively into an additional Lorentzian broadening. As a result, the HWHM exciton linewidth extracted from the linear one-phonon spectra is much smaller than that predicted by perturbation theory, see the inset of Fig.~\ref{fig:absorption}(a). The effect of high-energy phonons on the FWHM is overestimated by perturbation theory, while the effect of low-energy acoustic phonons is captured quite well. Therefore, the HWHM extracted from the spectrum more or less matches the perturbative result for only $K$- and $-K$-valleys shown in Fig.~\ref{fig:PT}(b). Remarkably, at elevated temperatures perturbation theory appears to be more accurate than the cluster expansion on doublet level, if the quadruplet (3-phonon) calculation is taken as a reference. We interpret this as a coincidence: The overestimation of the HWHM linewidth by perturbation theory accidentally mimics the significant increase of linewidth due to higher-order exciton-phonon scattering processes. 
\\
\\Experimental results on the temperature-dependent exciton linewidth in monolayer MoSe$_2$ have been obtained via combined reflectance and photoluminescence spectroscopy \cite{selig_excitonic_2016}, four-wave mixing spectroscopy \cite{jakubczyk_radiatively_2016} and multidimensional coherent spectroscopy \cite{martin_encapsulation_2020}. Only the latter experiment has been performed on hBN-encapsulated monolayers, which clearly leads to a reduction of the observed linewidths. Even more, as demonstrated in \cite{martin_encapsulation_2020}, even for high-quality encapsulated samples inhomogeneous and homogeneous contributions to the linewidth have to be carefully disentangled. Finally, as discussed above, the exciton is in general not characterized by a single linewidth parameter, but exhibits asymmetric PSB effects.
On the theory side, computational resources are often overstretched by converging the many-body hierarchy per se and its numerical solution at the same time. A quantitative comparison of the results in Fig.~\ref{fig:absorption} with the above experiments shows that our calculations underestimate phonon-induced broadening as measured by the HWHM: The linewidth reported in \cite{martin_encapsulation_2020} for $T=60$ K is comparable to our prediction for $T=100$ K. Assuming an approximately linear increase with temperature in this regime, this translates to an underestimation by a factor $\approx 2$.
It is plausible that our theory yields a lower bound for the phonon-induced exciton broadening for two reasons, which are imposed by the high computational demand of the method. First, we have limited ourselves to two dominant phonon branches. Second, we have reduced the available phase space by taking into account Bloch states in the vicinity of high-symmetry points. Although our method is nonlinear in terms of exciton-phonon scattering, it is likely that a reduction of scattering channels leads to an overall reduction of scattering efficiency.
\\
\\We relate our approach to the method used for monolayer MoS$_2$ in Ref.~\cite{chan_exciton_2023}. There, exciton-phonon scattering is evaluated perturbatively using energy-conserving delta functions, i.e., in the limit of vanishing phenomenologial broadening $\gamma_{\textrm{phen}}$. 
Chan et al. make the point that second-order phonon-photon scattering processes significantly increase exciton lifetimes, encoded within off-diagonal contributions to the exciton self-energy. Such off-diagonal terms correspond to off-diagonal coefficients in the Markov description in Eqs.~(\ref{eq:EOM_psi_X_Born_Markov}),(\ref{eq:EOM_psi_X_Born_Markov_coeff}). 
These terms are naturally included in our full computational approach, but not in the perturbative results shown in Fig.~\ref{fig:PT}.
Besides this, we argue that the assumption of exact energy conservation might underestimate the exciton linewidth due to the energetic distance of the $K$-valley to the $\Lambda$- and $-\Lambda$-valleys. The latter provide efficient scattering channels for K-valley excitons via absorption of a phonon. These scattering channels become (more) accessible in the presence of higher-order phonon-assisted processes, which remove strict energy conservation between an exciton and a single phonon. As discussed above, higher-order processes introduce microscopic dephasing of the phonon-assisted polarizations $\Xi$, while perfect energy conservation corresponds to infinitesimal dephasing. Another effect of the additional broadening due to higher-order processes is to reduce the sensitivity of exciton-phonon scattering to the fine structure L-T splitting introduced by electron-hole exchange \cite{qiu_nonanalyticity_2015}. 
\begin{figure*}
\centering
\includegraphics[width=1.\textwidth]{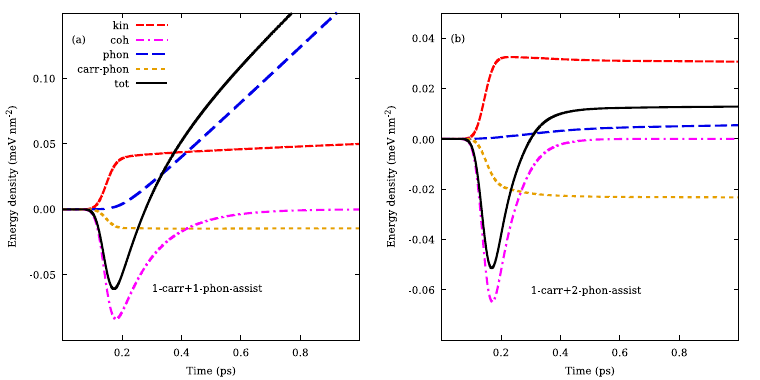}
\caption{Energy density of hBN-encapsulated monolayer MoSe$_2$ after resonant excitation of A$_{\textrm{1s}}$ with fluence $0.1$ $\mu$J cm$^{-2}$ at temperature $T=300$ K using correlations involving one carrier. We show the kinetic (kin, Eq.~(\ref{eq:energy_kin})), coherent (coh, Eq.~(\ref{eq:energy_coh})), phonon (phon, Eq.~(\ref{eq:energy_phon})), and carrier-phonon (c-ph, Eq.~(\ref{eq:energy_carr_phon})) contributions to the energy density as well as the total result (tot). (a) includes carrier singlets, pure phonon correlations and phonon-assisted polarizations (\textit{1-carr+1-phon-assist}). (b) additionally includes two-phonon-assisted polarizations (\textit{1-carr+2-phon-assist}).}
\label{fig:phonon_dyn1}
\end{figure*}

\subsection{Nonequilibrium dynamics of carrier singlets}

As a complementary experimental situation, we discuss the case of resonant exciton pumping with a relatively strong, spectrally broad electric field.
We choose a Gaussian envelope with an intensity FWHM of $\tau_{\textrm{pump}}=50$ fs (corresponding to $\sqrt{2}\times 50$ fs field FWHM) and circular polarization $\boldsymbol{e}_{\textrm{P}}=1/\sqrt{2}(\boldsymbol{e}_x+i\boldsymbol{e}_y)$ that selects the $K$-valley. 
The pump fluence is $0.1$ $\mu$J cm$^{-2}$.
To study the influence of different classes of correlation functions on the dynamics, we perform calculations using the cluster hierarchy at various levels of truncation, see Fig.~\ref{fig:sketch_cluster}. First, we stick to carrier correlations on the singlet level as discussed for the case of linear optical spectra. This means that we either solve 
the SBE together with the EOM for the phonon-assisted polarizations with a phenomenologial damping of the latter (\textit{1-carr+1-phon-assist}), or we include two-phonon-assisted polarizations as a source of microscopic scattering and dephasing in the $\Xi$-equations (\textit{1-carr+2-phon-assist}). In both cases, EOM for phonon singlets, Eq.~(\ref{eq:EOM_D}), and pure phonon doublets, Eq.~(\ref{eq:EOM_DD_main}), are included to capture the dynamics of phonons. As we will see in the following, the phonon doublets are particularly sensitive to the truncation of the cluster hierarchy.
\begin{figure*}
\centering
\includegraphics[width=1.\textwidth]{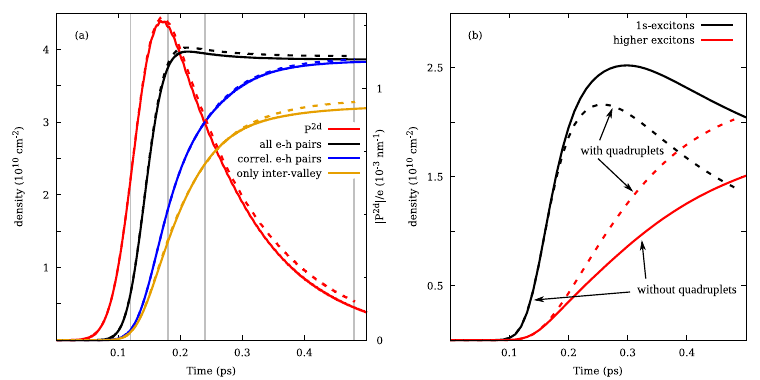}
\caption{Different contributions to electron-hole pair density of hBN-encapsulated monolayer MoSe$_2$ after resonant excitation of A$_{\textrm{1s}}$ with fluence $0.1$ $\mu$J cm$^{-2}$ at temperature $T=300$ K using correlations involving up to two carriers. Solid lines: without quadruplets (\textit{2-carr+2-phon-assist}), dashed lines: with quadruplets (\textit{2-carr+2-carr-2-phon+3-phon-assist (Markov)}) (a) Comparison of total density $n^{e/h}$ (Eq.~(\ref{eq:carr_dens_band_main})) (all e-h pairs), correlated contribution $n_{\mathrm{corr,tot}}$ ( (Eq.~(\ref{eq:n_corr})) (correl. e-h pairs) and inter-valley correlated contribution (only inter-valley). As an orientation, we also show the modulus of the macroscopic polarization $P^{\textrm{2d}}$ (normalized to the elementary charge) and mark the times $t=120$ fs, $t=180$ fs, $t=240$ fs, and $t=480$ fs by vertical grey lines for comparison with Fig.~\ref{fig:E_field}. (b) Contributions of 1s-excitons and higher excitons to the correlated part of the density (includes both, intra- and inter-valley correlations).}
\label{fig:s_d_dens}
\end{figure*}
\\
\\We begin by demonstrating the interplay of the electromagnetic field and the TMD carriers as obtained by solving the coupled cluster and Maxwell equations. We include two-phonon-assisted polarizations in this case.
The spatio-temporal evolution of the electric field is shown in Fig.~\ref{fig:E_field}. Note that oscillations with the carrier frequency are transformed out by means of a rotating frame, leaving only the pulse envelope for the initial field at time $t=0$. After $120$ fs, about half of the pulse has propagated through the heterostructure. To the left of the heterostructure, oscillations are caused by the interference of incident and reflected fields. On the right, oscillations of the transmitted field are due to the higher refractive index of the Si substrate.
The corresponding carrier dynamics can be observed in the inset of Fig.~\ref{fig:E_field}. The macroscopic polarization of the TMD layer has reached about half of its maximum value after $120$ fs. The buildup of excited-carrier density, being of at least second order in the electric field, is slightly delayed. At $t=240$ fs, the pulse has passed the heterostructure, corresponding to the carrier density having reached its maximum value. At this point, the polarization is already decaying due to two separate processes. On the one hand, dephasing processes induced by carrier-phonon interaction drive a transition from a coherent to an incoherent state. 
The dephasing is described microscopically by the coupling of coherent singlets $\psi$ to phonon-assisted polarizations $\Xi$ (doublets) in the SBE (\ref{eq:EOM_SBE}). On the other hand, the very coupling to the electromagnetic field causes a radiative decay of the polarization \cite{kira_many-body_2006}. This process becomes visible as on ongoing stream of electric field to both sides of the TMD layer even after the original pulse has passed. At $t=480$ fs, the electric field has almost vanished around the heterostructure.
\\
\\The coupling between excited-carrier and phonon dynamics can be visualized via the time-dependence of different contributions to the energy density:
The kinetic energy density $\mathcal{E}_{\textrm{kin}}$ is the contribution due to single-particle populations including Hartree-Fock-type energy renormalizations, while the inter-band polarizations yield a coherent single-particle contribution $\mathcal{E}_{\textrm{coh}}$. The correlated two-particle energy density splits into intra-band contributions $\mathcal{E}_{\textrm{corr,e}}$ and $\mathcal{E}_{\textrm{corr,h}}$ and the inter-band contribution $\mathcal{E}_{\textrm{corr,X}}$. The energy density of the phonons (relative to the thermal state) is $\mathcal{E}_{\textrm{phon}}$. The energy density stored in the carrier-phonon interaction is given by $\mathcal{E}_{\textrm{c-ph}}$. Details are given in Appendix~\ref{sec:energy_dens}. In Fig.~\ref{fig:phonon_dyn1} we compare the dynamics including (a) (artificially damped) phonon-assisted polarizations only and (b) including one- and two-phonon-assisted polarizations together. The most striking observation is the unlimited increase of the phonon energy in the former case. At the same time, the kinetic energy of carriers increases as well, which clearly leads to a violation of energy conservation. The reason for this artificial heating of the system is the Lorentzian spectral broadening of phonon-assisted polarizations caused by their phenomenological damping, see also the discussions in \cite{bonitz_non-lorentzian_1999,erben_optical_2022}. The Lorentzian broadening leads to long-range spectral tails that open up artificial carrier-phonon scattering channels. As soon as we replace the phenomenological damping by a coupling to two-phonon-assisted polarizations, carrier-phonon interaction leads to a well-behaved exchange of energy between the carrier and phonon subsystems, such that the total energy density is conserved after the external electric field has passed (around $t=250$ fs). We observe a slight cooling of carriers and corresponding heating of the phonons after optical excitation. This example shows how a microscopic description of scattering and dephasing, in this case by coupling of phonon-assisted doublets to phonon-assisted triplets, is necessary to ensure a physical behavior of the nonequilibrium system. One could say that, on a given level of the cluster hierarchy, coupling to the next higher clusters provides a proper spectral ``lineshape'' for scattering processes. 
The lineshape problem could in principle also be mitigated by 
using delta functions (or other model functions with no long-range tails) as lineshape functions, such as in Fermi's golden rule. The latter emerges from a strictly perturbative treatment of carrier-phonon interaction and yields scattering with exact energy conservation between unrenormalized energies.
We would like to emphasize that in the time-domain formulation of the cluster hierarchy, a spectral lineshape function never appears explicitly, but only in a Fourier sense develops from the dynamics in the long-time limit.
A lineshape function can only be applied explicitly when a correlation is eliminated adiabatically in Markov approximation, see Eq.~(\ref{eq:EOM_Q_dummy_sol_2}).
From Fig.~\ref{fig:phonon_dyn1}(b), we also find that the dephasing of interband coherences is faster in the case of two-phonon-assisted polarizations, which is consistent with a larger broadening of A$_{\textrm{1s}}$ in the linear spectrum, see Fig.~\ref{fig:absorption}(c). We also note that carrier-phonon interaction leads to a lowering of the energy density.
Since the above description takes place in terms of correlations involving only one carrier, resonantly pumped excitons are described as projections onto single-particle populations instead of ``true'' exciton populations. The discussion of energy density therefore only contains a kinetic contribution, but no direct signature of the exciton binding energy. This is changed in the next step, where we also include correlations involving two carriers.

\subsection{Nonequilibrium dynamics of carrier doublets}

Excitons are properly taken into account by including the EOM of carrier doublets, or carrier-carrier correlations, Eq.~(\ref{eq:EOM_C}). The main source of scattering and dephasing of carrier doublets in the low-density regime is provided by the coupling to carrier-carrier-phonon triplets $T$, whose dynamics is governed by the EOM (\ref{eq:EOM_Daaaa}). At the same time, these triplets open up scattering channels for phonon-assisted polarizations via carrier-carrier Coulomb interaction. By adding this whole complex of clusters to the subset of correlations involving one carrier, we obtain the approximation level \textit{2-carr+2-phon-assist}, see the scheme in Fig.~\ref{fig:sketch_cluster}. The mechanism for exciton formation after resonant optical excitation has been discussed in literature \cite{thranhardt_quantum_2000,kira_many-body_2006,selig_dark_2018}. Incoherent excitons, represented by the doublets $C_{v,c,v,c}$, emerge out of coherent excitons $\psi^{v,c}$ via carrier-phonon interaction. Microscopically, this process is described by the singlet-doublet source terms proportional to $\psi$ and $\Xi$ given by Eq.~(\ref{eq:EOM_C_doublet_phon}). After the $C_{v,c,v,c}$ are generated by the source terms, they relax due to coupling to the triplets $T$ according to Eq.~(\ref{eq:EOM_C_triplet}). In general, coupling to higher-level correlations is necessary to facilitate relaxation of any type of cluster. Fig.~\ref{fig:s_d_dens} shows numerical results for the exciton formation using two different levels of approximation: We either include a phenomenological damping $\frac{d }{dt}T^{\bl,\bq,\bk',\bk}_{1,2,3,4,j}\Big|_{\textrm{phen}}=-2\gamma_{\textrm{phen}}T^{\bl,\bq,\bk',\bk}_{1,2,3,4,j}$ (the factor $2$ reflecting the two-carrier character), or we take into account scattering and dephasing of the triplets via coupling to quadruplets in Markov approximation, see Eqs.~(\ref{eq:EOM_Daaaa_Q}) and (\ref{eq:EOM_Daaaa_Q_final}), which yields the highest level of approximation \textit{2-carr+2-carr-2-phon+3-phon-assist (Markov)}. Fig.~\ref{fig:s_d_dens}(a) shows that the electron-hole pair density $n^{e/h}$ builds up similar to the result shown in Fig.~\ref{fig:E_field}. It is worth noting that $n^{e/h}$, which is given by singlet populations, Eq.~(\ref{eq:carr_dens_band_main}), is not very sensitive to the inclusion of carrier doublets. 
\begin{figure}
\centering
\includegraphics[width=1.\columnwidth]{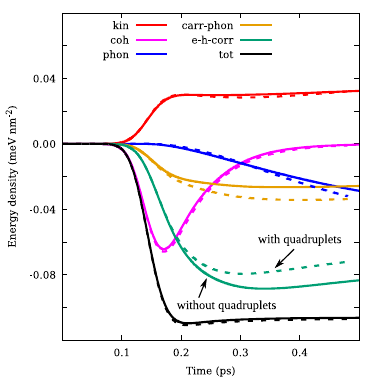}
\caption{Energy density of hBN-encapsulated monolayer MoSe$_2$ after resonant excitation of A$_{\textrm{1s}}$ with fluence $0.1$ $\mu$J cm$^{-2}$ at temperature $T=300$ K using correlations involving up to two carriers. We show the kinetic (kin, Eq.~(\ref{eq:energy_kin})), coherent (coh, Eq.~(\ref{eq:energy_coh})), phonon (phon, Eq.~(\ref{eq:energy_phon})), carrier-phonon (c-ph, Eq.~(\ref{eq:energy_carr_phon})), and correlated inter-band (e-h-corr, Eq.~(\ref{eq:energy_corr_inter})) contributions to the energy density as well as the total result (tot). Correlated intra-band terms are practically zero at this density. Solid lines: without quadruplets (\textit{2-carr+2-phon-assist}), dashed lines: with quadruplets (\textit{2-carr+2-carr-2-phon+3-phon-assist (Markov)}).}
\label{fig:phonon_dyn2}
\end{figure}
\\
\\The correlated electron-hole pair density $n_{\mathrm{corr,tot}}$ as defined in Eq.~(\ref{eq:n_corr}) builds up out of coherent polarizations with a delay according to the microscopic mechanism discussed above. After about $400$ fs, almost all electron-pairs all correlated. With an additional delay, inter-valley excitons, which are represented by doublets $c_{v,c,v,c}^{\bq,\bk',\bk}$ with the total momentum $\bq$ connecting two different valleys in the Brillouin zone (BZ), emerge due to efficient exciton-phonon scattering \cite{selig_dark_2018}. We will investigate this process in more detail later on. All of these results are almost insensitive to the approximation applied to the triplets $T$. Significant differences appear in a further analysis of the correlated density in Fig.~\ref{fig:s_d_dens}(b), where $n_{\mathrm{corr,tot}}$ is split into a contribution of 1s-excitons and the rest according to the projection (\ref{eq:n_corr_proj}). In general, the 1s-excitons, which form from the corresponding resonantly excited polarizations, are partly excited into higher internal states (2s, 2p, ...) via exciton-phonon coupling. This process of exciton dissociation has been discussed in Ref.~\cite{perea-causin_phonon-assisted_2020}. We find that including proper scattering of triplets via coupling to quadruplets in Markov approximation strongly influences the microscopic distribution of excitons, leading to a larger fraction of internally excited electron-hole pairs.
\\
\\The impact of carrier doublets can be further elaborated with the help of energy density contributions, which are shown in Fig.~\ref{fig:phonon_dyn2}. In addition to the kinetic and coherent contributions, which are given by carrier singlets, a pronounced negative contribution due to electron-hole correlations emerges, according to the formation of excitonic doublets. As a result, the total energy density is negative relative to the bandgap, stabilizing after the pump pulse passed the heterostructure at about $t=250$ fs. In contrast to the one-carrier calculations shown in Fig.~\ref{fig:phonon_dyn1}, the phonon energy density is negative due to net absorption of phonon energy by the excitons. The absorption corresponds to the internal excitation of electron-hole pairs discussed above and leads to an increase of correlated energy at later times. As we take into account microscopic scattering of triplets $T$ due to coupling to quadruplets, the correlated energy becomes larger, consistent with the results shown in Fig.~\ref{fig:s_d_dens}(b). To some degree, this increase of $\mathcal{E}_{\textrm{corr,X}}$ is compensated by a stronger cooling of phonons, but also the amount of (negative) carrier-phonon energy density is increased.
As a general trend, we again find that single-particle quantities are barely affected by the approximation made to the triplets $T$, which points to a certain convergence of the cluster expansion: The higher above truncation takes place, the smaller the truncation error of lower-level clusters becomes. Also, one can see that the effects of clusters on observables build up successively in time according to the order of the clusters. This is again typical for the cluster expansion, which might be truncated at lower orders, the shorter the considered time scales are \cite{fricke_transport_1996}. 

\subsection{Momentum-resolved carrier and phonon dynamics}

\begin{figure}
\centering
\includegraphics[width=1.\columnwidth]{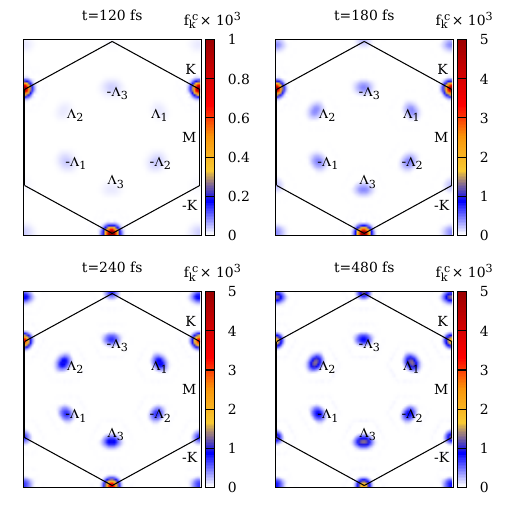}
\caption{Single-particle conduction-band occupancy in the first BZ at different times for resonant pump of the A$_{\textrm{1s}}$-exciton with fluence $0.1$ $\mu$J cm$^{-2}$ at $T=300$ K. Calculations are performed on the quadruplet level. Note the different scale in the first time step.}
\label{fig:sp_occ}
\end{figure}

We finally discuss carrier and phonon dynamics in momentum space after resonant optical excitation at the best level of approximation (\textit{2-carr+2-carr-2-phon+3-phon-assist (Markov)}), i.e. using dephasing of triplets $T$ due to quadruplets in Markov approximation. Fig.~\ref{fig:sp_occ} shows the singlet population of conduction-band states at different times. At $t=120$ fs, corresponding to about half of the pulse duration, an electron population emerges selectively at the $K$-valley due to the circular polarization of the pump field. As Fig.~\ref{fig:s_d_dens}(a) shows, most of the electrons are not yet correlated with a hole at this time, which means that they behave as unbound particles. Efficient carrier-phonon scattering leads to population of $\Lambda$- and $-\Lambda$-valleys already during the pulse, causing a decrease of the $K$-valley electron population between $t=120$ fs and $t=480$ fs. In this time window, 
an increasing fraction of electron-hole pairs is correlated, such that the displayed populations are a mixture of uncorrelated and correlated electrons, projected into the single-particle representation.
\begin{figure}
\centering
\includegraphics[width=1.\columnwidth]{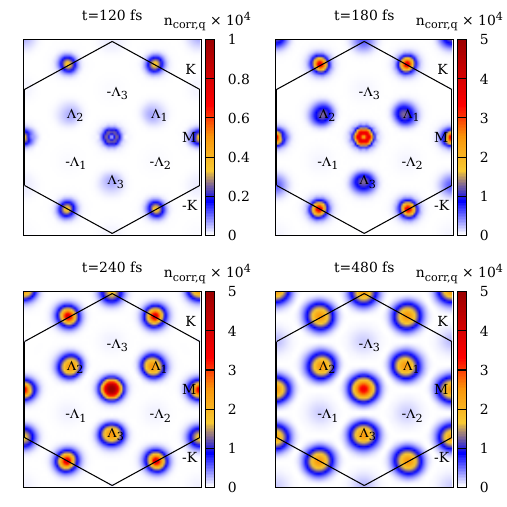}
\caption{Electron-hole correlations (excitonic occupancies) in the first BZ at different times for resonant pump of the A$_{\textrm{1s}}$-exciton with fluence $0.1$ $\mu$J cm$^{-2}$ at $T=300$ K. Calculations are performed on the quadruplet level. Note the different scale in the first time step.}
\label{fig:X_occ}
\end{figure}
\begin{figure*}
\centering
\includegraphics[width=1.\textwidth]{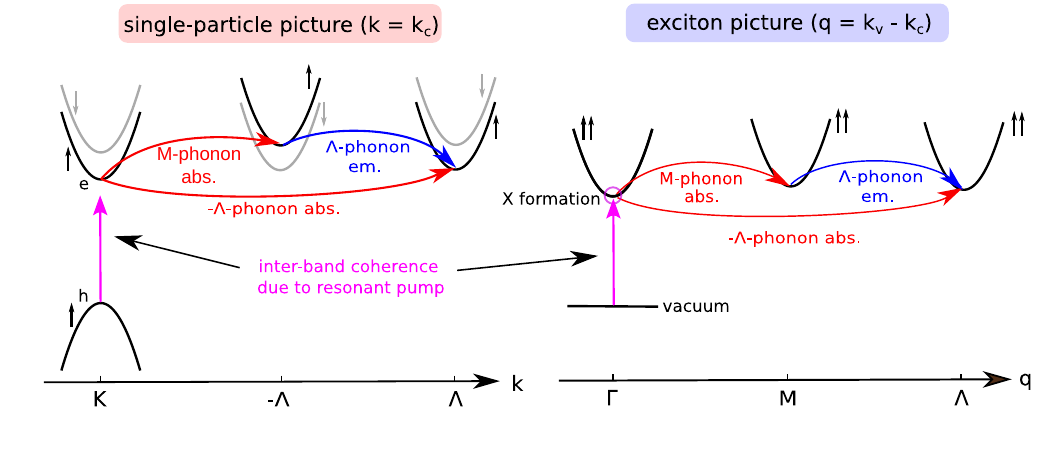}
\caption{Schematic representation of the dominant scattering channels in the single-particle and in the exciton picture after resonant pump of the lowest excitonic transition at the $K$-point, which generates incoherent spin-up electrons (e) and holes (h) and corresponding $\Gamma$-excitons (X).
$\bk_{\textrm{c}}$ and $\bk_{\textrm{v}}$ denote conduction-band and valence-band momentum, respectively, and $\bq$ denotes exciton momentum. For clarity, we only show optically active exciton configurations. In the single-particle picture, excitation takes place below the bandgap.}
\label{fig:scattering_scheme}
\end{figure*}
At $t=480$ fs, the $\Lambda$-valleys exhibit a similar population as $K$, with a slightly lower $-\Lambda$-population due to the higher energy of the valleys in the spin-up conduction band. Similarly, the $-K$-valley is less populated than the $K$-valley. 
\\
\\Fig.~\ref{fig:X_occ} shows the exciton (correlated two-particle) population as defined by Eq.~(\ref{eq:n_corr_q}) at different times. Note that the used BZ is that of two-particle states, spanned by the total momentum $\bq=\bk_v-\bk_c$ as opposed to the single-particle electron momentum $\bk=\bk_c$. At $t=120$ fs, weak exciton populations emerge around $\Gamma$ and $M$ points due to formation out of coherent interband polarizations. While $\bq=\boldsymbol{\Gamma}$ corresponds to momentum-direct excitons at the $K$-point, $\bq=\boldsymbol{M}=\boldsymbol{K}-(\boldsymbol{-\Lambda_3})$ corresponds to excitons with the hole at the $K$-point and the electron at the $-\Lambda_3$-point. These indirect excitons are hence formed out of coherent polarizations by scattering of electrons from $K$ to $-\Lambda_3$. 
Such inter-valley scattering processes are reflected by an electron population at $-\Lambda_3$, which is however too small to be visible in Fig.~\ref{fig:sp_occ} at $t=120$ fs.
Also, $\bq=\boldsymbol{-K}$-excitons are formed, which contain electrons at the $-K$-point.
\begin{figure}
\centering
\includegraphics[width=1.\columnwidth]{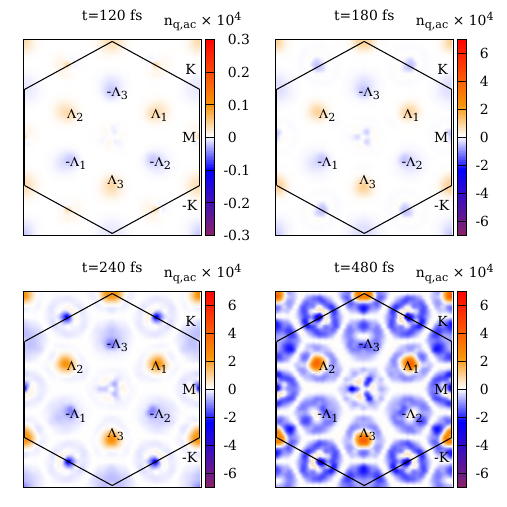}
\caption{Acoustic phonon populations relative to thermal equilibrium in the first BZ at different times for resonant pump of the A$_{\textrm{1s}}$-exciton with fluence $0.1$ $\mu$J cm$^{-2}$ at $T=300$ K. Calculations are performed on the quadruplet level. Positive values correspond to net phonon emission, negative values correspond to net absorption. Note the different scale in the first time step.}
\label{fig:phon_occ_ac}
\end{figure}
\begin{figure}
\centering
\includegraphics[width=1.\columnwidth]{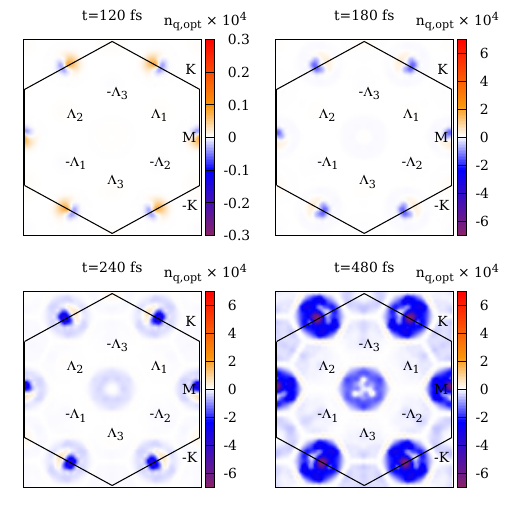}
\caption{Optical phonon populations relative to thermal equilibrium in the first BZ at different times for resonant pump of the A$_{\textrm{1s}}$-exciton with fluence $0.1$ $\mu$J cm$^{-2}$ at $T=300$ K. Calculations are performed on the quadruplet level. Positive values correspond to net phonon emission, negative values correspond to net absorption. Note the different scale in the first time step.}
\label{fig:phon_occ_opt}
\end{figure}
Later on, equilibration between the different exciton species takes place, leaving $\bq=\Gamma$-excitons with the largest occupancy since they have the lowest energy. 
The dominant scattering processes are summarized in Fig.~\ref{fig:scattering_scheme}.
Efficient scattering of excitons from $\boldsymbol{\Gamma}$ to $\boldsymbol{M}$ has also been predicted for MoS$_2$ based on first-principle calculations \cite{chan_exciton_2023}.
\\
\\The solution of the cluster equations also gives access to the time dependence of phonon populations throughout the BZ. We show acoustic and optical phonon populations at different times in Figs.~\ref{fig:phon_occ_ac} and \ref{fig:phon_occ_opt}, respectively. It turns out that some phonon modes might switch between net absorption and emission in the course of time. At early times, acoustic phonons with $\bq=\boldsymbol{M}$, $\bq=\boldsymbol{-K}$ and $\bq=\boldsymbol{\Lambda_i}$ are emitted, while $\boldsymbol{-\Lambda}$-phonons are absorbed. 
Later, $\boldsymbol{M}$-phonons switch to absorption.
Optical phonons are dominantly active around $\bq=\boldsymbol{M}$, with net emission at early times and absorption at later times.
An analysis of momentum conservation shows that the emission of $\boldsymbol{\Lambda}$-phonons and the absorption of $\boldsymbol{-\Lambda}$-phonons either correspond to electron scattering between $\boldsymbol{K}$ and the $\boldsymbol{\Lambda}$-valleys or to electron scattering between $\boldsymbol{-\Lambda}$- and $\boldsymbol{\Lambda}$-valleys (since e.g. $\boldsymbol{K} = \boldsymbol{\Lambda_1} + \boldsymbol{\Lambda_1}$ and $\boldsymbol{-\Lambda_2} = \boldsymbol{\Lambda_1} + \boldsymbol{\Lambda_3}$). The emission and absorption of $\boldsymbol{M}$-phonons occurs during electron scattering between $\boldsymbol{K}$ and $-\boldsymbol{\Lambda_i}$, while $\boldsymbol{-K}$-phonons connect the $K$- and $-K$-valleys.
Hence, the dynamics of phonon populations appears to be consistent with the carrier dynamics discussed before. Our findings of strongly anisotropic phonon distributions are also consistent with earlier \textit{ab initio} investigations of nonequilibrium lattice dynamics in MoS$_2$ \cite{caruso_nonequilibrium_2021}.
The analysis of inter-valley scattering can also be connected to the linear optical spectra shown in Fig.~\ref{fig:absorption}. As the spectra reveal, second-order exciton-phonon scattering processes play a major role at $T=300$ K. The lineshape of optical resonances reflects dephasing in connection with scattering processes of bright $K$$K$-excitons with phonons. In the light of the above discussion, it is very plausible that second-order scattering processes from $K$$K$- to $K$$(-\Lambda)$- to $K$$\Lambda$-excitons are a major source of dephasing.
Note that while momentum conservation is a strict requirement, individual scattering processes are not necessarily energy-conserving due to the non-Markovian character of dynamics, which is particularly important at early times.

\section{Conclusion}

We have shown how a first-principle description of coupled carrier-phonon-photon dynamics in two-dimensional semiconductors can be obtained by means of the cluster expansion technique. Consistency between various many-body interaction mechanisms is achieved by constructing all coupling matrix elements from the same localized set of Wannier functions. 
The microscopic insight into the nonequilibrium dynamics, demonstrated for the model case of hBN-encapsulated monolayer MoSe$_2$, is twofold:
\\(i) For weak photoexcitation, the cluster equations give access to the linear optical response of the (dielectrically embedded) two-dimensional material under the influence of many-body effects, such as exciton-phonon scattering. Systematically analyzing the impact of higher-order processes, we find that exciton-two-phonon scattering strongly renormalizes optical spectra, especially at elevated temperatures.
\\(ii) For strong photoexcitation, non-linear dynamics of carriers and phonons are described and the exchange of energy between the two subsystems is revealed. In this regime, the cluster expansion allows to disentangle, to some degree, effects due to correlated and uncorrelated carriers. We demonstrate that a consistent way to capture phonon dynamics and avoid artificial heating of the system is to include correlations between single charge carriers and two phonons. 
In a broader context, these results show that if the cluster expansion is used to describe observables or phenomena at a certain hierarchy level, the microscopic scattering and dephasing of correlations at this level should not be replaced by an artificial damping.
By introducing two-carrier correlations, we study the exciton formation and relaxation dynamics under the influence of phonon-assisted correlations. In particular, the phonon-induced ionization of 1s-excitons into higher internal states is significantly influenced by correlations of electron-hole pairs with two phonons. Moreover, access to the temporal evolution of phonon mode occupancies allows us to identify efficient scattering cascades of $K$$K$- to $K$$(-\Lambda)$- to $K$$\Lambda$-excitons.
\\
\\In summary, we revise the paradigm of exactly energy-conserving exciton-phonon coupling in first order, both for the description of optical spectra and carrier kinetics.
Our work lays the foundation for a consistent, material-realistic description of many-body correlations in photoexcited two-dimensional semiconductors in a wide range of fluence regimes. The presented approach closes the gap between pure single-particle and pure exciton-based descriptions and allows to monitor the transition from an uncorrelated electron-hole plasma to correlated pairs at very early times after photoexcitation.
In general, the cluster equations describe the successive buildup of higher-order correlations in time, causing scattering and dephasing of the respective lower-order correlations. Thus, the dynamical buildup of screening and many-body renormalization effects of carriers and phonons is automatically taken into account. The method's high accuracy in terms of quantum many-body effects comes at considerable numerical costs, since EOM of high-dimensional correlations need to be propagated explicitly. In this sense, the method is similar to the - technically related - time-dependent coupled cluster technique of quantum chemistry, delivering a high level of predictive power while being limited to ultrashort times $\lessapprox 1$ ps after excitation.
\\
\\\textbf{Acknowledgement}
\\
The authors gratefully acknowledge the computing time provided on the supercomputer Emmy/Grete at NHR@Göttingen as part of the NHR infrastructure. M.F. was supported in part through the computational resources and services provided by the Advanced Research Computing at the University of Michigan. A.S. would like to thank Daniel Erben and Monty Drechsler for fruitful discussions.


%

\newpage

\section{Appendix}

\renewcommand\thefigure{S\arabic{figure}}
\setcounter{figure}{0}
\renewcommand\theequation{S\arabic{equation}}
\setcounter{equation}{0}

\subsection{Density functional theory calculations, spin-orbit coupling and Coulomb matrix elements}\label{sec:abinitio}

\begin{figure}
\centering
\includegraphics[width=\columnwidth]{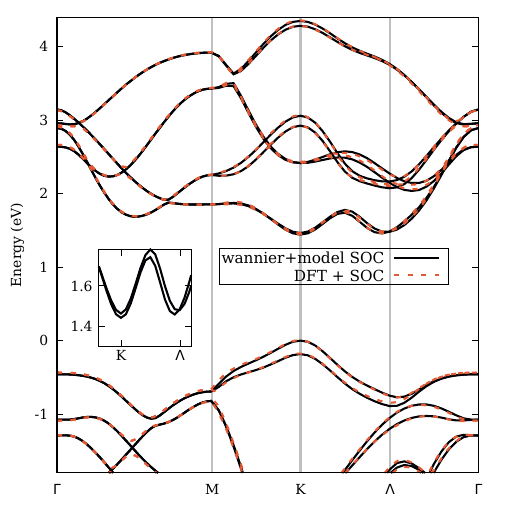}
\caption{Band structure of freestanding monolayer MoSe$_2$ as obtained from DFT+SOC calculation (dashed red lines) compared to a non-relativistic Wannier construction augmented by a nonlocal $\boldsymbol{L\cdot S}$-Hamiltonian (solid black lines). Inset: local minima of two lowest conduction bands.}
\label{fig:DFT_bands}
\end{figure}

Density functional theory (DFT) calculations for freestanding monolayer MoSe$_2$ are carried out using {\sc Quantum ESPRESSO}~\cite{giannozzi_quantum_2009, giannozzi_advanced_2017}. We apply the 
generalized gradient approximation (GGA) by Perdew, Burke, and Ernzerhof (PBE) \cite{perdew_generalized_1996, perdew_generalized_1997}
and use projector-augmented wave (PAW) pseudopotentials from the PSLibrary \cite{dal_corso_pseudopotentials_2014} at a plane-wave cutoff of $80$~Ry. 
Uniform meshes (including the $\Gamma$-point) with $18\times18\times1$ k-points are combined with a Fermi-Dirac smearing of $5$~mRy. 
Using a fixed lattice constant of $a=3.29$~\AA \cite{chen_highly_2017}
and a fixed cell height of $45$~\AA, forces are minimized below $10^{-3}$~eV/\AA.
\\
\\We use RESPACK \cite{nakamura_respack:_2021} to construct a lattice Hamiltonian $H_0(\bR)$ for an appropriate set of lattice vectors $\bR$ in an 11-dimensional localized basis of Wannier orbitals (d$_{z^2}$, d$_{xz}$, d$_{yz}$, d$_{x^2-y^2}$ and d$_{xy}$ for Mo, p$_x$, p$_y$ and p$_z$ for Se) from the DFT results. 
The momentum representation of the lattice Hamiltonian is obtained by Fourier transformation:
\begin{equation}
\begin{split}
H_0(\bk)=\sum_{\bR} e^{i\bk\cdot\bR} H_0(\bR) \,.
\end{split}
\label{eq:H_k_0}
\end{equation}
We also calculate the dielectric function as well as bare and screened Coulomb matrix elements in the localized basis. For the polarization function, a cutoff energy of $5$~Ry, $96$~bands as well as $70$~frequency points on a logarithmic grid are taken into account.
\\
\\Spin-orbit interaction is included using an on-site $\boldsymbol{L\cdot S}$-coupling Hamiltonian along the lines of \cite{liu_three-band_2013}, which we augment by a phenomenological non-local term. The spin-orbit Hamiltonian is added to the non-relativistic Wannier Hamiltonian:
\begin{equation}
\begin{split}
H_{\textrm{lat}}(\bk)=I_2 \otimes H_0(\bk)+H_{\textrm{SOC}}(\bk)\,.
\end{split}
\label{eq:H_tot}
\end{equation}
Here, $I_2$ is the $2\times2$ identity matrix in the Hilbert space spanned by eigenstates $\ket{\uparrow}$ and $\ket{\downarrow}$ of the spin z component (perpendicular to the monolayer). 
By treating spatial degrees of freedom and spin separately, we reduce the size of the Wannier Hamiltonian and
thereby the Coulomb matrix. We assume that the Coulomb matrix is spin-independent.
Diagonalization of $H(\bk)$ yields the band structure $\varepsilon_{\bk}^{\lambda}$ and the Bloch states $\ket{\bk,\lambda}=\sum_{\alpha} c^{\lambda}_{\alpha,\bk}\ket{\bk,\alpha} $,
where the coefficients $c^{\lambda}_{\alpha,\bk}$ describe the momentum-dependent contribution of the orbital $\alpha$ to the Bloch band $\lambda$.
The Bloch sums $\ket{\bk,\alpha}$ are connected to the localized basis via $\ket{\bk,\alpha}=\frac{1}{\sqrt{N}}\sum_{\bR}e^{i\bk\cdot\bR}\ket{\bR,\alpha}$ with the number of unit cells $N$ and lattice vectors $\bR$. The localized states are normalized according to $\big<\bR',\alpha'\big|\bR,\alpha\big>=\delta_{\bR,\bR'}\delta_{\alpha,\alpha'}$.
\\
\\The SOC-Hamiltonian is given by 
\begin{equation}
\begin{split}
H_{\textrm{SOC}}(\bk)=\frac{1}{\hbar^2}\boldsymbol{\tilde{L}}(\bk)\cdot\boldsymbol{S}=\frac{1}{2\hbar}\boldsymbol{\tilde{L}}(\bk)\cdot\boldsymbol{\sigma}
\end{split}
\label{eq:H_SOC}
\end{equation}
with the Pauli matrices $\boldsymbol{\sigma}=(\sigma_x,\sigma_y,\sigma_z)$. The modified angular momentum operator
\begin{equation}
\boldsymbol{\tilde{L}}(\bk)=
\begin{pmatrix}
      \lambda_{\textrm{Mo}}\boldsymbol{L}_{l=2} & 0   &  0 \\
      0 & \lambda_{\textrm{Se}}(\bk)\boldsymbol{L}_{l=1}   & 0 \\
      0 & 0 &  \lambda_{\textrm{Se}}(\bk)\boldsymbol{L}_{l=1}
    \end{pmatrix}
\label{eq:L_op}
\end{equation}
contains intra-atomic coupling parameters $\lambda_{\textrm{Mo}}$ for the $(l=2)$-subspace (d$_{z^2}$, d$_{xz}$, d$_{yz}$, d$_{x^2-y^2}$ and d$_{xy}$) and
a non-local function $\lambda_{\textrm{Se}}(\bk)=\lambda^0_{\textrm{Se}}+2\lambda^{\textrm{NN}}_{\textrm{Se}}\xi(\bk) $ for the $(l=1)$-subspace (p$_x$, p$_y$ and p$_z$). The nearest-neighbour coupling for Se-atoms is described by the momentum-dependent function 
\begin{equation}
\xi(\bk)= \textrm{cos}\Big(\frac{\pi}{2}\frac{|\boldsymbol{k}-\boldsymbol{\Gamma}|}{|\boldsymbol{K}-\boldsymbol{\Gamma}|}\Big)\,
\label{eq:NN_coupling}
\end{equation}
that vanishes a the K-point and reaches its maximum at $\Gamma$.
In the given basis, the angular momentum algebra for spherical harmonics yields: 
\begin{equation}
\begin{split}
L_{x,l=1}&=
\begin{pmatrix}
      0 & 0   &  0 \\
      0 & 0   & -i\hbar \\
      0 & i\hbar & 0
    \end{pmatrix} \,,
L_{y,l=1}=
\begin{pmatrix}
      0 & 0   &  i\hbar \\
      0 & 0   & 0 \\
      -i\hbar & 0 & 0
    \end{pmatrix} \,, \\
L_{z,l=1}&= 
\begin{pmatrix}
      0 & -i\hbar   &  0 \\
      i\hbar & 0   & 0 \\
      0 & 0 & 0
    \end{pmatrix} 
\end{split}
\label{eq:L_1}
\end{equation}
and 
\begin{equation}
\begin{split}
L_{x,l=2}&=
\begin{pmatrix}
      0 & 0   &  \sqrt{3}i\hbar & 0 & 0 \\
      0 & 0   & 0 & 0 & i\hbar\\
      -\sqrt{3}i\hbar & 0   & 0 & -i\hbar & 0\\
      0 & 0   & i\hbar & 0 & 0\\
      0 & -i\hbar &  0 & 0 & 0
    \end{pmatrix}\,,
    \\
L_{y,l=2}&=
\begin{pmatrix}
      0 & -\sqrt{3}i\hbar   &  0 & 0 & 0 \\
      \sqrt{3}i\hbar & 0   & 0 & -i\hbar & 0\\
      0 & 0   & 0 & 0 & -i\hbar\\
      0 & i\hbar  & 0 & 0 & 0\\
      0 & 0 &  i\hbar & 0 & 0
    \end{pmatrix}\,,
    \\
L_{z,l=2}&= 
\begin{pmatrix}
      0 & 0   &  0 & 0 & 0 \\
      0 & 0   & -i\hbar & 0 & 0\\
      0 & i\hbar   & 0 & 0 & 0\\
      0 & 0   & 0 & 0 & -2i\hbar\\
      0 & 0 &  0 & 2i\hbar & 0
    \end{pmatrix}\,.
\end{split}
\label{eq:L_2}
\end{equation}
We choose the coupling constants $\lambda_{\textrm{Mo}}=103.5$ meV, $\lambda^0_{\textrm{Se}}=83.5$ meV and $\lambda^{\textrm{NN}}_{\textrm{Se}}=173$ meV to reproduce the spin-orbit splittings at the K-point and at the conduction-band minimum close to the $\Lambda$-point
as obtained from DFT calculations including spin-orbit coupling using fully relativistic pseudopotentials. The valence-band and conduction-band splittings at K are $182$ meV and $21$ meV, respectively, with a like-spin ground state. Fig.~\ref{fig:DFT_bands} shows the excellent agreement between band structures directly from fully relativistic DFT calculations and from diagonalization of the spin-augmented Wannier Hamiltonian (\ref{eq:H_tot}). We find that the conduction-band minimum in the $\Lambda$-valley shown in the inset of Fig.~\ref{fig:DFT_bands} is about $15$ meV higher than the K-valley minimum, which makes monolayer MoSe$_2$ a direct semiconductor in our calculation.
We use a corrected bandgap of $2.4$ eV at the K-point, as predicted by G$_0$W$_0$ calculations reported in Ref.\cite{gillen_light-matter_2017}. To obtain a reasonable estimate of exciton energies in optical spectra, we apply a corresponding rigid shift to all conduction bands.
%
\\
\\Bloch electrons are described by the single-particle Hamiltonian
\begin{equation}
\begin{split}
H_{\textrm{carr}}=\sum_{\bk,\lambda}\varepsilon_{\bk}^{\lambda} a^{\dagger}_{\bk,\lambda}a^{\phantom\dagger}_{\bk,\lambda}\,,
\end{split}
\label{eq:H_Bloch}
\end{equation}
where $a^{\dagger}_{\bk,\lambda}$ and $a^{\phantom\dagger}_{\bk,\lambda}$ denote creation and annihilation operators, respectively, 
for a carrier with momentum $\bk$ in band $\lambda$.
The density-density-like bare Coulomb interaction matrix elements in the Wannier basis,
%
\begin{equation}
\begin{split}
U_{\alpha\beta}(\bq)&=\sum_{\bR}e^{i\bq\cdot\bR}U_{\alpha\beta\beta\alpha}(\bR) \\
&=\sum_{\bR}e^{i\bq\cdot\bR}\bra{\bn,\alpha}\bra{\bR,\beta} U(\br,\br') \ket{\bR,\beta}\ket{\bn,\alpha}\,,
\end{split}
\label{eq:U_1}
\end{equation}
and the corresponding (statically) screened matrix elements $V_{\alpha\beta}(\bq)$, are treated along the same lines as detailed in Ref.~\cite{steinhoff_microscopic_2021}: The bare Coulomb matrix $\boldsymbol{U}(\bq)$ is diagonalized to obtain eigenvalues $U_i(\bq)$ and eigenvectors $\boldsymbol{e}_i(\bq)$. The matrix elements of the screened interaction $\boldsymbol{V}(\bq)$ in the eigenbasis of the bare interaction are then obtained via
\begin{equation}
\begin{split}
V_i(\bq)=\varepsilon^{-1}_i(\bq)\,U_i(\bq)\,,
\end{split}
\label{eq:V_from_U}
\end{equation}
where the dielectric matrix $\boldsymbol{\varepsilon}(\bq)$ accounts for both the material-specific internal polarizability and the screening by the environment.
First, we introduce an analytic description for the freestanding monolayer dielectric function, i.e. in the absence of external screening.
The leading eigenvalue is expressed by
\begin{equation}
\begin{split}
\varepsilon_1(q)=\varepsilon_{\infty}(q)\frac{1-\beta^2 e^{-2qd}}{1+2\beta e^{-qd}+\beta^2 e^{-2qd}}\,,
\end{split}
\label{eq:eps_1}
\end{equation}
with 
$\beta= \frac{\varepsilon_{\infty}(q)-1}{\varepsilon_{\infty}(q)+1}$ 
\cite{rosner_wannier_2015} 
and the bulk dielectric constant given by a modified Resta model \cite{resta_thomas-fermi_1977}
\begin{equation}
\begin{split}
\varepsilon_{\infty}(q)=\frac{a+q^2}{\frac{a\, \textrm{sin}(qc)}{qbc}+q^2}+e \,.
\end{split}
\label{eq:eps_inf}
\end{equation}
As layer thickness, we use the layer separation in bulk $d=0.65$ nm \cite{kylanpaa_binding_2015}. The other eigenvalues are well described by third-order polynomials. Environmental screening can be taken into account according to the Wannier function continuum electrostatics approach \cite{rosner_wannier_2015} that combines a macroscopic electrostatic model for 
the screening by the dielectric environment with a localized description of Coulomb interaction.
The leading eigenvalue $\varepsilon_1(q)$, which is most sensitive to macroscopic screening, is modified by replacing the dielectric function of a freestanding monolayer 
with that of an arbitrary vertical heterostructure. The latter is obtained by solving Poisson's equation for a test charge in a slab with thickness $d$ and dielectric 
function $\varepsilon_{\infty}(q)$ embedded in a z-dependent dielectric profile \cite{florian_dielectric_2018}.
After calculating the eigenvalues $V_i(\bq)$, the screened Coulomb matrix in the Wannier basis $V_{\alpha\beta}(\bq)$ is obtained by using the eigenvectors $\boldsymbol{e}_i(\bq)$. 
We finally compute screened Coulomb matrix elements 
in the Bloch-state representation by a unitary transformation using the coefficients $c^{\lambda}_{\alpha,\bk}$:
\begin{equation}
\begin{split}
V^{\lambda,\nu,\nu',\lambda'}_{\bk_1, \bk_2, \bk_3, \bk_4} = \frac{1}{N}\sum_{\alpha, \beta} \big( c_{\alpha, \bk_1}^{\lambda} \big)^{*} \big( c_{\beta, \bk_2}^{\nu} \big)^{*} V^{\alpha\beta}_{|\bk_1-\bk_4|}c_{\beta, \bk_3}^{\nu'} c_{\alpha, \bk_4}^{\lambda'}  \, ,
\end{split}
\label{eq:Coul_ME}
\end{equation}
where $\bk_4=\bk_1+\bk_2-\bk_3+\bG$ due to momentum conservation.
The carrier-carrier interaction Hamiltonian is thus given by:
\begin{equation}
 \begin{split}
H_{\textrm{Coul}} =
\frac{1}{2}\sum_{\substack{\bk,\bk',\bq \\ \lambda,\nu,\nu',\lambda'}}V^{\lambda,\nu,\nu',\lambda'}_{\bk,\bk',\bk'+\bq,\bk-\bq} 
a^{\dagger}_{\bk,\lambda}a^{\dagger}_{\bk',\nu}a^{\phantom\dagger}_{\bk'+\bq,\nu'}a^{\phantom\dagger}_{\bk-\bq,\lambda'}\,.
\end{split}
\label{eq:carr_carr_Hamiltonian}
\end{equation}
We assume that every Bloch band can be assigned to $\ket{\uparrow}$ or $\ket{\downarrow}$, making the spin z component a good quantum number. Furthermore, we make use of the fact that Coulomb interaction is spin-conserving, so that we can set Coulomb matrix elements $V^{\lambda,\nu,\nu',\lambda'}_{\bk_1 \bk_2 \bk_3 \bk_4}$ to zero if $\lambda$ and $\lambda'$ or $\nu$ and $\nu'$ belong to different spins.
\\
\\When a TMD monolayer is encapsulated in a heterostructure, its band structure (as obtained from DFT/GW-calculations) is modified due to environmental screening of the long-range Coulomb interaction. This effect is captured by a GdW self-energy, which was first brought up by Rohlfing \cite{rohlfing_electronic_2010} and worked out for van der Waals heterostructures in Refs.~\cite{winther_band_2017,florian_dielectric_2018}. The idea is to approximately split the self-energy of the heterostructure into a part
describing the isolated TMD monolayer that is treated on a full \textit{ab initio} level, and a correction term containing environmental
screening effects via a continuum-electrostatics model. The resulting GdW self-energy that corrects single-particle energies of the TMD monolayer in the Hamiltonian (\ref{eq:H_Bloch}) is given in static approximation by:
\begin{equation}
\begin{split}
\Sigma^{\textrm{GdW},c}_{\bk}&=\frac{1}{2}\sum_{\bk',c'} \Delta V^{c,c',c,c'}_{\bk, \bk', \bk, \bk'}\,,\\
\Sigma^{\textrm{GdW},v}_{\bk}&=\frac{1}{2}\sum_{\bk',v'} \Delta V^{v,v',v,v'}_{\bk, \bk', \bk, \bk'}\,, \\
\Delta V &= V - V^{\textrm{freest.}}\,.
\end{split}
\label{eq:GdW}
\end{equation}
Here, $V^{\textrm{freest.}}$ corresponds to the Coulomb interaction in the freestanding monolayer, while $V$ contains environmental screening. 
In this work, we take into account screening from adjacent hBN layers (see Fig.~\ref{fig:sketch_HS}) with a high-frequency dielectric constant of $\varepsilon_{\textrm{hBN}} = \sqrt{4.95 \cdot 2.86}$ \cite{artus_natural_2018} and a narrow gap of $0.3$ nm between the layers \cite{florian_dielectric_2018}. Using a $36\times 36\times 1$ Monkhorst-Pack mesh for BZ sampling, we obtain a GdW bandgap correction of $-410$ meV at the K point. 

\subsection{Density functional perturbation theory calculations and carrier-phonon matrix elements}\label{sec:abinitio_cp}
\begin{figure}
\centering
\includegraphics[width=\columnwidth]{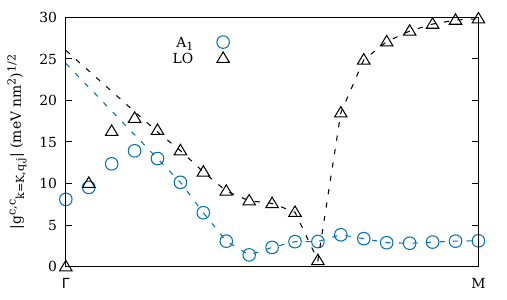}
\caption{Modulus of carrier-phonon matrix elements $g_{\bk,\bq, j}^{\lambda,\lambda'} $ (rescaled by $\sqrt{A_{\textrm{UC}}}$) for conduction-band electrons at the $K$-point depending on the transferred momentum $\bq$ for LO and $A_1$ branches. We correct the original data (symbols) at small $\bq$ by a linear interpolaton (dashed lines).}
\label{fig:phonon_correction}
\end{figure}

The phonon eigenenergies and eigenmodes as well as the carrier-phonon coupling are obtained from first principles using density function perturbation theory (DFPT). To achieve a consistent treatment of electronic and vibronic properties along with their mutual coupling, we follow the procedure in Refs.~\cite{giustino_electron-phonon_2007,berges_phonon_2023} to obtain the phononic data in the same localized Wannier representation as the electronic data, see the previous section.
\\
DFPT calculations are carried out using {\sc Quantum ESPRESSO}~\cite{giannozzi_quantum_2009, giannozzi_advanced_2017}. For the transformation of the electron-phonon coupling to the Wannier basis, we use WANNIER90~\cite{pizzi_wannier90_2020} and the EPW code~\cite{giustino_electron-phonon_2007, ponce_epw:_2016}. 
\\ The dynamical matrix $\tilde{\omega}_{\bR}^{i,i'}$ is given in a basis of atomic displacements $i,i'$ for an appropriate set of lattice vectors $\bR$. A momentum representation is obtained by Fourier transformation:
\begin{equation}
\begin{split}
\tilde{\omega}_{\bq}^{i,i'}=\sum_{\bR} e^{i\bq\cdot\bR} \tilde{\omega}_{\bR}^{i,i'} \,.
\end{split}
\label{eq:dyn_mat}
\end{equation}
\begin{figure}
\centering
\includegraphics[width=\columnwidth]{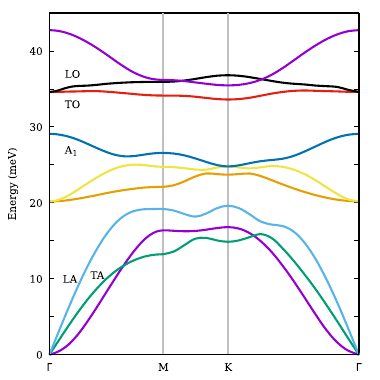}
\caption{Phonon dispersion of freestanding monolayer MoSe$_2$ as obtained from density-functional perturbation theory (DFPT). The most important branches, as identified in the main text, are labeled.}
\label{fig:phonon_dispersion}
\end{figure}
Diagonalization of the dynamical matrix yields the phonon frequencies $\omega_{\bq,j} $ and corresponding eigenvectors $\varepsilon^{i}_{\bq,j}=\big<i \big|\bq,j\big>$. The eigenvectors translate atomic displacements $i$ into phonon modes $j$, with all momentum states belonging to the same mode forming a phonon branch. Fig.~\ref{fig:phonon_dispersion} shows all $9$ branches of the phonon dispersion of freestanding monolayer MoSe$_2$. Note that we have disentangled the branches by maximizing the eigenvector overlap of neighbouring momentum states on the same branch. The disentanglement is essential to obtain carrier-phonon interaction matrix elements as continuous functions of momentum later on. 
The most important branches are given by transverse acoustic (TA), longitudinal acoustic (LA), out-of-plane optical ($A_1$), transverse optical (TA) and longitudinal optical (LO) modes. The phonons are described in second quantization by the Hamiltonian
\begin{equation}
\begin{split}
H_{\textrm{phon}}=\sum_{\bq,j}\hbar\omega_{\bq,j} \Big(D^{\dagger}_{\bq,j}D^{\phantom\dagger}_{\bq,j}+\frac{1}{2}\Big)\,,
\end{split}
\label{eq:H_phon}
\end{equation}
where $D^{\dagger}_{\bq,j}$ and $D^{\phantom\dagger}_{\bq,j}$ denote creation and annihilation operators, respectively, 
for a phonon with momentum $\bq$ in branch $j$. 
\\
\\The first-principle deformation potentials $D_{\bR,\bR'}^{i,\alpha,\beta}$ additionally contain information about the electronic Wannier orbitals $\alpha,\beta$. A representation in terms of momentum states and phonon eigenmodes is derived as:
\begin{equation}
\begin{split}
D_{\bk,\bq, j}^{\alpha,\beta}
&= \sum_{i} \varepsilon^{i}_{\bq,j} D_{\bk,\bq, i}^{\alpha,\beta} \\
&= \sum_{i} \varepsilon^{i}_{\bq,j} \frac{1}{\sqrt{M_i}}\left< \bk+\bq,\alpha \right| \frac{\partial V}{\partial u_{\bq,i}} \left| \bk,\beta \right> \\
&=\sum_{\bR,\bR',i} \varepsilon^{i}_{\bq,j} e^{i(\bq\cdot\bR+\bk\cdot\bR')} D_{\bR,\bR'}^{i,\alpha,\beta} \,,
\end{split}
\label{eq:def_pot_momentum_eigenmodes}
\end{equation}
where $\partial V/\partial u_{\bq,i}$ denotes the derivatives of the (self-consistent) crystal potential with respect to atomic displacements and $M_i$ is the corresponding atomic mass. 
The carrier-phonon interaction matrix elements in terms of Wannier orbitals are given by:
\begin{equation}
\begin{split}
g_{\bk,\bq, j}^{\alpha,\beta}=\frac{D_{\bk,\bq, j}^{\alpha,\beta}}{\sqrt{2\omega_{\bq,j}}} \,.
\end{split}
\label{eq:carr_phon_ME_wannier}
\end{equation}
Finally, we obtain the matrix elements in the Bloch representation using the coefficients $c^{\lambda}_{\alpha,\bk}$. As a convention, $g_{\bk,\bq, j}^{\lambda,\lambda'}$ shall denote the matrix element $\left< \bk,\lambda \right| \cdot \left| \bk-\bq,\lambda' \right> $:
\begin{equation}
\begin{split}
g_{\bk,\bq, j}^{\lambda,\lambda'}=\frac{1}{\sqrt{N}}\sum_{\alpha,\beta}\big(c^{\lambda}_{\alpha,\bk}\big)^*c^{\lambda'}_{\beta,\bk-\bq}  g_{\bk-\bq,\bq, j}^{\alpha,\beta} \,.
\end{split}
\label{eq:carr_phon_ME_bloch}
\end{equation}
Here, the factor $ 1/\sqrt{N}$ is due to the Wannier function normalization. 
The carrier-phonon interaction Hamiltonian is thus given by:
\begin{equation}
 \begin{split}
H_{\textrm{c-ph}} =
\sum_{\substack{\bk,\bq \\ \lambda,\lambda'}}g_{\bk,\bq, j}^{\lambda,\lambda'} 
a^{\dagger}_{\bk,\lambda}a^{\phantom\dagger}_{\bk-\bq,\lambda'}\big(D^{\phantom\dagger}_{\bq,j}+D^{\dagger}_{-\bq,j}  \big)\,.
\end{split}
\label{eq:carr_phon_Hamiltonian}
\end{equation}
\\Carrier-phonon interaction is assumed to be spin-independent and spin-conserving just as carrier-carrier Coulomb interaction to reduce numerical complexity, although our formalism would allow to include phonon-induced spin-flip processes \cite{molina-sanchez_ab_2017,jiang_real-time_2021}. Care has to be taken when treating the long-wavelength limit of the longitudinal optical phonon branch and the out-of-plane $A_1$ branch, where the Coulomb-type behavior of the interaction is not captured well by first-principle methods~\cite{kaasbjerg_phonon-limited_2012}. Inspired by Ref.~\cite{kaasbjerg_phonon-limited_2012}, we enforce the correct analytical behavior by replacing the numerically obtained band-diagonal interaction matrix elements (\ref{eq:carr_phon_ME_bloch}) for small $\bq$ ($\bq$=0 and adjacent mesh points) with a linear extrapolation of the data at surrounding mesh points, as examplarily shown in Fig.~\ref{fig:phonon_correction}.

\subsection{Light-matter coupling}\label{sec:LM}
We start from the N-carrier Hamiltonian with minimal coupling to the vector potential $\boldsymbol{A}$ in Coulomb gauge \cite{kira_quantum_1999}:
\begin{equation}
 \begin{split}
H_{\textrm{min}} = H_{\textrm{field}}
&+\sum_{i=1}^N\Big\{ \frac{1}{2m_i}\big(\boldsymbol{p}_i-q_i\boldsymbol{A}(\br_i)\big)^2+U(\br_i)\Big\} \\
&+\sum_{i<i'}^N V(|\br_i-\br_{i'}|)\,,
\end{split}
\label{eq:minimal_coupling_Hamiltonian}
\end{equation}
where $H_{\textrm{field}}$ describes the transversal electromagnetic field, $q_i$ is the charge of the $i$-th particle, $U(\br)$ is the periodic lattice potential and $V(\br)$ is the longitudinal Coulomb interaction potential. The Hamiltonian leads to a light-matter coupling term of the form $\bp\cdot\boldsymbol{A}$ as well as a term containing the squared vector potential $\boldsymbol{A}^2$. The latter is responsible for a Drude like response to THz fields \cite{kira_microscopic_2003}. We apply a Göppert-Mayer-type gauge transformation with the unitary operator $T=e^{S}$ and
\begin{equation}
 \begin{split}
S=-\frac{i}{\hbar}\sum_{i=1}^N q_i\br_i\cdot\boldsymbol{A}(\br_i)\,,
\end{split}
\label{eq:GM_op}
\end{equation}
assuming that the spatial variation of $\boldsymbol{A}$ is weak on the scale of the crystal unit cell (dipole approximation). We then obtain the transformed Hamiltonian
\begin{equation}
 \begin{split}
H'_{\textrm{min}} &= H_{\textrm{field}}+H_{\textrm{carr}}+H_{\textrm{Coul}}\\
&-\frac{1}{\varepsilon_0}\sum_{i=1}^N q_i\br_i\cdot \boldsymbol{D}'(\br_i)+H_{\textrm{dip-dip}}
\end{split}
\label{eq:minimal_coupling_Hamiltonian_transf}
\end{equation}
with the Hamiltonian $H_{\textrm{carr}}$ for Bloch electrons and the dipole-dipole interaction term $H_{\textrm{dip-dip}}$.
The transformed displacement field is given in second quantization as \cite{kira_quantum_1999}:
\begin{equation}
 \begin{split}
\boldsymbol{D}'(\br)= \varepsilon_0 \sum_{\bq,\sigma}iE_{q}\Big[\boldsymbol{U}_{\bq,\sigma}(\br)b_{\bq,\sigma} -\textrm{h.c.}\Big] \,,
\end{split}
\label{eq:displacement}
\end{equation}
where $E_q=\sqrt{\hbar\Omega_q/(2\varepsilon_0)}$ with the photon frequency $\Omega_q$. Here, $\bq$ is the photon momentum, $\sigma$ is the photon polarization,  $\boldsymbol{U}_{\bq,\sigma}(\br)$ are photon mode functions and $b_{\bq,\sigma}$ is the corresponding photon annihilation operator. 
The Hamiltonian of the free electromagnetic field is given by:
\begin{equation}
 \begin{split}
H_{\textrm{field}}=\sum_{\bq,\sigma}\hbar\Omega_{\bq}\Big(b^{\dagger}_{\bq,\sigma}b^{\phantom\dagger}_{\bq,\sigma}+\frac{1}{2}\Big)\,.
\end{split}
\label{eq:H_field}
\end{equation}
While the unitary transformation removed the $\boldsymbol{A}^2$-term, it introduced the dipole-dipole term, which formally is a two-particle interaction operator:
\begin{equation}
 \begin{split}
H_{\textrm{dip-dip}}= \frac{1}{2\varepsilon_0}\sum_{\bq,\sigma,i,i'}\big(q_i\br_i\cdot\boldsymbol{U}_{\bq,\sigma}(\br_i) \big)
\big(q_{i'}\br_{i'}\cdot\boldsymbol{U}_{\bq,\sigma}(\br_{i'})\big)^*\,.
\end{split}
\label{eq:H_dip-dip}
\end{equation}
The mode functions are decomposed into plane waves in the plane of the two-dimensional layer and one-dimensional solutions describing the propagation perpendicular to the layer (in z direction):
\begin{equation}
 \begin{split}
\boldsymbol{U}_{\bq,\sigma}(\br)=\boldsymbol{u}_{\bq,\sigma}(z)\frac{e^{i\bq_{\parallel}\cdot \boldsymbol{\rho}}}{\sqrt{\mathcal{A}}}=\boldsymbol{e}_{\sigma}(\bq)u_{\bq,\sigma}(z)\frac{e^{i\bq_{\parallel}\cdot \boldsymbol{\rho}}}{\sqrt{\mathcal{A}}}\,.
\end{split}
\label{eq:mode_fct}
\end{equation}
Here, we have given the polarization vector of the photon $\boldsymbol{e}_{\sigma}(\bq)$ explicitly. Note that we use the crystal area as quantization volume of the plane waves. The perpendicular component is normalized according to
\begin{equation}
 \begin{split}
\int dz\, n(z)^2 \big(u_{\bq',\sigma'}(z) \big)^* u_{\bq,\sigma}(z) = \delta_{\bq,\bq'}\delta_{\sigma,\sigma'}
\end{split}
\label{eq:mode_fct_norm}
\end{equation}
with the refractive index $n(z)$.
Introducing the second quantization for carriers in Eq.~(\ref{eq:minimal_coupling_Hamiltonian_transf}) using the field operators
\begin{equation}
 \begin{split}
\Psi(\br)=\sum_{\bk,\lambda} a_{\bk,\lambda} \big<\br\big|\bk,\lambda\big> \,,
\end{split}
\label{eq:field_op}
\end{equation}
we can identify the single-particle carrier Hamiltonian (\ref{eq:H_Bloch}) and the longitudinal Coulomb interaction Hamiltonian (\ref{eq:carr_carr_Hamiltonian}). The dipole Hamiltonian becomes
\begin{equation}
 \begin{split}
&H_{\textrm{dip}}=-\sum_{\substack{\bk,\bq,\sigma \\ \lambda,\lambda'}}i\frac{E_{q}}{\sqrt{\mathcal{A}}}
\big< \bk,\lambda \big| (-e)\br \big| \bk-\bq_{\parallel},\lambda' \big> \\ &\cdot
\Big[\boldsymbol{u}_{\bq,\sigma}(z_0) b^{\phantom\dagger}_{\bq,\sigma} - \big(\boldsymbol{u}_{-\bq,\sigma}(z_0)\big)^* b^{\dagger}_{-\bq,\sigma}\Big]
a^{\dagger}_{\bk,\lambda}a^{\phantom\dagger}_{\bk-\bq_{\parallel},\lambda'}\,,
\end{split}
\label{eq:H_dip}
\end{equation}
with $z_0$ the position of the two-dimensional layer and the electric charge of the electron $-e$.
Assuming that the in-plane component $\bq_{\parallel}$ of the photon momentum is negligible compared to Bloch electron momenta, we can use the dipole matrix element
\begin{equation}
 \begin{split}
\boldsymbol{d}^{\lambda,\lambda'}_{\bk}&=
\left< \bk,\lambda \right| (-e)\br \left| \bk,\lambda' \right> 
\\ &= 
\frac{-e}{i}\frac{1}{\varepsilon^{\lambda}_{\bk}-\varepsilon^{\lambda'}_{\bk}}
\sum_{\alpha\beta}(c^{\lambda}_{\alpha,\bk})^*c^{\lambda'}_{\beta,\bk} 
\nabla_{\bk}
H_{\textrm{lat}}^{\alpha\beta}(\bk)\,.
\end{split}
\label{eq:dip_ME}
\end{equation}
In the last line, we have applied a Peierls transformation \cite{tomczak_optical_2009,steinhoff_influence_2014} to express the dipole matrix elements directly in terms of the lattice Hamiltonian (\ref{eq:H_tot}). We limit ourselves to optical transitions where $\lambda$,$\lambda'$ always combine one valence and one conduction band. Next, we derive the dipole-dipole Hamiltonian (\ref{eq:H_dip-dip}) in second quantization: 
\begin{equation}
 \begin{split}
H_{\textrm{dip-dip}}&= \frac{1}{2\varepsilon_0}\int d^3\br\int d^3\br'\Psi^{\dagger}(\br)\Psi^{\dagger}(\br') \\
&\sum_{\bq,\sigma}\big(e\br\cdot\boldsymbol{U}_{\bq,\sigma}(\br) \big)
\big(e\br'\cdot\boldsymbol{U}_{\bq,\sigma}(\br')\big)^* \Psi(\br')\Psi(\br)\,.
\end{split}
\label{eq:H_dip-dip_2Q_start}
\end{equation}
We assume that the $q_z$-dependence of polarization vectors can be neglected, which becomes exact in the limit of perpendicular propagation. Then we can apply the completeness relation \cite{kira_quantum_1999}
\begin{equation}
 \begin{split}
\sum_{q_z} u_{\bq,\sigma}(z) \big(u_{\bq,\sigma}(z') \big)^*  = \frac{\delta(z-z')}{n^2(z)}\,.
\end{split}
\label{eq:mode_fct_complete}
\end{equation}
To disentangle the spatial dependence of photon modes and electronic states, we assume that a confinement wave function $\xi(z)$, which does not influence the in-plane dipoles, can be factorized out of the Bloch functions. This yields:
\begin{equation}
 \begin{split}
H_{\textrm{dip-dip}}&=\frac{w}{2\varepsilon_0n^2(z_0)\mathcal{A}}\\&\times\sum_{\substack{\bk,\bk' \\ \lambda,\nu,\nu',\lambda'}}\sum_{\bq_{\parallel},\sigma}
\boldsymbol{d}^{\lambda,\lambda'}_{\bk}\cdot \boldsymbol{e}_{\sigma}(\bq_{\parallel})\Big(\boldsymbol{d}^{\nu',\nu}_{\bk'}\cdot\boldsymbol{e}_{\sigma}(\bq_{\parallel})\Big)^* \\
&\times a^{\dagger}_{\bk,\lambda}a^{\dagger}_{\bk',\nu}a^{\phantom\dagger}_{\bk'+\bq_{\parallel},\nu'}a^{\phantom\dagger}_{\bk-\bq_{\parallel},\lambda'}
\end{split}
\label{eq:H_dip-dip_2Q}
\end{equation}
with $w=\int\,dz |\xi(z)|^4$. Note that if we again limit ourselves to dipole matrix elements of the types $\boldsymbol{d}^{cv}_{\bk}$ and $\boldsymbol{d}^{vc}_{\bk}$, 
the dipole-dipole term assumes the form of an electron-hole exchange interaction well-known for the longitudinal Coulomb interaction 
(with $\boldsymbol{e}_{\sigma}(\bq)\propto \bq $) \cite{qiu_nonanalyticity_2015,glazov_spin_2015}.
The final light-matter interaction Hamiltonian is obtained by adding $H_{\textrm{dip}}$ and $H_{\textrm{dip-dip}}$, keeping only ``resonant'' processes in $H_{\textrm{dip}}$, where the net sum of inter-gap excitations (either given by $b^{\dagger}_{\bq,\sigma}$ or $a^{\dagger}_{\bk,c}a^{\phantom\dagger}_{\bk-\bq_{\parallel},v}$) is zero:
%
\begin{equation}
 \begin{split}
H_{\textrm{LM}}&=-\sum_{\substack{\bk,\bq,\sigma \\ c,v}}i\frac{E_{q}}{\sqrt{\mathcal{A}}}
\boldsymbol{d}^{c,v}_{\bk}\cdot
\boldsymbol{u}_{\bq,\sigma}(z_0) b^{\phantom\dagger}_{\bq,\sigma}
a^{\dagger}_{\bk,c}a^{\phantom\dagger}_{\bk-\bq_{\parallel},v}\\
+&\sum_{\substack{\bk,\bq,\sigma \\ v,c}}i\frac{E_{q}}{\sqrt{\mathcal{A}}}
\boldsymbol{d}^{v,c}_{\bk} \cdot
\big(\boldsymbol{u}_{\bq,\sigma}(z_0)\big)^* b^{\dagger}_{\bq,\sigma}
a^{\dagger}_{\bk,v}a^{\phantom\dagger}_{\bk+\bq_{\parallel},c} \\
+&\frac{w}{2\varepsilon_0 n^2(z_0)\mathcal{A}}\sum_{\substack{\bk,\bk' \\ \mu,\bar{\mu},\nu,\bar{\nu}}}\sum_{\bq_{\parallel},\sigma}
\boldsymbol{d}^{\mu,\bar{\mu}}_{\bk}\cdot \boldsymbol{e}_{\sigma}(\bq_{\parallel})\Big(\boldsymbol{d}^{\nu,\bar{\nu}}_{\bk'}\cdot\boldsymbol{e}_{\sigma}(\bq_{\parallel})\Big)^* \\
&\times a^{\dagger}_{\bk,\mu}a^{\dagger}_{\bk',\bar{\nu}}a^{\phantom\dagger}_{\bk'+\bq_{\parallel},\nu}a^{\phantom\dagger}_{\bk-\bq_{\parallel},\bar{\mu}}
\,.
\end{split}
\label{eq:H_LM}
\end{equation}
Here, barred indices $\bar{\nu}$ denote a conduction band if $\nu$ is a valence band and vice versa.

\subsection{Derivation of EOM for correlation functions}

We start from a Hamiltonian for an interacting system of Bloch electrons, phonons and photons:
\begin{equation}
 \begin{split}
H&=H_{\textrm{carr}}+H_{\textrm{phon}}+H_{\textrm{field}}+H_{\textrm{Coul}}+H_{\textrm{c-ph}}+H_{\textrm{LM}}\,.
\end{split}
\label{eq:Hamiltonian}
\end{equation}
The Hamiltonians of free carriers, phonons, and photons are given by Eqs.~(\ref{eq:H_Bloch}), (\ref{eq:H_phon}) and (\ref{eq:H_field}), respectively.
The mutual interaction of carriers is given by $H_{\textrm{Coul}}$ (\ref{eq:carr_carr_Hamiltonian}), the carrier-phonon interaction is given by $H_{\textrm{c-ph}}$ (\ref{eq:carr_phon_Hamiltonian}) and the light-matter interaction is given by $H_{\textrm{LM}}$ (\ref{eq:H_LM}). In the following, we derive equations of motion (EOM) for various operators that can be used to represent observables such as optically induced polarizations or occupation functions of Bloch electrons and phonons. The dynamics of operators is governed by the Heisenberg EOM:
\begin{equation}
 \begin{split}
i\hbar\frac{d}{dt} \hat{A}= \left[ \hat{A}, H \right]\,.
\end{split}
\label{eq:heisenberg}
\end{equation}
We apply the so-called cluster expansion technique \cite{fricke_transport_1996, kira_many-body_2006} to formulate the dynamical equations in terms of correlation functions instead of expectation values.
To this end, we make use of the anti-commutation relations for fermionic operators,
\begin{equation}
 \begin{split}
&\left[a^{\phantom\dagger}_{\bk,\lambda},a^{\dagger}_{\bk',\lambda'} \right]_{+}=\delta_{\bk,\bk'}\delta_{\lambda,\lambda'}, \\ &\left[a^{\phantom\dagger}_{\bk,\lambda},a^{\phantom\dagger}_{\bk',\lambda'}\right]_{+}=\left[a^{\dagger}_{\bk,\lambda},a^{\dagger}_{\bk',\lambda'}\right]_{+}=0 \,,
\end{split}
\label{eq:anticomm_op}
\end{equation}
and commutation relations for bosonic operators:
\begin{equation}
 \begin{split}
&\left[D^{\phantom\dagger}_{\bq,j},D^{\dagger}_{\bq',j'} \right]=\delta_{\bq,\bq'}\delta_{j,j'}, \\ &\left[D^{\phantom\dagger}_{\bq,j},D^{\phantom\dagger}_{\bq',j'}\right]=\left[D^{\dagger}_{\bq,j},D^{\dagger}_{\bq',j'}\right]=0 \,.
\end{split}
\label{eq:comm_op}
\end{equation}
Moreover, we apply the identity
\begin{equation}
 \begin{split}
\left[ \hat{A}\hat{B}, \hat{C} \right]&=\hat{A}\left[ \hat{B}, \hat{C} \right]+\left[ \hat{A}, \hat{C} \right]\hat{B} \\
                                      &=\hat{A}\left[ \hat{B}, \hat{C} \right]_{+}-\left[ \hat{A}, \hat{C} \right]_+\hat{B}\,.
\end{split}
\label{eq:ABC}
\end{equation}
As discussed in the main text, correlation functions are defined recursively as the difference between an expectation value and all of its possible factorizations
into smaller correlation functions, i.e. with less operators. For example, a general correlation of four carrier operators is given by
\begin{equation}
\begin{split}
\Delta &\big\langle a_{\bk-\bq,\textrm{1}}^{\dagger}a_{\bk'+\bq,\textrm{2}}^{\dagger}a_{\bk',\textrm{3}}^{\phantom\dagger} a_{\bk,\textrm{4}}^{\phantom\dagger} \big\rangle \\
=&\big\langle a_{\bk-\bq,\textrm{1}}^{\dagger}a_{\bk'+\bq,\textrm{2}}^{\dagger}a_{\bk',\textrm{3}}^{\phantom\dagger} a_{\bk,\textrm{4}}^{\phantom\dagger} \big\rangle \\
-&
\Delta \big\langle a_{\bk-\bq,\textrm{1}}^{\dagger}a_{\bk,\textrm{4}}^{\phantom\dagger}\big\rangle \Delta\big\langle a_{\bk'+\bq,\textrm{2}}^{\dagger}a_{\bk',\textrm{3}}^{\phantom\dagger}\big\rangle\delta_{\bq,\bn} \\
+&\Delta \big\langle a_{\bk-\bq,\textrm{1}}^{\dagger}a_{\bk',\textrm{3}}^{\phantom\dagger}\big\rangle \Delta\big\langle a_{\bk'+\bq,\textrm{2}}^{\dagger}a_{\bk,\textrm{4}}^{\phantom\dagger}\big\rangle
\delta_{\bq,\bk-\bk'}\,.
\end{split}
\label{eq:factorize}
\end{equation}
Note that translational invariance of the crystal enforces momentum conservation within any expectation value and that transposition of fermionic operators
yields a minus sign. 
\\
\\Before we proceed with the systematic derivation, we discuss some implications of our formulation of the light-matter interaction Hamiltonian in Appendix~\ref{sec:LM} along the lines of \cite{kira_quantum_1999}. The EOM for a general polarization operator 
$\hat{\psi}^{v,c}_{\bk,\bq}=a^{\dagger}_{\bk,v}a^{\phantom\dagger}_{\bk+\bq,c}$ 
is given by:
%
\begin{equation}
\begin{split}
&i\hbar\frac{d}{dt}\hat{\psi}^{v,c}_{\bk,\bq}\Big|_{\textrm{LM}}=\big[a^{\dagger}_{\bk,v}, H_{\textrm{LM}}\big]a^{\phantom\dagger}_{\bk+\bq,c} +a^{\dagger}_{\bk,v}\big[a^{\phantom\dagger}_{\bk+\bq,c}, H_{\textrm{LM}}\big] \\
&=\sum_{\bq_{\parallel}',\bar{v}}a^{\dagger}_{\bk+\bq_{\parallel}',\bar{v}}\boldsymbol{d}^{\bar{v},v}_{\bk}\cdot\boldsymbol{E}(\bq_{\parallel}')a^{\phantom\dagger}_{\bk+\bq_{\parallel},c}\\
&-\sum_{\bq_{\parallel}',\bar{c}}a^{\dagger}_{\bk,v}\boldsymbol{d}^{c,\bar{c}}_{\bk}\cdot\boldsymbol{E}(\bq_{\parallel}')a^{\phantom\dagger}_{\bk+\bq_{\parallel}-\bq_{\parallel}',\bar{c}}
\end{split}
\label{eq:pol_EOM_LM}
\end{equation}
%
with the operator 
\begin{equation}
\begin{split}
\boldsymbol{E}(\bq_{\parallel}')&=\sum_{q_z',\sigma}i\frac{E_{q'}}{\sqrt{\mathcal{A}}}\big( \boldsymbol{u}_{\bq',\sigma}(z_0)b^{\phantom\dagger}_{\bq',\sigma}-
                        \big(\boldsymbol{u}_{-\bq',\sigma}(z_0))^*b^{\dagger}_{-\bq',\sigma} \big) \\
                        -&\frac{w}{\varepsilon_0 n^2(z_0)\mathcal{A}}\sum_{\bk',\nu,\bar{\nu},\sigma}\boldsymbol{e}_{\sigma}(\bq_{\parallel}')
                        \big( \boldsymbol{d}^{\bar{\nu},\nu}_{\bk'}\cdot \boldsymbol{e}_{\sigma}(\bq_{\parallel}') \big)^*
                        a^{\dagger}_{\bk',\nu}a^{\phantom\dagger}_{\bk'+\bq_{\parallel}',\bar{\nu}} \\
                        &=\frac{1}{\varepsilon_0}\boldsymbol{D}'(\bq_{\parallel}',z_0)-\frac{w}{\varepsilon_0 n^2(z_0)\mathcal{A}}\boldsymbol{P}'(\bq_{\parallel}')  \,
\end{split}
\label{eq:E_field_op}
\end{equation}
that fulfils $\boldsymbol{E}(-\bq_{\parallel}')=\big( \boldsymbol{E}(\bq_{\parallel}') \big)^* $.
In the first term, we introduced the in-plane Fourier transform of the displacement field (\ref{eq:displacement}),
\begin{equation}
\begin{split}
\boldsymbol{D}'(\bq_{\parallel}',z)=\frac{1}{\mathcal{A}}\int d^2\boldsymbol{\rho} e^{-i\bq_{\parallel}'\cdot\boldsymbol{\rho}} \boldsymbol{D}'(\boldsymbol{\rho},z)\,,
\end{split}
\label{eq:FT_displacement}
\end{equation}
while the second term contains the two-dimensional polarization function $\boldsymbol{P}'(\bq_{\parallel}')$ expanded in terms of the polarization vectors $\boldsymbol{e}_{\sigma}(\bq_{\parallel}')$. Using the same assumptions as for the derivation of the dipole-dipole term (\ref{eq:H_dip-dip_2Q}), one can show that $\boldsymbol{P}'(\bq_{\parallel}')$ is connected to the Fourier transform of the transversal polarization $\boldsymbol{P}_{\textrm{T}}(\br) $:
\begin{equation}
\begin{split}
\frac{|\xi(z)|^2}{n^2(z)\mathcal{A}} \boldsymbol{P}'(\bq_{\parallel}') = \frac{1}{\mathcal{A}}\int d^2\boldsymbol{\rho} e^{-i\bq_{\parallel}'\cdot\boldsymbol{\rho}} \boldsymbol{P}_{\textrm{T}}(\boldsymbol{\rho},z)\,.
\end{split}
\label{eq:FT_P}
\end{equation}
The components of $\boldsymbol{P}_{\textrm{T}}(\br) $ are given by:
\begin{equation}
\begin{split}
P_{\textrm{T},\mu}(\br)=\sum_{i=1}^N\frac{q_i}{n_i^2}\delta^{\textrm{T}}_{\mu\nu}(\br-\br_i)r_{i,\nu}
\end{split}
\label{eq:P_T}
\end{equation}
with the transversal delta function $\delta^{\textrm{T}}_{\mu\nu}(\br-\br')$, which can be represented by the completeness of mode functions \cite{kira_quantum_1999}:
\begin{equation}
\begin{split}
\frac{\delta^{\textrm{T}}_{\mu\nu}(\br-\br')}{n^2(z)}=\Big[\sum_{\bq,\sigma}\big(\boldsymbol{U}_{\bq,\sigma}(\br) \big)^* \otimes \boldsymbol{U}_{\bq,\sigma}(\br') \Big]_{\mu\nu}\,.
\end{split}
\label{eq:delta_T}
\end{equation}
By means of the Heisenberg EOM for $\boldsymbol{A}'(\br,t)=\boldsymbol{A}(\br,t)$, it can be shown that the displacement field and the transversal polarization constitute the transversal electric field \cite{kira_quantum_1999}:
\begin{equation}
\begin{split}
\boldsymbol{E}_{\textrm{T}}'(\br,t)=-\frac{\partial}{\partial t} \boldsymbol{A}'(\br,t) =
\frac{1}{\varepsilon_0} \boldsymbol{D}'(\br,t)-\frac{1}{\varepsilon_0} \boldsymbol{P}_{\textrm{T}}(\br,t) \,,
\end{split}
\label{eq:E_T}
\end{equation}
which obeys the operator wave equation
\begin{equation}
\begin{split}
\Big[\Delta-\frac{n^2(z)}{c^2} \frac{\partial^2}{\partial t^2}\Big]\boldsymbol{E}_{\textrm{T}}'(\br,t)=\mu_0 n^2(z) \frac{\partial^2}{\partial t^2} \boldsymbol{P}_{\textrm{T}}(\br,t) \,
\end{split}
\label{eq:E_T_wave}
\end{equation}
consistent with Maxwell's equations for classical fields. Combining Eqs.~(\ref{eq:E_field_op}), (\ref{eq:FT_displacement}), (\ref{eq:FT_P}) and (\ref{eq:E_T}) we conclude that the operator $\boldsymbol{E}(\bq_{\parallel}',t)$ is the in-plane Fourier transform of $\boldsymbol{E}_{\textrm{T}}'(\br,t)$ averaged over the TMD layer in z direction with the confinement factor $|\xi(z)|^2$: 
\begin{equation}
\begin{split}
\boldsymbol{E}(\bq_{\parallel}',t)&=\int dz |\xi(z)|^2 \boldsymbol{E}(\bq_{\parallel}',z,t) \\
&= \frac{1}{\mathcal{A}}\int dz |\xi(z)|^2 \int d^2\boldsymbol{\rho} e^{-i\bq_{\parallel}'\cdot\boldsymbol{\rho}} \boldsymbol{E}_{\textrm{T}}'(\boldsymbol{\rho},z,t)\,.
\end{split}
\label{eq:FT_av_E}
\end{equation}
We therefore find the wave equation in Fourier space:
\begin{equation}
\begin{split}
\Big[\frac{\partial^2}{\partial z^2}-q^2_{\parallel}-\frac{n^2(z)}{c^2} \frac{\partial^2}{\partial t^2}\Big]\boldsymbol{E}(\bq_{\parallel}',z,t)=\mu_0 \frac{\partial^2}{\partial t^2} \frac{1}{\mathcal{A}} \boldsymbol{P}'(\bq_{\parallel}') |\xi(z)|^2\,.
\end{split}
\label{eq:E_T_wave_FT}
\end{equation}
\\
\\We now derive the hierarchy of EOM for correlation functions. 
Correlations of operators that do not allow for further factorization and therefore equal the corresponding expectation values are classified as \textit{singlets} \cite{kira_many-body_2006}. 
For electrons, these are the two-operator functions constituting the single-particle density matrix. Its matrix elements are the single-particle occupation functions $f_{\bk}^{\lambda}=\ev{a_{\bk,\lambda}^{\dagger}a_{\bk,\lambda}^{\phantom\dagger}}$ and microscopic polarizations $\psi_{\bk}^{\lambda,\lambda'}=\ev{a_{\bk,\lambda}^{\dagger}a_{\bk,\lambda'}^{\phantom\dagger}}$. While the occupation functions determine the carrier density in every Bloch band,
\begin{equation}
\begin{split}
n^{\lambda}=\frac{1}{\mathcal{A}}\sum_{\bk} f_{\bk}^{\lambda}\,,
\end{split}
\label{eq:carr_dens_band}
\end{equation}
the off-diagonal terms determine the optical response, as we show below. The EOM for the single-particle density matrix are:
\begin{equation}
\begin{split}
&i\hbar\frac{d }{dt}f^{\nu}_{\bk}\\&=2i\,\textrm{Im}\Big\{ \sum_{\substack{\bk',\bq \\ \beta,\lambda,\lambda'}} 
V^{\nu,\lambda,\lambda',\beta}_{\bk, \bk', \bk'+\bq, \bk-\bq} \ev{a_{\bk,\nu}^{\dagger}a_{\bk',\lambda}^{\dagger}a_{\bk'+\bq,\lambda'}^{\phantom\dagger}a_{\bk-\bq,\beta}^{\phantom\dagger}  }  \Big\} \\
&+2i\,\textrm{Im}\Big\{\sum_{\bq,j,\lambda}g^{\nu,\lambda}_{\bk,\bq,j}\ev{Q^{\phantom\dagger}_{\bq,j}a_{\bk,\nu}^{\dagger}a_{\bk-\bq,\lambda}^{\phantom\dagger}} \Big\} \\
&+2i\,\textrm{Im}\Big\{ 
\sum_{\bq_{\parallel}',\bar{\nu}}\ev{a^{\dagger}_{\bk+\bq_{\parallel}',\bar{\nu}}\boldsymbol{d}^{\bar{\nu},\nu}_{\bk}\cdot\boldsymbol{E}(\bq_{\parallel}')a^{\phantom\dagger}_{\bk,\nu}}
\Big\}
\end{split}
\label{eq:EOM_f_k_ev}
\end{equation}
and
\begin{equation}
\begin{split}
&i\hbar\frac{d }{dt}\psi^{v,c}_{\bk}=(\en{c}{\bk}-\en{v}{\bk})\psi^{v,c}_{\bk} \\
&+\sum_{\substack{\bk',\bq \\ \beta,\lambda,\lambda'}} 
\Big(V^{c,\lambda,\lambda',\beta}_{\bk, \bk', \bk'+\bq, \bk-\bq} \ev{a_{\bk,v}^{\dagger}a_{\bk',\lambda}^{\dagger}a_{\bk'+\bq,\lambda'}^{\phantom\dagger}a_{\bk-\bq,\beta}^{\phantom\dagger}  }   \\
&-\big(V^{v,\lambda,\lambda',\beta}_{\bk, \bk', \bk'+\bq, \bk-\bq}\big)^* \big(\ev{a_{\bk,c}^{\dagger}a_{\bk',\lambda}^{\dagger}a_{\bk'+\bq,\lambda'}^{\phantom\dagger}a_{\bk-\bq,\beta}^{\phantom\dagger}  }\big)^*
\Big) \\
&+\sum_{\bq,j,\lambda}\Big( 
g^{c,\lambda}_{\bk,\bq,j}\ev{Q^{\phantom\dagger}_{\bq,j}a_{\bk,v}^{\dagger}a_{\bk-\bq,\lambda}^{\phantom\dagger}} \\
&-\big(g^{v,\lambda}_{\bk,\bq,j}\big)^*\big(\ev{Q^{\phantom\dagger}_{\bq,j}a_{\bk,c}^{\dagger}a_{\bk-\bq,\lambda}^{\phantom\dagger}}\big)^* \Big) \\
&+\sum_{\bq_{\parallel}',\bar{v}}\ev{a^{\dagger}_{\bk+\bq_{\parallel}',\bar{v}}\boldsymbol{d}^{\bar{v},v}_{\bk}\cdot\boldsymbol{E}(\bq_{\parallel}')a^{\phantom\dagger}_{\bk,c}}\\
&-\sum_{\bq_{\parallel}',\bar{c}}\ev{a^{\dagger}_{\bk,v}\boldsymbol{d}^{c,\bar{c}}_{\bk}\cdot\boldsymbol{E}(\bq_{\parallel}')a^{\phantom\dagger}_{\bk-\bq_{\parallel}',\bar{c}}}\,,
\end{split}
\label{eq:EOM_psi_k_ev}
\end{equation}
where we introduced the combined phonon operator $Q^{\phantom\dagger}_{\bq,j}=D^{\phantom\dagger}_{\bq,j}+D^{\dagger}_{-\bq,j}$.
Note that we limit ourselves to polarizations between valence and conduction bands, which are connected to the fundamental optical transitions of the semiconductor around the bandgap. We assume that polarizations between conduction bands and between valence bands, respectively, are not efficiently driven and can therefore be neglected. Moreover, we do not take into account phonon-assisted or Auger-type scattering processes across the bandgap. While the former are far off-resonant, the latter take place on time scales much longer than those considered in this work \cite{kumar_exciton-exciton_2014,mouri_nonlinear_2014,yuan_exciton_2017}.
\\
\\Due to the many-body interaction terms in the Hamiltonian (\ref{eq:Hamiltonian}), the singlets couple to higher-order expectation values. We now replace these expectation values by the sum over all possible factorizations into smaller correlation functions, or \textit{clusters}, and the correlation on the same order, see Eqs.~(\ref{eq:cluster_hierarchy}) and (\ref{eq:factorize}):
\begin{equation}
\begin{split}
\frac{d }{dt}f^{\nu}_{\bk}&=\frac{2}{\hbar}\,\textrm{Im}\Big\{ 
\sum_{\bar{\nu}} \big(\Omega_{\bk}^{\bar{\nu},\nu}\big)^*\psi^{\bar{\nu},\nu}_{\bk}
\Big\} \\
&+\frac{2}{\hbar}\,\textrm{Im}\Big\{ \sum_{\substack{\bk',\bq \\ \beta,\lambda,\lambda'}} 
V^{\nu,\lambda,\lambda',\beta}_{\bk, \bk', \bk'+\bq, \bk-\bq} c_{\nu,\lambda,\lambda'\beta}^{\bq,\bk',\bk}  \Big\} \\
&+\frac{2}{\hbar}\,\textrm{Im}\Big\{\sum_{\bq,j,\lambda}g^{\nu,\lambda}_{\bk,\bq,j}\Big(   
\big(\Xi^{\lambda,\nu}_{\bk,\bq,j}\big)^*+\Xi^{\nu,\lambda}_{\bk-\bq,-\bq,j}
\Big) \Big\} \\
&+\frac{2}{\hbar}\,\textrm{Im}\Big\{ 
\sum_{\bq_{\parallel}',\bar{\nu}}\boldsymbol{d}^{\bar{\nu},\nu}_{\bk}\cdot\Delta\ev{a^{\dagger}_{\bk+\bq_{\parallel}',\bar{\nu}}\boldsymbol{E}(\bq_{\parallel}')a^{\phantom\dagger}_{\bk,\nu}}
\Big\}
\end{split}
\label{eq:EOM_f_k_corr}
\end{equation}
and
\begin{equation}
\begin{split}
&i\hbar\frac{d }{dt}\psi^{v,c}_{\bk}=(\ent{c}{\bk}-\ent{v}{\bk})\psi^{v,c}_{\bk} 
+\sum_{\bar{v}} \Omega_{\bk}^{v,\bar{v}} \psi^{\bar{v},c}_{\bk}
-\sum_{\bar{c}} \psi^{v,\bar{c}}_{\bk} \Omega_{\bk}^{\bar{c},c} 
\\
&+\sum_{\substack{\bk',\bq \\ \beta,\lambda,\lambda'}} 
\Big(V^{c,\lambda,\lambda',\beta}_{\bk, \bk', \bk'+\bq, \bk-\bq} c_{v,\lambda,\lambda'\beta}^{\bq,\bk',\bk}  -\big(V^{v,\lambda,\lambda',\beta}_{\bk, \bk', \bk'+\bq, \bk-\bq}c_{c,\lambda,\lambda'\beta}^{\bq,\bk',\bk}\big)^*
\Big) \\
&+\sum_{j,\lambda}2\,\textrm{Re}\big\{\ev{D^{\phantom\dagger}_{\bn,j}}\big\}\Big( 
g^{c,\lambda}_{\bk,\bn,j}\psi^{v,\lambda}_{\bk} -\big(g^{v,\lambda}_{\bk,\bn,j}\psi^{c,\lambda}_{\bk}\big)^* \Big) \\
&+\sum_{\bq,j,\lambda}\Big( 
g^{c,\lambda}_{\bk,\bq,j}\big( \big(\Xi^{\lambda,v}_{\bk,\bq,j} \big)^* + \Xi^{v,\lambda}_{\bk-\bq,-\bq,j} \big) \\
&-\big(g^{v,\lambda}_{\bk,\bq,j}\big)^*\big( \Xi^{\lambda,c}_{\bk,\bq,j}  +   \big( \Xi^{c,\lambda}_{\bk-\bq,-\bq,j} \big)^* \Big) \\
&+\sum_{\bq_{\parallel}',\bar{v}}\boldsymbol{d}^{\bar{v},v}_{\bk}\cdot\Delta\ev{a^{\dagger}_{\bk+\bq_{\parallel}',\bar{v}}\boldsymbol{E}(\bq_{\parallel}')a^{\phantom\dagger}_{\bk,c}}\\
&-\sum_{\bq_{\parallel}',\bar{c}}\boldsymbol{d}^{c,\bar{c}}_{\bk}\cdot\Delta\ev{a^{\dagger}_{\bk,v}\boldsymbol{E}(\bq_{\parallel}')a^{\phantom\dagger}_{\bk-\bq_{\parallel}',\bar{c}}}\,.
\end{split}
\label{eq:EOM_psi_k_corr}
\end{equation}
Here, we have introduced correlations of four carrier operators, which are classified as \textit{doublets}:
\begin{equation}
\begin{split}
c_{\lambda_1,\lambda_2,\lambda_3\lambda_4}^{\bq,\bk',\bk}= 
\Delta\ev{a_{\bk,\lambda_1}^{\dagger}a_{\bk',\lambda_2}^{\dagger}a_{\bk'+\bq,\lambda_3}^{\phantom\dagger}a_{\bk-\bq,\lambda_4}^{\phantom\dagger}  }
\,.
\end{split}
\label{eq:def_C}
\end{equation}
We will discuss their EOM later on.
A part of the factorization of corresponding expectation values into singlets leads to Hartree-Fock terms that renormalize the single-particle band structure appearing in the noninteracting term in Eq.~(\ref{eq:EOM_psi_k_corr}). We treat Hartree-Fock renormalizations in the electron-hole picture with $\en{h}{\bk}=-\en{v}{\bk}$ and $f^{h}_{\bk}=1-f^{v}_{\bk}$ (while keeping the Coulomb matrix elements unchanged):
\begin{equation}
\begin{split}
\ent{e}{\bk}=\en{e}{\bk}&-\sum_{\bk',e'} V^{e,e',e,e'}_{\bk, \bk', \bk, \bk'} f^{e'}_{\bk'}+\sum_{\bk',h'} U^{e,h',e,h'}_{\bk, \bk', \bk, \bk'} f^{h'}_{\bk'} \\
&+\sum_{\bk',e'} V^{e,e',e',e}_{\bk, \bk', \bk', \bk} f^{e'}_{\bk'}                        -\sum_{\bk',h'} V^{e,h',h',e}_{\bk, \bk', \bk', \bk} f^{h'}_{\bk'} \\
=\en{e}{\bk}&+\Sigma^{\textrm{HF},e}_{\bk} \,,   \\                     
\ent{h}{\bk}=\en{h}{\bk}&-\sum_{\bk',h'} V^{h,h',h,h'}_{\bk, \bk', \bk, \bk'} f^{h'}_{\bk'} +\sum_{\bk',e'} U^{h,e',h,e'}_{\bk, \bk', \bk, \bk'} f^{e'}_{\bk'}\\                                     &+\sum_{\bk',h'} V^{h,h',h',h}_{\bk, \bk', \bk', \bk} f^{h'}_{\bk'}
        -\sum_{\bk',e'} V^{h,e',e',h}_{\bk, \bk', \bk', \bk} f^{e'}_{\bk'} \\
             =\en{h}{\bk}&+\Sigma^{\textrm{HF},h}_{\bk}                        
                                     \,.
\end{split}
\label{eq:def_eps_tilde}
\end{equation}
Electron-hole exchange contributions are calculated using bare matrix elements as discussed in Ref.~\cite{erben_excitation-induced_2018}. Hartree-type matrix elements with zero momentum transfer are obtained along the lines of \cite{hofmann_k-resolved_2025}.
\\
\\In the case of bosonic particles, single operators have finite expectation values and are therefore classified as singlets, which is consistent with the fact that they appear together with pairs of fermionic operators in the Hamiltonian. The electric field operator $\boldsymbol{E}(\bq_{\parallel}')$ contains photon and carrier operators on the singlet level, see Eq.~(\ref{eq:E_field_op}). Due to spatial homogeneity, the factorization of light-matter-interaction-induced expectation values yields electric-field singlets with zero in-plane momentum, corresponding to a propagation perpendicular to the two-dimensional layer:
\begin{equation}
\begin{split}
\ev{a^{\dagger}_{\bk+\bq_{\parallel}',\bar{\nu}}\boldsymbol{E}(\bq_{\parallel}')a^{\phantom\dagger}_{\bk,\nu}}&=
\ev{\boldsymbol{E}(\bn)}\psi^{\bar{\nu},\nu}_{\bk} \delta_{\bq_{\parallel}',\bn} \\
&+\Delta\ev{a^{\dagger}_{\bk+\bq_{\parallel}',\bar{\nu}}\boldsymbol{E}(\bq_{\parallel}')a^{\phantom\dagger}_{\bk,\nu}}
\,.
\end{split}
\label{eq:light_matter_fact}
\end{equation}
According to Eqs.~(\ref{eq:FT_av_E}) and (\ref{eq:E_T_wave_FT}), the dynamics of the electric-field singlets $\boldsymbol{E}^{\textrm{2d}}=\ev{\boldsymbol{E}(\bn)}$ follows from a semi-classical wave equation
\begin{equation}
\begin{split}
\Big[\frac{\partial^2}{\partial z^2}-\frac{n^2(z)}{c^2} \frac{\partial^2}{\partial t^2}\Big]\boldsymbol{E}(z,t)=\mu_0 \frac{\partial^2}{\partial t^2} \boldsymbol{P}^{\textrm{2d}}(t) |\xi(z)|^2
\end{split}
\label{eq:E_T_wave_FT_class}
\end{equation}
by averaging the solution over the two-dimensional layer in z-direction. Here, we have introduced the two-dimensional polarization
\begin{equation}
\begin{split}
\boldsymbol{P}^{\textrm{2d}}(t)&=\frac{1}{\mathcal{A}} \boldsymbol{P}'(\bn) \\ &=
\frac{1}{\mathcal{A}}\sum_{\bk',\nu,\bar{\nu},\sigma}\boldsymbol{e}_{\sigma}(\bn)
                        \big( \boldsymbol{d}^{\bar{\nu},\nu}_{\bk'}\cdot \boldsymbol{e}_{\sigma}(\bn) \big)^*
                        \psi_{\bk'}^{\nu,\bar{\nu}}
\,.
\end{split}
\label{eq:def_pol_2d}
\end{equation}
The factorizations of the type (\ref{eq:light_matter_fact}) together with the remaining singlet factorization of the Coulomb-induced term in Eqs.~(\ref{eq:EOM_f_k_ev}) and (\ref{eq:EOM_psi_k_ev}) are collected in the so-called renormalized Rabi energy
\begin{equation}
\begin{split}
\Omega_{\bk}^{\nu,\bar{\nu}}=\boldsymbol{d}^{\bar{\nu},\nu}_{\bk}\cdot \boldsymbol{E}^{\textrm{2d}}+\sum_{\bk',\lambda,\lambda'}
V^{\bar{\nu},\lambda,\nu,\lambda'}_{\bk, \bk', \bk, \bk'}\psi_{\bk'}^{\lambda,\lambda'}
\,.
\end{split}
\label{eq:def_Rabi}
\end{equation}
The "direct" Coulomb-induced term in Eq.~(\ref{eq:def_Rabi}) accounts for the attractive electron-hole interaction that renormalizes the optical interband polarizations $\psi^{v,c}_{\bk}$ \cite{haug_quantum_2004}. We have neglected a corresponding electron-hole-exchange-type term that introduces further fine-structure renormalizations of the exciton spectrum \cite{qiu_nonanalyticity_2015}. Similar exchange terms are also neglected in higher-order correlations.
In this work, we focus on semiclassical light-matter interaction by discarding correlations between electric-field operators and carriers:
\begin{equation}
\begin{split}
\Delta\ev{a^{\dagger}_{\bk+\bq_{\parallel}',\bar{\nu}}\boldsymbol{E}(\bq_{\parallel}')a^{\phantom\dagger}_{\bk,\nu}}\equiv 0
\,.
\end{split}
\label{eq:light_matter_neglect}
\end{equation}
Hence, we do not take into account quantum-electrodynamical effects such as spontaneous emission \cite{kira_quantum_1999, kira_many-body_2006} and transversal electron-hole exchange interaction induced by the dipole-dipole term (\ref{eq:H_dip-dip_2Q}).
In the absence of doublet correlations, the coupled EOM (\ref{eq:EOM_f_k_corr}) and (\ref{eq:EOM_psi_k_corr}) for carrier singlets are known as semiconductor Bloch equations \cite{haug_quantum_2004,kira_many-body_2006}. The microscopic polarizations also couple to singlets of the type $\ev{D^{\dagger}_{\bn,j}}$, which represent coherent phonons \cite{kira_many-body_2006}. They are governed by the EOM:
\begin{equation}
\begin{split}
i\hbar\frac{d }{dt}\ev{D^{\dagger}_{\bn,j}}&=(-\hbar\omega_{\bn,j}-i\hbar\gamma_{\textrm{phon}})\ev{D^{\dagger}_{\bn,j}}\\&-\sum_{\bk,\nu,\nu'}g^{\nu,\nu'}_{\bk,\bn,j}\psi_{\bk}^{\nu,\nu'}.
\end{split}
\label{eq:EOM_D}
\end{equation}
It has been shown in Ref.~\cite{perfetto_theory_2024} that coherent phonons couple to excitons via unscreened matrix elements. Since our DFPT-based matrix elements describe a freestanding monolayer, for which there is no screening at $\bq=\bn$ in the macroscopic limit, we assume that screening effects are weak. We therefore keep the matrix elements as they emerge from DFPT.
In Eq.~(\ref{eq:EOM_D}), we introduced a phenomenological phonon damping rate $\gamma_{\textrm{phon}}$, which is due to lattice anharmonicities. Phonon thermalization can be systematically included in the cluster formalism by means of Lindblad terms \cite{breuer_theory_2007}
\begin{equation}
\begin{split}
&\frac{d }{dt}\ev{\hat{A}}\Big|_{\textrm{Lindblad}}=\sum_{i}\gamma_{i}\ev{\left[\hat{O}^{\dagger}_{i},\hat{A}\right]\hat{O}^{\phantom\dagger}_{i}+\hat{O}^{\dagger}_{i}\left[\hat{A},\hat{O}^{\phantom\dagger}_{i}\right]  }\,,
\end{split}
\label{eq:Lindblad}
\end{equation}
where the set $\left\{i\right\}$ describes transitions within the phonon system induced by a bath of phonons, which is not part of the system Hamiltonian. The set contains excitation and decay processes of phonons, such that either $\hat{O}_{i}=D^{\phantom\dagger}_{\bq,j}$ with $\gamma_{i}=\gamma_{\textrm{phon}}(1+n^{\textrm{B}}_{\bq,j})$ or $\hat{O}_{i}=D^{\dagger}_{\bq,j}$ with $\gamma_{i}=\gamma_{\textrm{phon}}n^{\textrm{B}}_{\bq,j}$, where $n^{\textrm{B}}_{\bq,j}$ is a Bose function at the bath temperature. In this work, we use $\gamma_{\textrm{phon}}=10$ ps \cite{gu_phonon_2014,sood_quasi-ballistic_2019}.
\\
\\The carrier singlets also couple to so-called phonon-assisted polarizations
\begin{equation}
\begin{split}
\Xi^{\nu,\nu'}_{\bk,\bq,j}= 
\Delta\ev{D^{\dagger}_{\bq,j} a_{\bk-\bq,\nu}^{\dagger}a_{\bk,\nu'}^{\phantom\dagger}}
\,,
\end{split}
\label{eq:def_Xi}
\end{equation}
which are mixed fermion-boson doublets. Their EOM is:
\begin{equation}
\begin{split}
&i\hbar\frac{d }{dt}\Xi^{\nu,\nu'}_{\bk,\bq,j}=(\ent{\nu'}{\bk}-\ent{\nu}{\bk-\bq}-\hbar\omega_{\bq,j}-i\hbar\gamma_{\textrm{phon}})\Xi^{\nu,\nu'}_{\bk,\bq,j} \\ 
&+\sum_{\bar{\nu}} \Omega_{\bk-\bq}^{\nu,\bar{\nu}} \Xi^{\bar{\nu},\nu'}_{\bk,\bq,j}
-\sum_{\bar{\nu}'} \Xi^{\nu,\bar{\nu}'}_{\bk,\bq,j} \Omega_{\bk}^{\bar{\nu}',\nu'} 
\\
&+\sum_{j',\lambda}2\,\textrm{Re}\big\{\ev{D^{\phantom\dagger}_{\bn,j'}}\big\}\Big( 
g^{\nu',\lambda}_{\bk,\bn,j'}\Xi^{\nu,\lambda}_{\bk,\bq,j} -\big(g^{\nu,\lambda}_{\bk-\bq,\bn,j'}\big)^*\Xi^{\lambda,\nu'}_{\bk,\bq,j} \Big) \\
&+\sum_{\bar{\nu},\bar{\nu}'}g^{\bar{\nu}',\bar{\nu}}_{\bk,\bq,j}\psi^{\bar{\nu}',\nu'}_{\bk}\psi^{\nu,\bar{\nu}}_{\bk-\bq}
-\sum_{\bar{\nu}'}g^{\bar{\nu}',\nu}_{\bk,\bq,j}\psi^{\bar{\nu}',\nu'}_{\bk}(1-f^{\nu}_{\bk-\bq})
\\
&+\sum_{\bar{\nu}}g^{\nu',\bar{\nu}}_{\bk,\bq,j}f^{\nu'}_{\bk}\psi^{\nu,\bar{\nu}}_{\bk-\bq}
-g^{\nu',\nu}_{\bk,\bq,j}f^{\nu'}_{\bk}(1-f^{\nu}_{\bk-\bq}) \\
&-\sum_{\bk',\lambda,\lambda'}g^{\lambda,\lambda'}_{\bk'+\bq,\bq,j}\big(c^{\bq,\bk',\bk}_{\nu',\lambda,\lambda',\nu}\big)^*
\\
&+\sum_{j'}\Delta\ev{D^{\dagger}_{\bq,j}Q^{\phantom\dagger}_{\bq,j'}}\Big(\sum_{\bar{\nu}}  
g^{\nu',\bar{\nu}}_{\bk,\bq,j'}\psi^{\nu,\bar{\nu}}_{\bk-\bq}+g^{\nu',\nu}_{\bk,\bq,j'}f^{\nu}_{\bk-\bq} \Big)
\\
&-\sum_{j'}\Delta\ev{D^{\dagger}_{\bq,j}Q^{\phantom\dagger}_{\bq,j'}}\Big(\sum_{\bar{\nu}'}  
g^{\bar{\nu}',\nu}_{\bk,\bq,j'}\psi^{\bar{\nu}',\nu'}_{\bk}+g^{\nu',\nu}_{\bk,\bq,j'}f^{\nu'}_{\bk} \Big)
\\
&+\sum_{\bq_{\parallel}',\bar{\nu}}\boldsymbol{d}^{\bar{\nu},\nu}_{\bk-\bq}\cdot\Delta\ev{D^{\dagger}_{\bq,j}a^{\dagger}_{\bk-\bq+\bq_{\parallel}',\bar{\nu}}\boldsymbol{E}(\bq_{\parallel}')a^{\phantom\dagger}_{\bk,\nu'}}\\
&-\sum_{\bq_{\parallel}',\bar{\nu}'}\boldsymbol{d}^{\nu',\bar{\nu}'}_{\bk}\cdot\Delta\ev{D^{\dagger}_{\bq,j}a^{\dagger}_{\bk-\bq,\nu}\boldsymbol{E}(\bq_{\parallel}')a^{\phantom\dagger}_{\bk-\bq_{\parallel}',\bar{\nu}'}} \\
&+i\hbar\frac{d }{dt}\Xi^{\nu,\nu'}_{\bk,\bq,j}\Big|_{\textrm{c-c}}+i\hbar\frac{d }{dt}\Xi^{\nu,\nu'}_{\bk,\bq,j}\Big|_{\textrm{T}}
\,,
\end{split}
\label{eq:EOM_Xi}
\end{equation}
where we again discard correlations involving the electric field:
\begin{equation}
\begin{split}
\Delta\ev{D^{\dagger}_{\bq,j}a^{\dagger}_{\bk-\bq+\bq_{\parallel}',\bar{\nu}}\boldsymbol{E}(\bq_{\parallel}')a^{\phantom\dagger}_{\bk,\nu'}}\equiv 0
\,.
\end{split}
\label{eq:light_matter_neglect_Xi}
\end{equation}
Note that we also discarded correlations between phonons and the electric field in Eq.~(\ref{eq:EOM_Xi}).
Carrier-carrier interaction leads to renormalized two-particle energies by mixing phonon-assisted polarizations with different carrier quantum numbers, similar to the renormalized Rabi energy (\ref{eq:def_Rabi}) or a Bethe-Salpeter equation:
\begin{equation}
\begin{split}
&i\hbar\frac{d }{dt}\Xi^{\nu,\nu'}_{\bk,\bq,j}\Big|_{\textrm{c-c}} \\
&=(f^{\nu'}_{\bk}-f^{\nu}_{\bk-\bq})\sum_{\bl,\lambda,\lambda'}
\Big(V^{\nu',\lambda,\nu,\lambda'}_{\bk, \bl-\bq, \bk-\bq, \bl}-V^{\nu',\lambda,\lambda',\nu}_{\bk, \bl-\bq, \bl, \bk-\bq}\Big)\Xi^{\lambda,\lambda'}_{\bl,\bq,j} \\
&+ \sum_{\bar{\nu}'}\psi^{\bar{\nu}',\nu'}_{\bk}\sum_{\bl,\lambda,\lambda'}
\Big(V^{\bar{\nu}',\lambda,\nu,\lambda'}_{\bk, \bl-\bq, \bk-\bq, \bl}-V^{\bar{\nu}',\lambda,\lambda',\nu}_{\bk, \bl-\bq, \bl, \bk-\bq}\Big)\Xi^{\lambda,\lambda'}_{\bl,\bq,j} \\
&- \sum_{\bar{\nu}}\psi^{\nu,\bar{\nu}}_{\bk-\bq}\sum_{\bl,\lambda,\lambda'}
\Big(V^{\nu',\lambda,\bar{\nu},\lambda'}_{\bk, \bl-\bq, \bk-\bq, \bl}-V^{\nu',\lambda,\lambda',\bar{\nu}}_{\bk, \bl-\bq, \bl, \bk-\bq}\Big)\Xi^{\lambda,\lambda'}_{\bl,\bq,j}
\,.
\end{split}
\label{eq:EOM_Xi_cc}
\end{equation}
Such terms dynamically introduce exciton binding energies into higher-order correlations by "diagonalizing" the two-particle problem, without explicitly changing to an exciton representation.
We also obtain coupling to \textit{triplet} correlations via
\begin{equation}
\begin{split}
&i\hbar\frac{d }{dt}\Xi^{\nu,\nu'}_{\bk,\bq,j}\Big|_{\textrm{T}} \\
&=\sum_{\bq',j',\lambda}g^{\lambda,\nu}_{\bk-\bq+\bq',\bq',j'}\Delta\ev{D^{\dagger}_{\bq,j}Q^{\phantom\dagger}_{\bq',j'}a^{\dagger}_{\bk-\bq+\bq',\lambda}a^{\phantom\dagger}_{\bk,\nu'}} \\
&-\sum_{\bq',j',\lambda}g^{\nu',\lambda}_{\bk,\bq',j'}\Delta\ev{D^{\dagger}_{\bq,j}Q^{\phantom\dagger}_{\bq',j'}a^{\dagger}_{\bk-\bq,\nu}a^{\phantom\dagger}_{\bk-\bq',\lambda}}
\\
&+\sum_{\substack{\bk',\bl \\ \beta,\lambda,\lambda'}} 
\Big(V^{\nu',\lambda,\lambda',\beta}_{\bk, \bk', \bk'+\bl, \bk-\bl} \Delta\ev{D^{\dagger}_{\bq,j}
a_{\bk-\bq,\nu}^{\dagger}a_{\bk',\lambda}^{\dagger}a_{\bk'+\bl,\lambda'}^{\phantom\dagger}a_{\bk-\bl,\beta}^{\phantom\dagger}
}  \\
&-V^{\beta,\lambda,\lambda',\nu}_{\bk-\bl, \bk'+\bl, \bk'+\bq, \bk-\bq}\Delta\ev{D^{\dagger}_{\bq,j}
a_{\bk-\bl,\beta}^{\dagger}a_{\bk'+\bl,\lambda}^{\dagger}a_{\bk'+\bq,\lambda'}^{\phantom\dagger}a_{\bk,\nu'}^{\phantom\dagger}
}
\Big)
\,,
\end{split}
\label{eq:EOM_Xi_T}
\end{equation}
which will be discussed later on.
The phonon-assisted polarizations couple to phonon doublets $\Delta\ev{D^{\dagger}_{\bq,j}D^{\phantom\dagger}_{\bq,j'}}$ and $\Delta\ev{D^{\dagger}_{\bq,j}D^{\dagger}_{-\bq,j'}} $, the first one being connected to the phonon occupation $n_{\bq,j}=\ev{D^{\dagger}_{\bq,j}D^{\phantom\dagger}_{\bq,j}}$.
The EOM of phonon doublets are:
\begin{equation}
\begin{split}
&i\hbar\frac{d }{dt} \Delta\ev{D^{\dagger}_{\bq,j}D^{\phantom\dagger}_{\bq,j'}} = (\hbar\omega_{\bq,j'}-\hbar\omega_{\bq,j})\Delta\ev{D^{\dagger}_{\bq,j}D^{\phantom\dagger}_{\bq,j'}} \\ 
&-i\hbar2\gamma_{\textrm{phon}}(\Delta\ev{D^{\dagger}_{\bq,j}D^{\phantom\dagger}_{\bq,j'}}-n^{\textrm{B}}_{\bq,j}\delta_{j,j'}) \\
&+\sum_{\bk,\lambda,\lambda'} \Big(g^{\lambda,\lambda'}_{\bk,-\bq,j'}\Xi^{\lambda,\lambda'}_{\bk+\bq,\bq,j}
-g^{\lambda,\lambda'}_{\bk,\bq,j}\big(\Xi^{\lambda',\lambda}_{\bk,\bq,j'}\big)^*
\Big)
\\
&i\hbar\frac{d }{dt} \Delta\ev{D^{\dagger}_{\bq,j}D^{\dagger}_{-\bq,j'}} = -(\hbar\omega_{\bq,j}+\hbar\omega_{-\bq,j'})\Delta\ev{D^{\dagger}_{\bq,j}D^{\dagger}_{-\bq,j'}}\\
&-i\hbar2\gamma_{\textrm{phon}}\Delta\ev{D^{\dagger}_{\bq,j}D^{\dagger}_{-\bq,j'}} \\
&-\sum_{\bk,\lambda,\lambda'} \Big(g^{\lambda,\lambda'}_{\bk,-\bq,j'}\Xi^{\lambda,\lambda'}_{\bk+\bq,\bq,j}
+g^{\lambda,\lambda'}_{\bk,\bq,j}\Xi^{\lambda,\lambda'}_{\bk-\bq,-\bq,j'}
\Big)
\,.
\end{split}
\label{eq:EOM_DD}
\end{equation}
Note that phonon occupations $n_{\bq,j}$ relax towards Bose functions at the bath temperature with the relaxation time $\tau=1/(2\gamma_{\textrm{phon}})$. While the inclusion of a decay channel via Lindblad terms is essential for pure phonon correlations, it has little influence on correlations involving carrier operators such as $\Xi^{\nu,\nu'}_{\bk,\bq,j}$, for which much faster relaxation mechanisms dominate the dynamics.
\\
\\The phonon-assisted polarizations are also coupled to carrier-carrier doublets via carrier-phonon interaction. Discarding correlations involving the electric field, the dynamics of these doublets is governed by the following EOM:
\begin{widetext}
\begin{equation}
\begin{split}
&i\hbar\frac{d }{dt} c^{\bq,\bk',\bk}_{\lambda,\nu,\nu',\lambda'}=(\ent{\lambda'}{\bk-\bq}+\ent{\nu'}{\bk'+\bq}-\ent{\nu}{\bk'}-\ent{\lambda}{\bk})c^{\bq,\bk',\bk}_{\lambda,\nu,\nu',\lambda'} \\ 
&+\sum_{\bar{\lambda}} \Omega_{\bk}^{\lambda,\bar{\lambda}} c^{\bq,\bk',\bk}_{\bar{\lambda},\nu,\nu',\lambda'}
+\sum_{\bar{\nu}} \Omega_{\bk'}^{\nu,\bar{\nu}} c^{\bq,\bk',\bk}_{\lambda,\bar{\nu},\nu',\lambda'} 
-\sum_{\bar{\nu}'}  c^{\bq,\bk',\bk}_{\lambda,\nu,\bar{\nu}',\lambda'} \Omega_{\bk'+\bq}^{\bar{\nu}',\nu'}
-\sum_{\bar{\lambda}'}  c^{\bq,\bk',\bk}_{\lambda,\nu,\nu',\bar{\lambda}'} \Omega_{\bk-\bq}^{\bar{\lambda}',\lambda'}
\\
&+\sum_{j,\mu}2\,\textrm{Re}\big\{\ev{D^{\phantom\dagger}_{\bn,j}}\big\}\Big( 
g^{\nu',\mu}_{\bk'+\bq,\bn,j}c^{\bq,\bk',\bk}_{\lambda,\nu,\mu,\lambda'}
+g^{\lambda',\mu}_{\bk-\bq,\bn,j}c^{\bq,\bk',\bk}_{\lambda,\nu,\nu',\mu} 
-\big(g^{\lambda,\mu}_{\bk,\bn,j}\big)^*c^{\bq,\bk',\bk}_{\mu,\nu,\nu',\lambda'} 
-\big(g^{\nu,\mu}_{\bk',\bn,j}\big)^*c^{\bq,\bk',\bk}_{\lambda,\mu,\nu',\lambda'}
\Big)\\
&+i\hbar\frac{d }{dt} c^{\bq,\bk',\bk}_{\lambda,\nu,\nu',\lambda'}\Big|_{\textrm{S}}
+i\hbar\frac{d }{dt} c^{\bq,\bk',\bk}_{\lambda,\nu,\nu',\lambda'}\Big|_{\textrm{D,c-ph}} 
+i\hbar\frac{d }{dt} c^{\bq,\bk',\bk}_{\lambda,\nu,\nu',\lambda'}\Big|_{\textrm{D,c-c}}
+i\hbar\frac{d }{dt} c^{\bq,\bk',\bk}_{\lambda,\nu,\nu',\lambda'}\Big|_{\textrm{T}}
\,.
\end{split}
\label{eq:EOM_C}
\end{equation}
The singlet factorization, which leads to carrier-carrier scattering of the single-particle density matrix, is given by
\begin{equation}
\begin{split}
i\hbar\frac{d }{dt} &c^{\bq,\bk',\bk}_{\lambda,\nu,\nu',\lambda'}\Big|_{\textrm{S}}= \\
\sum_{\mu,\mu'}\Big(
&\big(V^{\mu,\mu',\nu',\lambda'}_{\bk, \bk',\bk'+\bq,\bk-\bq}-V^{\mu,\mu',\lambda',\nu'}_{\bk, \bk',\bk-\bq,\bk'+\bq}\big)^*\psi^{\lambda,\mu}_{\bk}\psi^{\nu,\mu'}_{\bk'}
-\big(V^{\lambda,\nu,\mu,\mu'}_{\bk, \bk',\bk'+\bq,\bk-\bq}-V^{\lambda,\nu,\mu',\mu}_{\bk, \bk',\bk-\bq,\bk'+\bq}\big)^*\psi^{\mu,\nu}_{\bk'+\bq}\psi^{\mu',\lambda'}_{\bk-\bq}
\Big) \\
-\sum_{\mu,\mu',\mu''}
\Big(
&\big(V^{\mu,\mu',\nu',\mu''}_{\bk, \bk',\bk'+\bq,\bk-\bq}-V^{\mu,\mu',\mu'',\nu'}_{\bk, \bk',\bk-\bq,\bk'+\bq}\big)^*\psi^{\lambda,\mu}_{\bk}\psi^{\nu,\mu'}_{\bk'}\psi^{\mu'',\lambda'}_{\bk-\bq} 
+\big(V^{\mu,\mu',\mu'',\lambda'}_{\bk, \bk',\bk'+\bq,\bk-\bq}-V^{\mu,\mu',\lambda',\mu''}_{\bk, \bk',\bk-\bq,\bk'+\bq}\big)^*\psi^{\lambda,\mu}_{\bk}\psi^{\nu,\mu'}_{\bk'}\psi^{\mu'',\nu'}_{\bk'+\bq} \\
-&\big(V^{\lambda,\mu,\mu',\mu''}_{\bk, \bk',\bk'+\bq,\bk-\bq}-V^{\lambda,\mu,\mu'',\mu'}_{\bk, \bk',\bk-\bq,\bk'+\bq}\big)^*\psi^{\nu,\mu}_{\bk'}\psi^{\mu',\nu'}_{\bk'+\bq}\psi^{\mu'',\lambda'}_{\bk-\bq} 
-\big(V^{\mu,\nu,\mu',\mu''}_{\bk, \bk',\bk'+\bq,\bk-\bq}-V^{\mu,\nu,\mu'',\mu'}_{\bk, \bk',\bk-\bq,\bk'+\bq}\big)^*\psi^{\lambda,\mu}_{\bk}\psi^{\mu',\nu'}_{\bk'+\bq}\psi^{\mu'',\lambda'}_{\bk-\bq}
\Big)
\,.
\end{split}
\label{eq:EOM_C_singlet}
\end{equation}
Note that here we included carrier populations $f^{\lambda}_{\bk}$ as band-diagonal polarizations for the sake of notational simplicity.
Carrier-carrier doublets couple to phonon-assisted polarizations via
\begin{equation}
\begin{split}
&i\hbar\frac{d }{dt} c^{\bq,\bk',\bk}_{\lambda,\nu,\nu',\lambda'}\Big|_{\textrm{D,c-ph}}=\\
\sum_{\mu,j}\Big[
&\psi^{\mu,\nu'}_{\bk'+\bq}g^{\mu,\lambda}_{\bk'+\bq,\bk'+\bq-\bk,j}\big(
\pap{\nu}{\lambda'}{\bk-\bq}{\bk-\bk'-\bq}{j}+(\pap{\lambda'}{\nu}{\bk'}{\bk'+\bq-\bk}{j})^*
\big)
-\psi^{\mu,\lambda'}_{\bk-\bq}g^{\mu,\lambda}_{\bk-\bq,-\bq,j}\big(
\pap{\nu}{\nu'}{\bk'+\bq}{\bq}{j}+(\pap{\nu'}{\nu}{\bk'}{-\bq}{j})^*
\big) \\
-&\psi^{\mu,\nu'}_{\bk'+\bq}g^{\mu,\nu}_{\bk'+\bq,\bq,j}\big(
\pap{\lambda}{\lambda'}{\bk-\bq}{-\bq}{j}+(\pap{\lambda'}{\lambda}{\bk}{\bq}{j})^*
\big)
+\psi^{\mu,\lambda'}_{\bk-\bq}g^{\mu,\nu}_{\bk-\bq,\bk-\bk'-\bq,j}\big(
\pap{\lambda}{\nu'}{\bk'+\bq}{\bk'+\bq-\bk}{j}+(\pap{\nu'}{\lambda}{\bk}{\bk-\bk'-\bq}{j})^*
\big) \\
-&\psi^{\lambda,\mu}_{\bk}g^{\nu',\mu}_{\bk'+\bq,\bk'+\bq-\bk,j}\big(
\pap{\nu}{\lambda'}{\bk-\bq}{\bk-\bk'-\bq}{j}+(\pap{\lambda'}{\nu}{\bk'}{\bk'+\bq-\bk}{j})^*
\big) 
+\psi^{\nu,\mu}_{\bk'}g^{\nu',\mu}_{\bk'+\bq,\bq,j}\big(
\pap{\lambda}{\lambda'}{\bk-\bq}{-\bq}{j}+(\pap{\lambda'}{\lambda}{\bk}{\bq}{j})^*
\big) \\
+&\psi^{\lambda,\mu}_{\bk}g^{\lambda',\mu}_{\bk-\bq,-\bq,j}\big(
\pap{\nu}{\nu'}{\bk'+\bq}{\bq}{j}+(\pap{\nu'}{\nu}{\bk'}{-\bq}{j})^*
\big) 
-\psi^{\nu,\mu}_{\bk'}g^{\lambda',\mu}_{\bk-\bq,\bk-\bk'-\bq,j}\big(
\pap{\lambda}{\nu'}{\bk'+\bq}{\bk'+\bq-\bk}{j}+(\pap{\nu'}{\lambda}{\bk}{\bk-\bk'-\bq}{j})^*
\big)
\Big]
\end{split}
\label{eq:EOM_C_doublet_phon}
\end{equation}
and among each other via
\begin{equation}
\begin{split}
&i\hbar\frac{d }{dt} c^{\bq,\bk',\bk}_{1,2,3,4}\Big|_{\textrm{D,c-c}}=\\
\sum_{\bl,5,6,7}\Big[
&\big(V^{5,6,7,1}_{\bl,\bk-\bq,\bl-\bq,\bk}-V^{5,6,1,7}_{\bl,\bk-\bq,\bk,\bl-\bq} \big)\psi^{6,4}_{\bk-\bq}c^{\bq,\bk',\bl}_{5,2,3,7} 
-\big(V^{4,5,6,7}_{\bk-\bq,\bl,\bk,\bl-\bq}-V^{4,5,7,6}_{\bk-\bq,\bl,\bl-\bq,\bk} \big)\psi^{1,6}_{\bk}c^{\bq,\bk',\bl}_{5,2,3,7} \\
+&\big(V^{5,6,7,2}_{\bl,\bk'+\bq,\bl+\bq,\bk'}-V^{5,6,2,7}_{\bl,\bk'+\bq,\bk',\bl+\bq} \big)\psi^{6,3}_{\bk'+\bq}c^{\bq,\bl,\bk}_{1,5,7,4} 
-\big(V^{3,5,6,7}_{\bk'+\bq,\bl,\bk',\bl+\bq}-V^{3,5,7,6}_{\bk'+\bq,\bl,\bl+\bq,\bk'} \big)\psi^{2,6}_{\bk'}c^{\bq,\bl,\bk}_{1,5,7,4}¸\\
-&\big(V^{5,6,7,1}_{\bl,\bk'+\bq,\bl-\bk+\bk'+\bq,\bk}-V^{5,6,1,7}_{\bl,\bk'+\bq,\bk,\bl-\bk+\bk'+\bq} \big)\psi^{6,3}_{\bk'+\bq}c^{\bk-\bk'-\bq,\bk',\bl}_{5,2,4,7} 
+\big(V^{3,5,6,7}_{\bk'+\bq,\bl,\bk,\bl-\bk+\bk'+\bq}-V^{3,5,7,6}_{\bk'+\bq,\bl,\bl-\bk+\bk'+\bq,\bk} \big)\psi^{1,6}_{\bk}c^{\bk-\bk'-\bq,\bk',\bl}_{5,2,4,7}\\
-&\big(V^{5,6,7,2}_{\bl,\bk-\bq,\bl+\bk-\bk'-\bq,\bk'}-V^{5,6,2,7}_{\bl,\bk-\bq,\bk',\bl+\bk-\bk'-\bq} \big)\psi^{6,4}_{\bk-\bq}c^{\bk-\bk'-\bq,\bl,\bk}_{1,5,7,3} 
+\big(V^{4,5,6,7}_{\bk-\bq,\bl,\bk',\bl+\bk-\bk'-\bq}-V^{4,5,7,6}_{\bk-\bq,\bl,\bl+\bk-\bk'-\bq,\bk'} \big)\psi^{2,6}_{\bk'}c^{\bk-\bk'-\bq,\bl,\bk}_{1,5,7,3} \\
+& V^{5,6,2,7}_{\bk-\bq+\bl,\bk'+\bq-\bl,\bk',\bk}(\psi^{1,7}_{\bk}-\delta_{1,7})(c^{-\bl,\bk'+\bq,\bk-\bq}_{4,3,6,5})^* 
- V^{5,6,1,7}_{\bk-\bq+\bl,\bk'+\bq-\bl,\bk,\bk'}\psi^{2,7}_{\bk'}(c^{-\bl,\bk'+\bq,\bk-\bq}_{4,3,6,5})^* \\
-& V^{3,5,6,7}_{\bk'+\bq,\bk-\bq,\bk-\bl,\bk'+\bl}(\psi^{5,4}_{\bk-\bq}-\delta_{5,4})c^{\bl,\bk',\bk}_{1,2,7,6} 
+ V^{4,5,6,7}_{\bk-\bq,\bk'+\bq,\bk-\bl,\bk'+\bl}\psi^{5,3}_{\bk'+\bq}c^{\bl,\bk',\bk}_{1,2,7,6}\Big]\,.
\end{split}
\label{eq:EOM_C_doublet_coul}
\end{equation}
While coupling to phonon-assisted polarizations via Eq.~(\ref{eq:EOM_C_doublet_phon}) leads to the formation of incoherent excitons out of coherent inter-band polarizations \cite{thranhardt_quantum_2000,kira_many-body_2006}, Coulomb interaction between carrier-carrier doublets results in screening of the singlet scattering terms (\ref{eq:EOM_C_singlet}) as well as the buildup of two-particle complexes. For a detailed discussion of all terms we refer to Ref.~\cite{kira_many-body_2006}.
\\
\\The carrier-carrier doublets couple to triplet correlations via carrier-phonon and carrier-carrier interaction:
\begin{equation}
\begin{split}
i\hbar\frac{d }{dt} c^{\bq,\bk',\bk}_{1,2,3,4}\Big|_{\textrm{T}}&=\\
-\sum_{\bl,j,5}\Big[
&g^{5,1}_{\bk+\bl,\bl,j}\Delta\ev{Q^{\phantom\dagger}_{\bl,j}a_{\bk+\bl,5}^{\dagger}a_{\bk',2}^{\dagger}a_{\bk'+\bq,3}^{\phantom\dagger}a_{\bk-\bq,4}^{\phantom\dagger}} 
+g^{5,2}_{\bk'+\bl,\bl,j}\Delta\ev{Q^{\phantom\dagger}_{\bl,j}a_{\bk,1}^{\dagger}a_{\bk'+\bl,5}^{\dagger}a_{\bk'+\bq,3}^{\phantom\dagger}a_{\bk-\bq,4}^{\phantom\dagger}} \\
-&g^{3,5}_{\bk'+\bq,\bl,j}\Delta\ev{Q^{\phantom\dagger}_{\bl,j}a_{\bk,1}^{\dagger}a_{\bk',2}^{\dagger}a_{\bk'+\bq-\bl,5}^{\phantom\dagger}a_{\bk-\bq,4}^{\phantom\dagger}} 
-g^{4,5}_{\bk-\bq,\bl,j}\Delta\ev{Q^{\phantom\dagger}_{\bl,j}a_{\bk,1}^{\dagger}a_{\bk',2}^{\dagger}a_{\bk'+\bq,3}^{\phantom\dagger}a_{\bk-\bq-\bl,5}^{\phantom\dagger}}
\Big] \\
+\sum_{\bl,\bl',5,6,7}\Big[&
V^{4567}_{\bk-\bq,\bl,\bl+\bl',\bk-\bq-\bl'}
\Delta\ev{a_{\bk,1}^{\dagger}a_{\bk',2}^{\dagger}a_{\bl,5}^{\dagger}a_{\bl+\bl',6}^{\phantom\dagger}a_{\bk'+\bq,3}^{\phantom\dagger}a_{\bk-\bq-\bl',7}^{\phantom\dagger}} \\
+&V^{3567}_{\bk'+\bq,\bl,\bl+\bl',\bk'+\bq-\bl'}
\Delta\ev{a_{\bk,1}^{\dagger}a_{\bk',2}^{\dagger}a_{\bl,5}^{\dagger}a_{\bl+\bl',6}^{\phantom\dagger}a_{\bk'+\bq-\bl',7}^{\phantom\dagger}a_{\bk-\bq,4}^{\phantom\dagger}} \\
-\big(&V^{2567}_{\bk',\bl,\bl+\bl',\bk'-\bl'}
\Delta\ev{a_{\bk-\bq,4}^{\dagger}a_{\bk'+\bq,3}^{\dagger}a_{\bl,5}^{\dagger}a_{\bl+\bl',6}^{\phantom\dagger}a_{\bk'-\bl',7}^{\phantom\dagger}a_{\bk,1}^{\phantom\dagger}}
\big)^* \\
-\big(&V^{1567}_{\bk,\bl,\bl+\bl',\bk-\bl'}
\Delta\ev{a_{\bk-\bq,4}^{\dagger}a_{\bk'+\bq,3}^{\dagger}a_{\bl,5}^{\dagger}a_{\bl+\bl',6}^{\phantom\dagger}a_{\bk',2}^{\phantom\dagger}a_{\bk-\bl',7}^{\phantom\dagger}}
\big)^*
\Big]\,.
\end{split}
\label{eq:EOM_C_triplet}
\end{equation}
As we limit ourselves to low electron-hole pair densities in this work, we discard the coupling to pure carrier triplets via carrier-carrier Coulomb interaction. To this end, we truncate the hierarchy of EOM in this direction by setting $\Delta\ev{a_{\bk,1}^{\dagger}a_{\bk',2}^{\dagger}a_{\bl,5}^{\dagger}a_{\bl+\bl',6}^{\phantom\dagger}a_{\bk'+\bq,3}^{\phantom\dagger}a_{\bk-\bq-\bl',4}^{\phantom\dagger}}\equiv 0$. The coupling to such triplets introduces excitation-induced scattering and dephasing of two-particle correlations. A specific example is Auger-type exciton-exciton annihilation as shown in Refs.~\cite{steinhoff_microscopic_2021,erkensten_dark_2021}. We therefore have to consider two types of triplets correlations: The carrier-carrier-phonon correlations 
\begin{equation}
T^{\bl,\bq,\bk',\bk}_{1,2,3,4,j}=\Delta\ev{D^{\phantom\dagger}_{\bl,j}a_{\bk,1}^{\dagger}a_{\bk',2}^{\dagger}a_{\bk'+\bq,3}^{\phantom\dagger}a_{\bk-\bq-\bl,4}^{\phantom\dagger}}
\label{eq:def_Daaaa}
\end{equation}
appearing in Eqs.~(\ref{eq:EOM_Xi_T}) and (\ref{eq:EOM_C_triplet}) and the two-phonon-assisted polarizations
\begin{equation}
\begin{split}
S^{(-),1,2}_{\bk,\bq,\bq',j,j'}&=\Delta\ev{D^{\dagger}_{\bq,j}D^{\phantom\dagger}_{\bq',j'}a^{\dagger}_{\bk-\bq+\bq',1}a^{\phantom\dagger}_{\bk,2}}\,, \\
S^{(+),1,2}_{\bk,\bq,\bq',j,j'}&=\Delta\ev{D^{\dagger}_{\bq,j}D^{\dagger}_{-\bq',j'}a^{\dagger}_{\bk-\bq+\bq',1}a^{\phantom\dagger}_{\bk,2}}
\end{split}
\label{eq:def_DDaa}
\end{equation}
appearing in Eq.~(\ref{eq:EOM_Xi_T}) only. The former fulfill the following EOM:
\begin{equation}
\begin{split}
&i\hbar\frac{d }{dt} T^{\bl,\bq,\bk',\bk}_{1,2,3,4,j} 
=(\ent{4}{\bk-\bq-\bl}+\ent{3}{\bk'+\bq}-\ent{2}{\bk'}-\ent{1}{\bk}+\hbar\omega_{\bl,j}-i\hbar\gamma_{\textrm{phon}})T^{\bl,\bq,\bk',\bk}_{1,2,3,4,j} \\ 
&+\sum_{j',5}2\,\textrm{Re}\big\{\ev{D^{\phantom\dagger}_{\bn,j'}}\big\}
\Big(
g^{3,5}_{\bk'+\bq,\bn,j'}T^{\bl,\bq,\bk',\bk}_{1,2,5,4,j}
+g^{4,5}_{\bk-\bq,\bn,j'}T^{\bl,\bq,\bk',\bk}_{1,2,3,5,j} 
-\big(g^{1,5}_{\bk,\bn,j'}\big)^*T^{\bl,\bq,\bk',\bk}_{5,2,3,4,j} 
-\big(g^{2,5}_{\bk',\bn,j'}\big)^*T^{\bl,\bq,\bk',\bk}_{1,5,3,4,j}
\Big)\\
&+i\hbar\frac{d }{dt} T^{\bl,\bq,\bk',\bk}_{1,2,3,4,j}\Big|_{\textrm{c-ph}}
+i\hbar\frac{d }{dt} T^{\bl,\bq,\bk',\bk}_{1,2,3,4,j}\Big|_{\textrm{SSD,c-c}}
+i\hbar\frac{d }{dt} T^{\bl,\bq,\bk',\bk}_{1,2,3,4,j}\Big|_{\textrm{DD,c-c}} 
+i\hbar\frac{d }{dt} T^{\bl,\bq,\bk',\bk}_{1,2,3,4,j}\Big|_{\textrm{ST,c-c}} \\
&+i\hbar\frac{d }{dt} T^{\bl,\bq,\bk',\bk}_{1,2,3,4,j}\Big|_{\textrm{LM}} 
+i\hbar\frac{d }{dt} T^{\bl,\bq,\bk',\bk}_{1,2,3,4,j}\Big|_{\textrm{Q}}
\,.
\end{split}
\label{eq:EOM_Daaaa}
\end{equation}
Carrier-phonon interaction leads to the following factorizations into singlets, doublets, and triplets: 
\begin{equation}
\begin{split}
i\hbar\frac{d }{dt} T^{\bl,\bq,\bk',\bk}_{1,2,3,4,j}\Big|_{\textrm{c-ph}} =
-\sum_{5,6,j'}\Big[
&g^{5,6}_{\bk-\bl,-\bl,j'}(\delta_{1,6}\Delta\ev{Q^{\phantom\dagger}_{-\bl,j'}D^{\phantom\dagger}_{\bl,j}}
+\delta_{j,j'}\psi^{1,6}_{\bk})c^{\bq,\bk',\bk-\bl}_{5,2,3,4} \\
+&g^{5,6}_{\bk'-\bl,-\bl,j'}(\delta_{2,6}\Delta\ev{Q^{\phantom\dagger}_{-\bl,j'}D^{\phantom\dagger}_{\bl,j}}
+\delta_{j,j'}\psi^{2,6}_{\bk'})c^{\bq+\bl,\bk'-\bl,\bk}_{1,5,3,4} \\
-&g^{6,5}_{\bk'+\bq,-\bl,j'}(\delta_{3,6}(\delta_{j,j'}+\Delta\ev{Q^{\phantom\dagger}_{-\bl,j'}D^{\phantom\dagger}_{\bl,j}})
-\delta_{j,j'}\psi^{6,3}_{\bk'+\bq})c^{\bq+\bl,\bk',\bk}_{1,2,5,4} \\
-&g^{6,5}_{\bk-\bq-\bl,-\bl,j'}(\delta_{4,6}(\delta_{j,j'}+\Delta\ev{Q^{\phantom\dagger}_{-\bl,j'}D^{\phantom\dagger}_{\bl,j}})
-\delta_{j,j'}\psi^{6,4}_{\bk-\bq-\bl})c^{\bq,\bk',\bk}_{1,2,3,5}\Big] \\
+\sum_{\bl',5,6} &g^{5,6}_{\bl',-\bl,j}
\Delta\ev{a_{\bk,1}^{\dagger}a_{\bk',2}^{\dagger}a_{\bl',5}^{\dagger}a_{\bl'+\bl,6}^{\phantom\dagger}a_{\bk'+\bq,3}^{\phantom\dagger}a_{\bk-\bq-\bl,4}^{\phantom\dagger}} \\
+\mathcal{O}\Big(&g\Delta\ev{DQa^{\dagger}a}\ev{a^{\dagger}a}\Big)+i\hbar\frac{d }{dt} T^{\bl,\bq,\bk',\bk}_{1,2,3,4,j}\Big|_{\Xi^2}
\,.
\end{split}
\label{eq:EOM_Daaaa_carr_phon}
\end{equation}
The first four lines describe the coupling of phonon-driven triplets back to carrier-carrier correlations on the hierarchy level below, which induces via Eq.~(\ref{eq:EOM_C_triplet}) scattering and dephasing of the carrier-carrier doublets.
As discussed above, the pure carrier triplets are discarded. Moreover, to reduce numerical complexity, we neglect factorizations involving two-phonon-assisted polarizations in Eq.~(\ref{eq:EOM_Daaaa_carr_phon}): $\mathcal{O}\Big(g\Delta\ev{DQa^{\dagger}a}\ev{a^{\dagger}a}\Big)\equiv 0$.
Factorizations into phonon-assisted polarizations take the form:
\begin{equation}
\begin{split}
&i\hbar\frac{d }{dt} T^{\bl,\bq,\bk',\bk}_{1,2,3,4,j}\Big|_{\Xi^2}=i\hbar\frac{d }{dt}\Delta\ev{D^{\phantom\dagger}_{\bl,j}a_{\bk_1,1}^{\dagger}a_{\bk_2,2}^{\dagger}a_{\bk_3,3}^{\phantom\dagger}a_{\bk_4,4}^{\phantom\dagger}}\Big|_{\Xi^2}  \\
=\sum_{5,j'}\Big[&g^{5,1}_{\bk_3+\bl,\bk_2-\bk_4,j'} (\Xi^{3,5}_{\bk_3+\bl,\bl,j})^*\big((\Xi^{4,2}_{\bk_2,\bk_2-\bk_4,j'})^*+\Xi^{2,4}_{\bk_4,\bk_4-\bk_2,j'}  \big) \\
-&g^{5,1}_{\bk_4+\bl,\bk_2-\bk_3,j'} (\Xi^{4,5}_{\bk_4+\bl,\bl,j})^*\big((\Xi^{3,2}_{\bk_2,\bk_2-\bk_3,j'})^*+\Xi^{2,3}_{\bk_3,\bk_3-\bk_2,j'}  \big) \\
-&g^{5,2}_{\bk_3+\bl,\bk_1-\bk_4,j'} (\Xi^{3,5}_{\bk_3+\bl,\bl,j})^*\big((\Xi^{4,1}_{\bk_1,\bk_1-\bk_4,j'})^*+\Xi^{1,4}_{\bk_4,\bk_4-\bk_1,j'}  \big) \\
+&g^{5,2}_{\bk_4+\bl,\bk_1-\bk_3,j'} (\Xi^{4,5}_{\bk_4+\bl,\bl,j})^*\big((\Xi^{3,1}_{\bk_1,\bk_1-\bk_3,j'})^*+\Xi^{1,3}_{\bk_3,\bk_3-\bk_1,j'}  \big) \\
+&g^{3,5}_{\bk_3,\bk_1-\bk_4,j'}(\Xi^{5,2}_{\bk_2,\bl,j})^*\big((\Xi^{4,1}_{\bk_1,\bk_1-\bk_4,j'})^* + \Xi^{1,4}_{\bk_4,\bk_4-\bk_1,j'}  \big) \\
-&g^{3,5}_{\bk_3,\bk_2-\bk_4,j'}(\Xi^{5,1}_{\bk_1,\bl,j})^*\big((\Xi^{4,2}_{\bk_2,\bk_2-\bk_4,j'})^* + \Xi^{2,4}_{\bk_4,\bk_4-\bk_2,j'}  \big) \\
-&g^{4,5}_{\bk_4,\bk_1-\bk_3,j'}(\Xi^{5,2}_{\bk_2,\bl,j})^*\big((\Xi^{3,1}_{\bk_1,\bk_1-\bk_3,j'})^* + \Xi^{1,3}_{\bk_3,\bk_3-\bk_1,j'}  \big) \\
+&g^{4,5}_{\bk_4,\bk_2-\bk_3,j'}(\Xi^{5,1}_{\bk_1,\bl,j})^*\big((\Xi^{3,2}_{\bk_2,\bk_2-\bk_3,j'})^* + \Xi^{2,3}_{\bk_3,\bk_3-\bk_2,j'}  \big)
\Big]
\,.
\end{split}
\label{eq:EOM_Daaaa_carr_phon_Xi_square}
\end{equation}
The next two classes of factorizations in Eq.~(\ref{eq:EOM_Daaaa}) describe coupling to phonon-assisted polarizations via Coulomb interaction on a singlet-singlet-doublet and doublet-doublet level, respectively, inducing scattering and dephasing of phonon-assisted polarizations via Eq.~(\ref{eq:EOM_Xi_T}):
\begin{equation}
\begin{split}
&i\hbar\frac{d }{dt} T^{\bl,\bq,\bk',\bk}_{1,2,3,4,j}\Big|_{\textrm{SSD,c-c}} \\
=\sum_{5}\Big[
&\Pi^{(1),5,2,3,4}_{\bk',\bk'+\bq,\bk-\bq-\bl}(\Xi^{5,1}_{\bk,\bl,j})^*
-\Pi^{(1),5,1,3,4}_{\bk,\bk'+\bq,\bk-\bq-\bl}(\Xi^{5,2}_{\bk',\bl,j})^* 
-\Pi^{(2),5,1,2,4}_{\bk,\bk',\bk-\bq-\bl}(\Xi^{3,5}_{\bk'+\bq+\bl,\bl,j})^*
+\Pi^{(2),5,1,2,3}_{\bk,\bk',\bk'+\bq}(\Xi^{4,5}_{\bk-\bq,\bl,j})^*
\Big]
\,,
\end{split}
\label{eq:EOM_Daaaa_SSD_cc}
\end{equation}
\begin{equation}
\begin{split}
&i\hbar\frac{d }{dt} T^{\bl,\bq,\bk',\bk}_{1,2,3,4,j}\Big|_{\textrm{DD,c-c}} \\
=\sum_{5}\Big[
&\tilde{\Pi}^{(2),5,2,4,3}_{\bk'+\bq,\bk-\bl,\bk'}(\Xi^{5,1}_{\bk,\bl,j})^*
-\tilde{\Pi}^{(2),5,2,3,4}_{\bk-\bq-\bl,\bk-\bl,\bk'}(\Xi^{5,1}_{\bk,\bl,j})^* 
+\tilde{\Pi}^{(4),5,3,4,2}_{\bk',\bk-\bl,\bk'+\bq}(\Xi^{5,1}_{\bk,\bl,j})^*
-\tilde{\Pi}^{(2),5,1,4,3}_{\bk'+\bq,\bk'-\bl,\bk}(\Xi^{5,2}_{\bk',\bl,j})^* \\
+&\tilde{\Pi}^{(2),5,1,3,4}_{\bk-\bq-\bl,\bk'-\bl,\bk}(\Xi^{5,2}_{\bk',\bl,j})^*
-\tilde{\Pi}^{(4),5,3,4,1}_{\bk,\bk'-\bl,\bk'+\bq}(\Xi^{5,2}_{\bk',\bl,j})^* 
+\tilde{\Pi}^{(1),1,2,4,5}_{\bk,\bk'+\bq+\bl,\bk'}(\Xi^{3,5}_{\bk'+\bq+\bl,\bl,j})^*
-\tilde{\Pi}^{(1),2,1,4,5}_{\bk',\bk'+\bq+\bl,\bk}(\Xi^{3,5}_{\bk'+\bq+\bl,\bl,j})^* \\
-&\tilde{\Pi}^{(3),5,1,2,4}_{\bk-\bq-\bl,\bk'+\bq+\bl,\bk'}(\Xi^{3,5}_{\bk'+\bq+\bl,\bl,j})^*
-\tilde{\Pi}^{(1),1,2,3,5}_{\bk,\bk-\bq,\bk'}(\Xi^{4,5}_{\bk-\bq,\bl,j})^* 
+\tilde{\Pi}^{(1),2,1,3,5}_{\bk',\bk-\bq,\bk}(\Xi^{4,5}_{\bk-\bq,\bl,j})^*
+\tilde{\Pi}^{(3),5,1,2,3}_{\bk'+\bq,\bk-\bq,\bk'}(\Xi^{4,5}_{\bk-\bq,\bl,j})^*
\,.
\end{split}
\label{eq:EOM_Daaaa_DD_cc}
\end{equation}
Here we have introduced the scattering integrals:
\begin{equation}
\begin{split}
\Pi^{(1),5,2,3,4}_{\bk_2,\bk_3,\bk_4}=\sum_{6,7}
\Big[(&V^{5,2,6,7}_{\bk_4+\bk_3-\bk_2,\bk_2,\bk_3,\bk_4} - V^{5,2,7,6}_{\bk_4+\bk_3-\bk_2,\bk_2,\bk_4,\bk_3})^*
\psi^{6,3}_{\bk_3}\psi^{7,4}_{\bk_4} \\
-(&V^{5,6,3,7}_{\bk_4+\bk_3-\bk_2,\bk_2,\bk_3,\bk_4} - V^{5,6,7,3}_{\bk_4+\bk_3-\bk_2,\bk_2,\bk_4,\bk_3})^*
\psi^{2,6}_{\bk_2}(\psi^{7,4}_{\bk_4} - \delta_{7,4}) \\
-(&V^{5,6,7,4}_{\bk_4+\bk_3-\bk_2,\bk_2,\bk_3,\bk_4} - V^{5,6,4,7}_{\bk_4+\bk_3-\bk_2,\bk_2,\bk_4,\bk_3})^*
\psi^{2,6}_{\bk_2}\psi^{7,3}_{\bk_3}
\Big]
\,, \\
\Pi^{(2),5,1,2,3}_{\bk_1,\bk_2,\bk_3}=\sum_{6,7}
\Big[(&V^{1,7,6,5}_{\bk_1,\bk_2,\bk_3,\bk_1+\bk_2-\bk_3} - V^{1,7,5,6}_{\bk_1,\bk_2,\bk_1+\bk_2-\bk_3,\bk_3})^*
\psi^{6,3}_{\bk_3}(\psi^{2,7}_{\bk_2}-\delta_{2,7}) \\
+(&V^{7,2,6,5}_{\bk_1,\bk_2,\bk_3,\bk_1+\bk_2-\bk_3} - V^{7,2,5,6}_{\bk_1,\bk_2,\bk_1+\bk_2-\bk_3,\bk_3})^*
\psi^{6,3}_{\bk_3}\psi^{1,7}_{\bk_1} \\
-(&V^{6,7,3,5}_{\bk_1,\bk_2,\bk_3,\bk_1+\bk_2-\bk_3} - V^{6,7,5,3}_{\bk_1,\bk_2,\bk_1+\bk_2-\bk_3,\bk_3})^*
\psi^{1,6}_{\bk_1}\psi^{2,7}_{\bk_2}
\Big]\,,
\end{split}
\label{eq:EOM_Daaaa_SSD_cc_scatt_int1}
\end{equation}
\begin{equation}
\begin{split}
\tilde{\Pi}^{(1),1,2,3,4}_{\bk_1,\bk_2,\bk_3}=\sum_{\bl,6,7}
(&V^{4,6,1,7}_{\bk_2,\bl,\bk_1,\bk_2+\bl-\bk_1} - V^{6,4,1,7}_{\bl,\bk_2,\bk_1,\bk_2+\bl-\bk_1})
c^{\bk_2+\bl-\bk_1-\bk_3,\bk_3,\bl}_{6,2,7,3}, \\
\tilde{\Pi}^{(2),1,2,3,4}_{\bk_1,\bk_2,\bk_3}=\sum_{\bl,5,6}
(&V^{4,5,6,1}_{\bk_1,\bl,\bk_1-\bk_2+\bl,\bk_2} - V^{4,5,1,6}_{\bk_1,\bl,\bk_2,\bk_1-\bk_2+\bl})
c^{\bk_1+\bl-\bk_2-\bk_3,\bk_3,\bl}_{5,2,6,3}, \\
\tilde{\Pi}^{(3),1,2,3,4}_{\bk_1,\bk_2,\bk_3}=\sum_{\bl,6,7}
&V^{4,1,6,7}_{\bk_1,\bk_2,\bk_2+\bl,\bk_1-\bl}
(c^{\bk_3-\bl-\bk_2,\bk_2+\bl,\bk_1-\bl}_{7,6,3,2})^*, \\
\tilde{\Pi}^{(4),1,2,3,4}_{\bk_1,\bk_2,\bk_3}=\sum_{\bl,5,6}
&V^{5,6,4,1}_{\bk_1+\bk_2-\bl,\bl,\bk_1,\bk_2}
c^{\bk_3-\bl,\bl,\bk_1+\bk_2-\bl}_{5,6,2,3}\,. \\
\end{split}
\label{eq:EOM_Daaaa_SSD_cc_scatt_int2}
\end{equation}
The singlet-triplet factorization in Eq.~(\ref{eq:EOM_Daaaa}) couples carrier-carrier-phonon triplets among each other via Coulomb interaction, similar to the terms (\ref{eq:EOM_Xi_cc}) and (\ref{eq:EOM_C_doublet_coul}). As shown in Eq.~(\ref{eq:EOM_Xi_cc}), there are terms involving either coherent or incoherent carrier singlets. To reduce numerical complexity, we retain only terms with incoherent singlets in $i\hbar\frac{d }{dt} T^{\bl,\bq,\bk',\bk}_{1,2,3,4,j}\Big|_{\textrm{ST,c-c}}$:
\begin{equation}
\begin{split}
i\hbar\frac{d }{dt} T^{\bl,\bq,\bk',\bk}_{1,2,3,4,j}\Big|_{\textrm{ST,c-c}}=
\sum_{\bl',5,6}\Big[
&\big(V^{5,4,6,1}_{\bl',\bk-\bq-\bl,\bl'-\bq-\bl,\bk}-V^{5,4,1,6}_{\bl',\bk-\bq-\bl,\bk,\bl'-\bq-\bl} \big)(f^{4}_{\bk-\bq-\bl}-f^{1}_{\bk})T^{\bl,\bq,\bk',\bl'}_{5,2,3,6} \\
+&\big(V^{5,3,6,2}_{\bl',\bk'+\bq,\bl'+\bq,\bk'}-V^{5,3,2,6}_{\bl',\bk'+\bq,\bk',\bl'+\bq} \big)(f^{3}_{\bk'+\bq}-f^{2}_{\bk'})T^{\bl,\bq,\bl',\bk}_{1,5,6,4} \\
-&\big(V^{5,3,6,1}_{\bl',\bk'+\bq,\bl'-\bk+\bk'+\bq,\bk}-V^{5,3,1,6}_{\bl',\bk'+\bq,\bk,\bl'-\bk+\bk'+\bq} \big)(f^{3}_{\bk'+\bq}-f^{1}_{\bk})T^{\bl,\bk-\bk'-\bq-\bl,\bk',\bl'}_{5,2,4,6} \\
-&\big(V^{5,4,6,2}_{\bl',\bk-\bq-\bl,\bl'+\bk-\bk'-\bq-\bl,\bk'}-V^{5,4,2,6}_{\bl',\bk-\bq-\bl,\bk',\bl'+\bk-\bk'-\bq-\bl} \big)(f^{4}_{\bk-\bq-\bl}-f^{2}_{\bk'})T^{\bl,\bk-\bk'-\bq-\bl,\bl',\bk}_{1,5,6,3} \\
+& V^{6,5,2,1}_{\bk+\bk'-\bl',\bl',\bk',\bk}(1-f^{1}_{\bk}-f^{2}_{\bk'})T^{\bl,\bq+\bl'-\bk,\bk+\bk'-\bl',\bl'}_{5,6,3,4} \\
+& V^{4,3,5,6}_{\bk-\bq-\bl,\bk'+\bq,\bl',\bk+\bk'-\bl'-\bl}(1-f^{3}_{\bk'+\bq}-f^{4}_{\bk-\bq-\bl})T^{\bl,\bl'-\bk',\bk',\bk}_{1,2,5,6}\Big]\,.
\end{split}
\label{eq:EOM_Daaaa_ST_coul}
\end{equation}
The light-matter coupling term $i\hbar\frac{d }{dt} T^{\bl,\bq,\bk',\bk}_{1,2,3,4,j}\Big|_{\textrm{LM}}$ also contains factorizations into coherent carrier singlets and carrier-carrier-phonon triplets via the renormalized Rabi energy, see Eq.~(\ref{eq:def_Rabi}). For consistency reasons, we drop these terms as well, such that the light-matter contribution is given by:
\begin{equation}
\begin{split}
&i\hbar\frac{d }{dt} T^{\bl,\bq,\bk',\bk}_{1,2,3,4,j}\Big|_{\textrm{LM}}= \\
&+\sum_{\bar{1}} \boldsymbol{d}^{\bar{1},1}_{\bk}\cdot \boldsymbol{E}^{\textrm{2d}} T^{\bl,\bq,\bk',\bk}_{\bar{1},2,3,4}
+\sum_{\bar{2}} \boldsymbol{d}^{\bar{2},2}_{\bk'}\cdot \boldsymbol{E}^{\textrm{2d}} T^{\bl,\bq,\bk',\bk}_{1,\bar{2},3,4} -\sum_{\bar{3}}  T^{\bl,\bq,\bk',\bk}_{1,2,\bar{3},4} \boldsymbol{d}^{3,\bar{3}}_{\bk'+\bq}\cdot \boldsymbol{E}^{\textrm{2d}}
-\sum_{\bar{4}}  T^{\bl,\bq,\bk',\bk}_{1,2,3,\bar{4}} \boldsymbol{d}^{4,\bar{4}}_{\bk-\bq-\bl}\cdot \boldsymbol{E}^{\textrm{2d}}
\,.
\end{split}
\label{eq:EOM_Daaaa_LM}
\end{equation}
Finally, carrier-carrier-phonon triplets couple to \textit{quadruplets} of the form $\Delta\ev{D^{\phantom\dagger}_{\bl,j}Q^{\phantom\dagger}_{\bl',j'}a_{\bk,1}^{\dagger}a_{\bk',2}^{\dagger}a_{\bk'+\bq,3}^{\phantom\dagger}a_{\bk-\bq-\bl-\bl',4}^{\phantom\dagger}}$ via carrier-phonon interaction and of the form $\Delta\ev{D^{\phantom\dagger}_{\bl,j}a_{\bk,1}^{\dagger}a_{\bk',2}^{\dagger}a_{\bl',5}^{\dagger}a_{\bl'+\bq',6}^{\phantom\dagger}a_{\bk'+\bq,3}^{\phantom\dagger}a_{\bk-\bq-\bl-\bq',4}^{\phantom\dagger}}$ via carrier-carrier interaction. We neglect excitation-induced dephasing of the triplets by discarding the latter type of quadruplets.
The remaining coupling terms to two-phonon-assisted quadruplets are given by:
\begin{equation}
\begin{split}
 i\hbar\frac{d }{dt} T^{\bl,\bq,\bk',\bk}_{1,2,3,4,j}\Big|_{\textrm{Q}} 
 = -\sum_{\bl',j',5}\Big[
&g^{5,1}_{\bk+\bl',\bl',j'}\Delta\ev{D^{\phantom\dagger}_{\bl,j}Q^{\phantom\dagger}_{\bl',j'}a_{\bk+\bl',5}^{\dagger}a_{\bk',2}^{\dagger}a_{\bk'+\bq,3}^{\phantom\dagger}a_{\bk-\bq-\bl,4}^{\phantom\dagger}} \\
+&g^{5,2}_{\bk'+\bl',\bl',j'}\Delta\ev{D^{\phantom\dagger}_{\bl,j}Q^{\phantom\dagger}_{\bl',j'}a_{\bk,1}^{\dagger}a_{\bk'+\bl',5}^{\dagger}a_{\bk'+\bq,3}^{\phantom\dagger}a_{\bk-\bq-\bl,4}^{\phantom\dagger}} \\
-&g^{3,5}_{\bk'+\bq,\bl',j'}\Delta\ev{D^{\phantom\dagger}_{\bl,j}Q^{\phantom\dagger}_{\bl',j'}a_{\bk,1}^{\dagger}a_{\bk',2}^{\dagger}a_{\bk'+\bq-\bl',5}^{\phantom\dagger}a_{\bk-\bq-\bl,4}^{\phantom\dagger}} \\
-&g^{4,5}_{\bk-\bq-\bl,\bl',j'}\Delta\ev{D^{\phantom\dagger}_{\bl,j}Q^{\phantom\dagger}_{\bl',j'}a_{\bk,1}^{\dagger}a_{\bk',2}^{\dagger}a_{\bk'+\bq,3}^{\phantom\dagger}a_{\bk-\bq-\bl-\bl',5}^{\phantom\dagger}}
\Big]
\,.
\end{split}
\label{eq:EOM_Daaaa_Q}
\end{equation}
It is straightforward to derive the EOM of such quadruplets generated by the Hamiltonian $H_{\textrm{carr}}+H_{\textrm{phon}}+H_{\textrm{c-ph}}$ and solve them adiabatically as shown in \cite{kira_many-body_2006}. To this end, we discard \textit{quintuplet} correlations and determine all factorizations into smaller clusters, which leads to equations of the form
\begin{equation}
\begin{split}
& i\hbar\frac{d }{dt} \ev{4}=\delta\ev{4}+\sum_{\nu} V_{\nu}(t)
\,.
\end{split}
\label{eq:EOM_Q_dummy}
\end{equation}
Here, $\ev{4}$ stands for a quadruplet, $\delta$ is the free-energy difference that determines the trivial oscillation and $V_{\nu}(t)$ are the factorizations into other types of correlations acting as source terms, or inhomogeneities, for $\ev{4}$. Eq.~(\ref{eq:EOM_Q_dummy}) can be formally solved by
\begin{equation}
\begin{split}
\ev{4}(t)=-\frac{i}{\hbar}\int^{t}_{-\infty}du\sum_{\nu}V_{\nu}(u)e^{\frac{i}{\hbar}(\delta-i\gamma)(u-t)}
\,,
\end{split}
\label{eq:EOM_Q_dummy_sol}
\end{equation}
where we have added a phenomenological damping $\gamma$. We further extract the trivial oscillations of each source term, similar to $\delta$, by setting $V_{\nu}(u)=\tilde{V}_{\nu}(u)e^{i\omega_{\nu}u}$. For the slowly varing part $\tilde{V}_{\nu}$, we apply a Markov approximation, replacing the function argument by the ``external'' time $t$:
\begin{equation}
\begin{split}
\ev{4}(t)&\approx -\frac{i}{\hbar}\sum_{\nu}e^{i\omega_{\nu}t} \int^{t}_{-\infty}du\sum_{\nu}\tilde{V}_{\nu}(t)e^{\frac{i}{\hbar}(\delta+\hbar\omega_{\nu}-i\gamma)(u-t)}
=-\sum_{\nu}\frac{V_{\nu}(t)}{\delta+\hbar\omega_{\nu}-i\gamma}
\,.
\end{split}
\label{eq:EOM_Q_dummy_sol_2}
\end{equation}
In the EOM of the quadruplets in Eq.~(\ref{eq:EOM_Daaaa_Q}), we keep only factorizations containing triplets with phonon mode $\bl,j$ as well as incoherent singlets and doublets. Limiting ourselves to intra-band scattering processes, we obtain the following form for the quadruplet term:
\begin{equation}
\begin{split}
 &i\hbar\frac{d }{dt} T^{\bl,\bq,\bk',\bk}_{1,2,3,4,j}\Big|_{\textrm{Q}} 
 =  -(\Gamma^{(D),3}_{\bk'+\bq}+\Gamma^{(D),4}_{\bk-\bq-\bl}-(\Gamma^{(D),1}_{\bk})^*-(\Gamma^{(D),2}_{\bk'})^*) T^{\bl,\bq,\bk',\bk}_{1,2,3,4,j} \\
 +\sum_{\bl'}\Big[&T^{\bl,\bq,\bk',\bl'}_{1,2,3,4,j}(\Gamma^{(O),4,1}_{\bk-\bq-\bl,\bk,\bk-\bl'} -(\Gamma^{(O),1,4}_{\bk,\bk-\bq-\bl,\bk-\bl'})^*) 
 +T^{\bl,\bq,\bl',\bk}_{1,2,3,4,j}(\Gamma^{(O),3,2}_{\bk'+\bq,\bk',\bk'-\bl'} -(\Gamma^{(O),2,3}_{\bk',\bk'+\bq,\bk'-\bl'})^*) \\
 +&T^{\bl,\bq+\bl'-\bk,\bk',\bl'}_{1,2,3,4,j}(\Gamma^{(O),3,1}_{\bk'+\bq,\bk,\bk-\bl'} -(\Gamma^{(O),1,3}_{\bk,\bk'+\bq,\bk-\bl'})^*) 
  +T^{\bl,\bk'+\bq-\bl',\bl',\bk}_{1,2,3,4,j}(\Gamma^{(O),4,2}_{\bk-\bq-\bl,\bk',\bk'-\bl'} -(\Gamma^{(O),2,4}_{\bk',\bk-\bq-\bl,\bk'-\bl'})^*) \\
  +&T^{\bl,\bq+\bl'-\bk,\bk+\bk'-\bl',\bl'}_{1,2,3,4,j}(\Gamma^{(O),2,1}_{\bl',\bk,\bk'-\bl'} +\Gamma^{(O),1,2}_{\bk+\bk'-\bl',\bk',\bk'-\bl'}) 
  +T^{\bl,\bl'-\bk',\bk',\bk}_{1,2,3,4,j}((\Gamma^{(O),4,3}_{\bk'+\bk-\bl-\bl',\bk'+\bq,¸\bk'+\bq-\bl'})^* +(\Gamma^{(O),3,4}_{\bl',\bk-\bq-\bl,\bl'-\bk'-\bq})^*)
\Big]
\end{split}
\label{eq:EOM_Daaaa_Q_final}
\end{equation}
with diagonal $(D)$ and off-diagonal $(O)$ dephasing rates in Markov approximation:
\begin{equation}
\begin{split}
& \Gamma^{(D),\lambda}_{\bk}= \sum_{\bq,j}
|g^{\lambda,\lambda}_{\bk,\bq,j}|^2 
\Big[
\frac{f^{\lambda}_{\bk-\bq} + n_{-\bq,j}}{\en{\lambda}{\bk-\bq}-\en{\lambda}{\bk}-\hbar\omega_{\bq,j}-i\gamma}
+\frac{1-f^{\lambda}_{\bk-\bq} + n_{\bq,j}}{\en{\lambda}{\bk-\bq}-\en{\lambda}{\bk}+\hbar\omega_{\bq,j}-i\gamma}
\Big]
\,,\\
& \Gamma^{(O),\lambda,\lambda'}_{\bk,\bk',\bq}=\sum_j
g^{\lambda,\lambda}_{\bk,\bq,j}(g^{\lambda',\lambda'}_{\bk',\bq,j})^* 
\Big[
\frac{f^{\lambda'}_{\bk'} + n_{\bq,j}}{\en{\lambda'}{\bk'-\bq}-\en{\lambda'}{\bk'}+\hbar\omega_{\bq,j}-i\gamma}
+\frac{1-f^{\lambda'}_{\bk'} + n_{-\bq,j}}{\en{\lambda'}{\bk'-\bq}-\en{\lambda'}{\bk'}-\hbar\omega_{\bq,j}-i\gamma}
\Big]
\,.
\end{split}
\label{eq:deph_carr_phon}
\end{equation}
For the phenomenological damping, we choose $\gamma=2\gamma_{\textrm{phen}}$, where $\gamma_{\textrm{phen}}$ is considered the "elementary" damping rate for a single pair of carrier operators (i.e., a fermionic singlet). This choice is in the spirit of the Lindblad formalism for phonon damping (\ref{eq:Lindblad}), which systematically leads to one unit of damping per phonon operator in a correlation function. 
\\
\\The two-phonon-assisted polarizations, Eq.~(\ref{eq:def_DDaa}), fulfill the following EOM:
\begin{equation}
\begin{split}
i\hbar\frac{d }{dt} S^{(\mp),1,2}_{\bk,\bq,\bq',j,j'} 
=&(\ent{2}{\bk}-\ent{1}{\bk-\bq+\bq'}-\hbar\omega_{\bq,j}\pm\hbar\omega_{\bq',j'}-i\hbar2\gamma_{\textrm{phon}})S^{(\mp),1,2}_{\bk,\bq,\bq',j,j'} \\ 
&+\sum_{j'',3}2\,\textrm{Re}\big\{\ev{D^{\phantom\dagger}_{\bn,j''}}\big\}\Big( 
g^{2,3}_{\bk,\bn,j''}S^{(\mp),1,3}_{\bk,\bq,\bq',j,j'}
-\big(g^{1,3}_{\bk-\bq+\bq',\bn,j''}\big)^*S^{(\mp),3,2}_{\bk,\bq,\bq',j,j'}
\Big)\\
&+i\hbar\frac{d }{dt} S^{(\mp),1,2}_{\bk,\bq,\bq',j,j'}\Big|_{\textrm{c-ph}}
+i\hbar\frac{d }{dt} S^{(\mp),1,2}_{\bk,\bq,\bq',j,j'}\Big|_{\Xi^2}
+i\hbar\frac{d }{dt} S^{(\mp),1,2}_{\bk,\bq,\bq',j,j'}\Big|_{\textrm{ST,c-c}}
+i\hbar\frac{d }{dt} S^{(\mp),1,2}_{\bk,\bq,\bq',j,j'}\Big|_{\textrm{LM}} \\
&+i\hbar\frac{d }{dt} S^{(\mp),1,2}_{\bk,\bq,\bq',j,j'}\Big|_{\textrm{Q}}
\,.
\end{split}
\label{eq:EOM_DDaa}
\end{equation}
The carrier-phonon-induced contributions are:
\begin{equation}
\begin{split}
i\hbar\frac{d }{dt} S^{(\mp),1,2}_{\bk,\bq,\bq',j,j'}\Big|_{\textrm{c-ph}}=\sum_{3,4,j''}\Big[
&g^{3,4}_{\bk+\bq',\bq,j''}
(\delta_{j,j''}\psi^{1,4}_{\bk-\bq+\bq'}-\delta_{1,4}(\delta_{j,j''}+\Delta\ev{D^{\dagger}_{\bq,j}Q^{\phantom\dagger}_{\bq,j''}}))
\left\{
\begin{array}{c}
(\Xi^{2,3}_{\bk+\bq',\bq',j'})^*\\
\Xi^{3,2}_{\bk,-\bq',j'}\\
\end{array}
\right\} \\
+&g^{3,4}_{\bk,\bq,j''}(\delta_{j,j''}\psi^{3,2}_{\bk}+\delta_{2,3}\Delta\ev{D^{\dagger}_{\bq,j}Q^{\phantom\dagger}_{\bq,j''}})
\left\{
\begin{array}{c}
(\Xi^{4,1}_{\bk-\bq+\bq',\bq',j'})^*\\
\Xi^{1,4}_{\bk-\bq,-\bq',j'}\\
\end{array}
\right\} \\
\mp&g^{3,4}_{\bk-\bq,-\bq',j''}(\delta_{j',j''}\psi^{1,4}_{\bk-\bq+\bq'}-\delta_{1,4}
\left\{
\begin{array}{c}
-\Delta\ev{Q^{\phantom\dagger}_{-\bq',j''}D^{\phantom\dagger}_{\bq',j'}} \\
\delta_{j',j''}+\Delta\ev{D^{\dagger}_{-\bq',j'}Q^{\phantom\dagger}_{-\bq',j''}}\\
\end{array}
\right\}
\Xi^{3,2}_{\bk,\bq,j} \\
\mp&g^{3,4}_{\bk,-\bq',j''}(\delta_{j',j''}\psi^{3,2}_{\bk}+\delta_{2,3}
\left\{
\begin{array}{c}
-\delta_{j',j''}-\Delta\ev{Q^{\phantom\dagger}_{-\bq',j''}D^{\phantom\dagger}_{\bq',j'}} \\
\Delta\ev{D^{\dagger}_{-\bq',j'}Q^{\phantom\dagger}_{-\bq',j''}}\\
\end{array}
\right\}
\Xi^{1,4}_{\bk+\bq',\bq,j}\\
+&\mathcal{O}\Big(g\Delta\ev{Da^{\dagger}a^{\dagger}aa}\Big)
\Big]
\,.
\end{split}
\label{eq:EOM_DDaa_carr_phon}
\end{equation}
Consistent with the treatment of Eq.~(\ref{eq:EOM_Daaaa_carr_phon}), we discard coupling to carrier-carrier-phonon triplets given by the last term.
Factorizations into products of phonon-assisted polarizations are introduced via carrier-carrier Coulomb interaction:
\begin{equation}
\begin{split}
i\hbar\frac{d }{dt} S^{(\mp),1,2}_{\bk,\bq,\bq',j,j'}\Big|_{\Xi^2}=\sum_{\bl,3,4,5}\Big[
&\Xi^{3,2}_{\bk,\bq,j}\Big(V^{3,4,1,5}_{\bk-\bq,\bl,\bk-\bq+\bq',\bl-\bq'}-V^{3,4,5,1}_{\bk-\bq,\bl,\bl-\bq',\bk-\bq+\bq'}\Big)
\left\{
\begin{array}{c}
(\Xi^{5,4}_{\bl,\bq',j'})^*\\
\Xi^{4,5}_{\bl-\bq',-\bq',j'}\\
\end{array}
\right\} \\
-&\Xi^{1,3}_{\bk+\bq',\bq,j}\Big(V^{2,4,3,5}_{\bk,\bl,\bk+\bq',\bl-\bq'}-V^{2,4,5,3}_{\bk,\bl,\bl-\bq',\bk+\bq'}\Big)
\left\{
\begin{array}{c}
(\Xi^{5,4}_{\bl,\bq',j'})^*\\
\Xi^{4,5}_{\bl-\bq',-\bq',j'}\\
\end{array}
\right\} \\
+&\left\{
\begin{array}{c}
(\Xi^{2,3}_{\bk+\bq',\bq',j'})^*\\
\Xi^{3,2}_{\bk,-\bq',j'}\\
\end{array}
\right\} 
\Big(V^{3,4,1,5}_{\bk+\bq',\bl,\bk-\bq+\bq',\bl+\bq}-V^{3,4,5,1}_{\bk+\bq',\bl,\bl+\bq,\bk-\bq+\bq'}\Big)
\Xi^{4,5}_{\bl+\bq,\bq,j}
\\
-&\left\{
\begin{array}{c}
(\Xi^{3,1}_{\bk-\bq+\bq',\bq',j'})^*\\
\Xi^{1,3}_{\bk-\bq,-\bq',j'}\\
\end{array}
\right\} 
\Big(V^{2,4,3,5}_{\bk,\bl,\bk-\bq,\bl+\bq}-V^{2,4,5,3}_{\bk,\bl,\bl+\bq,\bk-\bq}\Big)
\Xi^{4,5}_{\bl+\bq,\bq,j}
\Big]\,.
\end{split}
\label{eq:EOM_DDaa_Xi_square}
\end{equation}
As for the carrier-carrier-phonon triplets, we discard all singlet-triplet factorizations in $i\hbar\frac{d }{dt} S^{(\mp),1,2}_{\bk,\bq,\bq',j,j'}\Big|_{\textrm{ST,c-c}}$ and $i\hbar\frac{d }{dt} S^{(\mp),1,2}_{\bk,\bq,\bq',j,j'}\Big|_{\textrm{LM}}$ that contain coherent singlets:
\begin{equation}
\begin{split}
i\hbar\frac{d }{dt} S^{(\mp),1,2}_{\bk,\bq,\bq',j,j'}\Big|_{\textrm{ST,c-c}}=
\sum_{\bl',3,4}
\big(V^{3,2,4,1}_{\bl'-\bq+\bq',\bk,\bl',\bk-\bq+\bq'}-V^{3,2,1,4}_{\bl'-\bq+\bq',\bk,\bk-\bq+\bq',\bl'} \big)
(f^{2}_{\bk}-f^{1}_{\bk-\bq+\bq'})S^{(\mp),3,4}_{\bl',\bq,\bq',j,j'} \,,
\end{split}
\label{eq:EOM_DDaa_ST_coul}
\end{equation}
\begin{equation}
\begin{split}
i\hbar\frac{d }{dt} S^{(\mp),1,2}_{\bk,\bq,\bq',j,j'}\Big|_{\textrm{LM}}= 
+\sum_{\bar{1}} \boldsymbol{d}^{\bar{1},1}_{\bk-\bq+\bq'}\cdot \boldsymbol{E}^{\textrm{2d}} S^{(\mp),\bar{1},2}_{\bk,\bq,\bq',j,j'}
-\sum_{\bar{2}}  S^{(\mp),1,\bar{2}}_{\bk,\bq,\bq',j,j'} \boldsymbol{d}^{2,\bar{2}}_{\bk}\cdot \boldsymbol{E}^{\textrm{2d}}
\,.
\end{split}
\label{eq:EOM_DDaa_LM}
\end{equation}
Finally, we discuss the quadruplet contributions to the EOM of two-phonon-assisted polarizations.
We neglect excitation-induced dephasing by discarding Coulomb-assisted coupling to quadruplets of the type
$\Delta\ev{D^{\dagger}_{\bq,j}(D^{\phantom\dagger}_{\bq',j'}/D^{\dagger}_{-\bq',j'})a_{\bk-\bq+\bq',1}^{\dagger}a_{\bk',3}^{\dagger}a_{\bk'+\bl,4}^{\phantom\dagger}a_{\bk-\bl,5}^{\phantom\dagger}}$.
The remaining coupling terms to three-phonon-assisted quadruplets are given by:
\begin{equation}
\begin{split}
 i\hbar\frac{d }{dt} S^{(\mp),1,2}_{\bk,\bq,\bq',j,j'}\Big|_{\textrm{Q}}=-\sum_{\bq'',j'',3}\Big[ 
g^{3,1}_{\bk-\bq+\bq'+\bq'',\bq'',j''}&\Delta\ev{D^{\dagger}_{\bq,j}
\left\{
\begin{array}{c}
D^{\phantom\dagger}_{\bq',j'}\\
D^{\dagger}_{-\bq',j'}\\
\end{array}
\right\} 
Q^{\phantom\dagger}_{\bq'',j''}a_{\bk-\bq+\bq'+\bq'',3}^{\dagger}a_{\bk,2}^{\phantom\dagger}} \\
-g^{2,3}_{\bk,\bq'',j''}&\Delta\ev{D^{\dagger}_{\bq,j}
\left\{
\begin{array}{c}
D^{\phantom\dagger}_{\bq',j'}\\
D^{\dagger}_{-\bq',j'}\\
\end{array}
\right\} 
Q^{\phantom\dagger}_{\bq'',j''}a_{\bk-\bq+\bq',1}^{\dagger}a_{\bk-\bq'',3}^{\phantom\dagger}}
\Big]
\,.
\end{split}
\label{eq:EOM_DDaa_Q}
\end{equation}
We eliminate the three-phonon-assisted polarizations adiabatically by applying the Markov approximation as discussed above, retaining factorizations that can be grouped in terms of Markovian scattering and dephasing rates (\ref{eq:deph_carr_phon}):
\begin{equation}
\begin{split}
 i\hbar\frac{d }{dt} S^{(\mp),1,2}_{\bk,\bq,\bq',j,j'}\Big|_{\textrm{Q}} 
 = & -(\Gamma^{(D),2}_{\bk}-(\Gamma^{(D),1}_{\bk-\bq+\bq'})^*) S^{(\mp),1,2}_{\bk,\bq,\bq',j,j'} \\
 +&\sum_{\bl'}S^{(\mp),1,2}_{\bl',\bq,\bq',j,j'}(\Gamma^{(O),2,1}_{\bk,\bk-\bq+\bq',\bk-\bl'} -(\Gamma^{(O),1,2}_{\bk-\bq+\bq',\bk,\bk-\bl'})^*)
\,.
\end{split}
\label{eq:EOM_DDaa_Q_final}
\end{equation}
Since the two-phonon-assisted polarizations are singlets with respect to carrier operators, the Markov terms are computed with a damping $\gamma=\gamma_{\textrm{phen}}$.
Including two-phonon-assisted polarizations with a phonon-induced dephasing in Markov approximation has been termed \textit{damped 4th order Born approximation} in Ref.~\cite{lengers_theory_2020}.
\\
\\The coupled EOM of correlation functions are numerically solved using an adaptive Adams-Bashforth-Moulton (ABM) predictor-corrector method \cite{shampine_computer_1975}.
\end{widetext}

\subsection{Calculation of energy densities}\label{sec:energy_dens}

To discuss the pump-induced carrier and phonon dynamics, we investigate the time dependence of the energy density,
similar to the discussion in \cite{kira_many-body_2006}. To this end, we calculate expectation values of different contributions to the system Hamiltonian (\ref{eq:Hamiltonian_start}), normalized by the TMD area $\mathcal{A}$.
Combining the free-carrier energy and the carrier-carrier Coulomb interaction energy, we find:
\begin{equation}
 \begin{split}
\mathcal{E}_{\textrm{carr}}&=\frac{1}{\mathcal{A}}\ev{H_{\textrm{carr}}+ H_{\textrm{Coul}}} \\
&=\frac{1}{\mathcal{A}}\sum_{\bk,\lambda}\varepsilon_{\bk}^{\lambda} \ev{a^{\dagger}_{\bk,\lambda}a^{\phantom\dagger}_{\bk,\lambda}} \\
&+\frac{1}{2\mathcal{A}}\sum_{\substack{\bk,\bk',\bq \\ \lambda,\nu,\nu',\lambda'}}V^{\lambda,\nu,\nu',\lambda'}_{\bk,\bk',\bk'+\bq,\bk-\bq} 
\ev{a^{\dagger}_{\bk,\lambda}a^{\dagger}_{\bk',\nu}a^{\phantom\dagger}_{\bk'+\bq,\nu'}a^{\phantom\dagger}_{\bk-\bq,\lambda'}} \\
&=\mathcal{E}_{\textrm{kin}}+\mathcal{E}_{\textrm{coh}}+\mathcal{E}_{\textrm{corr,e}}+\mathcal{E}_{\textrm{corr,h}}+\mathcal{E}_{\textrm{corr,X}}\,.
\end{split}
\label{eq:energy_carr}
\end{equation}
The kinetic energy density, formulated in the electron-hole picture, is the contribution due to single-particle populations including Hartree-Fock-type energy renormalizations:
\begin{equation}
 \begin{split}
\mathcal{E}_{\textrm{kin}}= \frac{1}{\mathcal{A}}\sum_{\bk,\lambda}\Big(\varepsilon_{\bk}^{\lambda}+\frac{1}{2}\Sigma^{\textrm{HF},\lambda}_{\bk}\Big)f^{\lambda}_{\bk}
\,.
\end{split}
\label{eq:energy_kin}
\end{equation}
The Hartree-Fock self-energy $\Sigma^{\textrm{HF},\lambda}_{\bk}$ is defined in Eq.~(\ref{eq:def_eps_tilde})). The inter-band polarizations yield a coherent single-particle contribution
\begin{equation}
 \begin{split}
\mathcal{E}_{\textrm{coh}}=-\frac{1}{\mathcal{A}}\sum_{\bk,\bk',v,c,v',c'}
V^{c,v',v,c'}_{\bk,\bk',\bk,\bk'} \psi^{v',c'}_{\bk'}
(\psi^{v,c}_{\bk})^*
\,.
\end{split}
\label{eq:energy_coh}
\end{equation}
The correlated two-particle energy density also splits into intra-band contributions
\begin{equation}
 \begin{split}
\mathcal{E}_{\textrm{corr,e}}=\frac{1}{2\mathcal{A}}\sum_{\bk,\bk',\bq,c,c'}
V^{c,c',c,c'}_{\bk,\bk',\bk'+\bq,\bk-\bq} c_{c,c',c,c'}^{\bq,\bk',\bk}\,, \\
\mathcal{E}_{\textrm{corr,h}}=\frac{1}{2\mathcal{A}}\sum_{\bk,\bk',\bq,v,v'}
V^{v,v',v,v'}_{\bk,\bk',\bk'+\bq,\bk-\bq} c_{v,v',v,v'}^{\bq,\bk',\bk}
\end{split}
\label{eq:energy_corr_intra}
\end{equation}
and inter-band contributions
\begin{equation}
 \begin{split}
\mathcal{E}_{\textrm{corr,X}}=-\frac{1}{\mathcal{A}}\sum_{\bk,\bk',\bq,v,c,v',c'}
V^{v,c',c,v'}_{\bk,\bk',\bk'+\bq,\bk-\bq} c_{v,c',v',c}^{\bk-\bk'-\bq,\bk',\bk}\,,
\end{split}
\label{eq:energy_corr_inter}
\end{equation}
where we have neglected matrix elements of the electron-hole-exchange type.
The energy density of the phonons is defined relative to the thermal state:
\begin{equation}
 \begin{split}
\mathcal{E}_{\textrm{phon}}&=\frac{1}{\mathcal{A}}\Big(\ev{H_{\textrm{phon}}}- \ev{H_{\textrm{phon}}}\Big|_{\textrm{th}}\Big)
-\mathcal{E}_{\textrm{phon,relax}}\\
&=\frac{1}{\mathcal{A}}\sum_{\bq,j}\hbar\omega_{\bq,j}\Big(
\Delta\ev{D^{\dagger}_{\bq,j}D^{\phantom\dagger}_{\bq,j}}-n^{\textrm{B}}_{\bq,j}
\Big) \\
&+\frac{1}{\mathcal{A}}\sum_j \hbar\omega_{\bn,j} |\ev{D_{\bn,j}}|^2
-\mathcal{E}_{\textrm{phon,relax}}
\,.
\end{split}
\label{eq:energy_phon}
\end{equation}
Here, we have subtracted the energy density $\mathcal{E}_{\textrm{phon,relax}}$ transferred by the phenomenological phonon relaxation term in Eq.~(\ref{eq:EOM_DD}), since it violates the energy conservation of the carrier-phonon system.
It is given by the integral
\begin{equation}
\begin{split}
&\mathcal{E}_{\textrm{phon, relax}}(t)= \\
&-\int_{t_0}^{t} dt'
2\gamma_{\textrm{phon}}\frac{1}{\mathcal{A}}\sum_{\bq,j}\hbar\omega_{\bq,j}(\Delta\ev{D^{\dagger}_{\bq,j}D^{\phantom\dagger}_{\bq,j}}(t')-n^{\textrm{B}}_{\bq,j}) 
\,.
\end{split}
\label{eq:energy_phon_res}
\end{equation}
The energy density stored in the carrier-phonon interaction is given by:
\begin{equation}
 \begin{split}
\mathcal{E}_{\textrm{c-ph}}&=\frac{1}{\mathcal{A}}\ev{H_{\textrm{c-ph}}} \\
&=\frac{1}{\mathcal{A}}\sum_{\bk,\bq,\nu,\nu',j}
2\textrm{Re}\,\Big\{
g^{\nu,\nu'}_{\bk,\bq,j}\big(\Xi^{\nu',\nu}_{\bk,\bq,j}\big)^*
\Big\}
 \\
&+\frac{1}{\mathcal{A}}\sum_{\bk,\nu,j}g^{\nu,\nu}_{\bk,\bn,j}f^{\nu}_{\bk}
2\textrm{Re}\,\Big\{ \ev{D_{\bn,j}}\Big\}
\\
&+\frac{1}{\mathcal{A}}\sum_{\bk,\nu,\bar{\nu},j} g^{\nu,\bar{\nu}}_{\bk,\bn,j}\psi^{\nu,\bar{\nu}}_{\bk}
2\textrm{Re}\,\Big\{  \ev{D_{\bn,j}}         \Big\}
\,,
\end{split}
\label{eq:energy_carr_phon}
\end{equation}
where the single-particle populations $f^{\nu}_{\bk}$ are given in the electron-hole picture.
We do not take into account contributions from time-dependent light-matter interaction in the following, such that conservation of the total energy density $\mathcal{E}_{\textrm{tot}}=\mathcal{E}_{\textrm{carr}}+\mathcal{E}_{\textrm{phon}}+\mathcal{E}_{\textrm{c-ph}}$ can be expected after the external field has vanished.

\subsection{Maxwell's Equations}

As shown in the previous chapter, we describe light-matter interaction on a semiclassical level, neglecting correlations between photons and charge carriers. Therefore, the photonic properties are solely determined by the complex-valued electromagnetic field $\boldsymbol{E}(z,t)$, whose dynamics is governed by the wave equation (\ref{eq:E_T_wave_FT_class}). It couples as an external source to all correlation functions that involve carrier operators via the dipole matrix elements $\boldsymbol{d}^{\bar{\nu},\nu}_{\bk}$. On the other hand, excitations in the carrier system act back on the electromagnetic field via the two-dimensional polarization $\boldsymbol{P}^{\textrm{2d}}(t)$, which contains microscopic interband polarizations (\ref{eq:def_pol_2d}). This means that Maxwell's equations need to be solved self-consistently with the EOM for correlation functions.
In this paper, we treat two different scenarios of light-matter interaction. The first one is the regime of linear optical response, where a weak and spectrally broad electric field is used to probe the material response in (quasi-)equilibrium. The quantity of interest in this case is the linear optical susceptibility
\begin{equation}
\begin{split}
\chi(\omega)=\frac{\boldsymbol{P}^{\textrm{2d}}(\omega)\cdot \boldsymbol{e}_{\textrm{p}}}{\varepsilon_0 \boldsymbol{E}^{\textrm{2d}}(\omega)\cdot \boldsymbol{e}_{\textrm{p}}}=\frac{P(\omega)}{\varepsilon_0 E(\omega)}\, ,
\end{split}
\label{eq:def_chi}
\end{equation}
where the electric field and the macroscopic polarization in the frequency domain are projected onto the same polarization vector $\boldsymbol{e}_{\textrm{p}}$. The probe field drives only coherent interband polarizations $\psi^{v,c}_{\bk}$, but no single-particle populations $f^{\lambda}_{\bk}$ or many-particle correlations (doublets or higher-order clusters). Therefore, we can first compute $\chi(\omega)$ using a simple probe field and solve Maxwell's equations afterwards to obtain transmission, reflection, and absorption of a given heterostructure. In the second scenario, we investigate the material response to a strong, spectrally narrow pump field. Here, all kinds of many-body correlations build up, changing the optical response dynamically, such that the EOM of correlation functions and Maxwell's equations have to be propagated simultaneously.

\subsubsection{Linear optical response}\label{sec:linear_optics}

For the linear case, we start by introducing the electric field in the frequency domain,
\begin{equation}
\begin{split}
\boldsymbol{E}(z,\omega)=\int_{-\infty}^{\infty}\,dt e^{i\omega t}\boldsymbol{E}(z,t)\,,
\end{split}
\label{eq:E_FT_freq}
\end{equation}
as well as the corresponding macroscopic polarization. These are used to transform the wave equation (\ref{eq:E_T_wave_FT_class}), which yields:
\begin{equation}
\begin{split}
\Big[\frac{\partial^2}{\partial z^2}+q^2\Big]\boldsymbol{E}(z,\omega)=-\mu_0\omega^2\boldsymbol{P}^{\textrm{2d}}(\omega) |\xi(z)|^2
\end{split}
\label{eq:E_T_wave_FT_class_FT}
\end{equation}
with the wave number $q=\frac{n(z)\omega}{c}$. Since the thickness of TMD monolayers is much smaller than optical wavelengths, we can safely approximate the confinement functions by delta distributions, $|\xi(z)|^2=\delta(z-z_{\textrm{TMD}})$. Then it is $\boldsymbol{E}^{\textrm{2d}}(\omega)=\boldsymbol{E}(z_{\textrm{TMD}},\omega)$ and we can introduce the susceptibility:
\begin{equation}
\begin{split}
\Big[\frac{\partial^2}{\partial z^2}+q^2\Big]E(z,\omega)=-\mu_0\omega^2\chi(\omega)E(z_{\textrm{TMD}},\omega)\delta(z-z_{\textrm{TMD}})\,.
\end{split}
\label{eq:E_T_wave_FT_class_FT2}
\end{equation}
\begin{figure}
\centering
\includegraphics[width=\columnwidth]{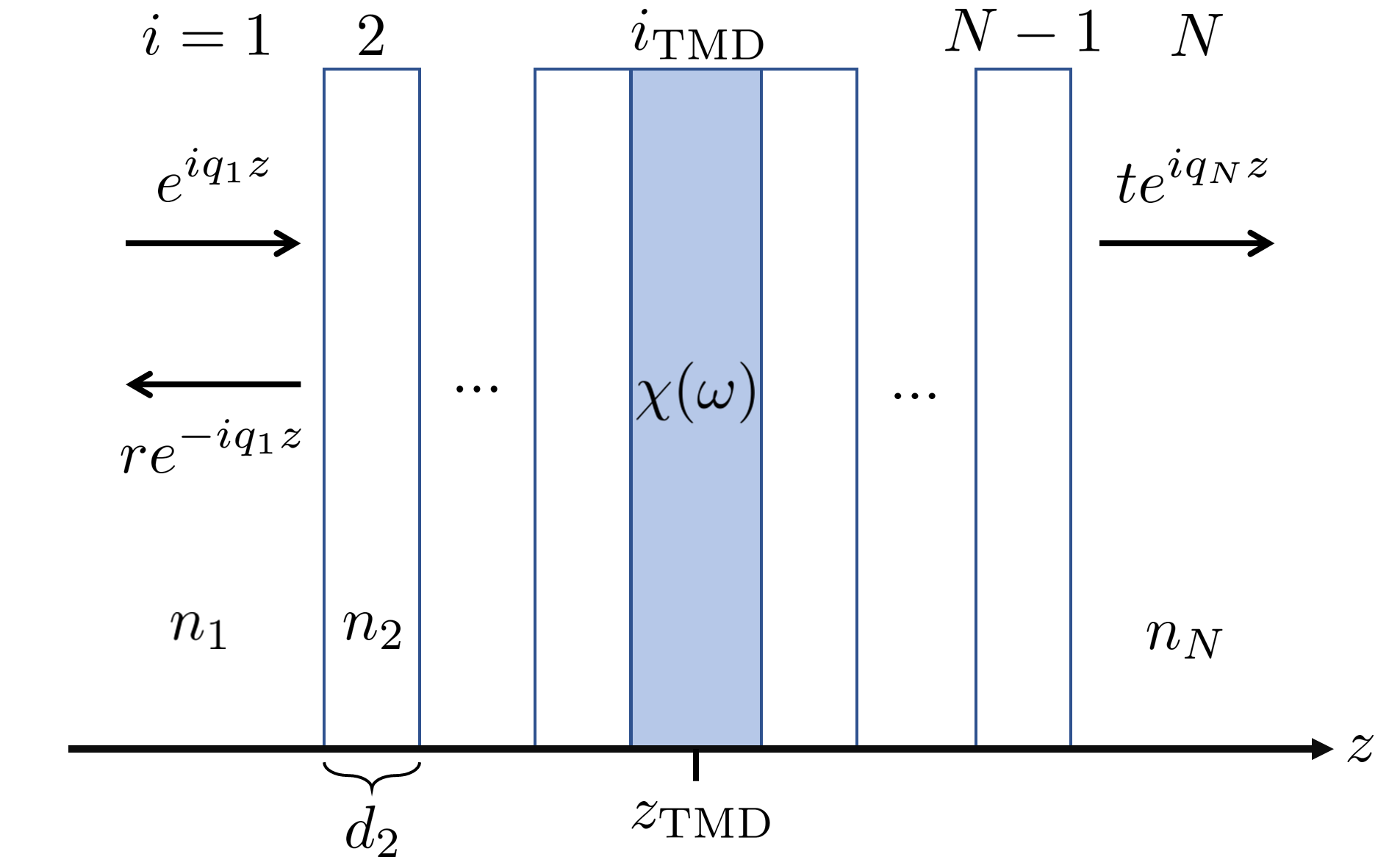}
\caption{Sketch of the dielectric heterostructure encapsulating the active TMD layer at position $z=z_{\textrm{TMD}}$ with linear susceptibility $\chi(\omega)$. The refractive index and thickness of layer $i$ are denoted by $n_i$ and $d_i$, respectively, with corresponding wave numbers $q_i=\frac{n_i\omega}{c}$. An incoming plane wave with unity amplitude propagates from left to right. The amplitude $r$ of the reflected wave and the amplitude $t$ of the transmitted wave are obtained by using transfer matrices as described in the text.}
\label{fig:sketch_TM}
\end{figure}
This equation has the homogeneous solutions
\begin{equation}
\begin{split}
&E_{\textrm{L}}(z)=L^+ e^{iq_{\textrm{L}}z}+L^- e^{-iq_{\textrm{L}}z}, \\
&E_{\textrm{R}}(z)=R^+ e^{iq_{\textrm{R}}z}+R^- e^{-iq_{\textrm{R}}z}
\end{split}
\label{eq:E_T_wave_hom}
\end{equation}
to the left (L) and to the right (R) of the TMD layer, where the wave numbers $q_i=\frac{n_i\omega}{c}$ depend on the respective refractive indices, which are assumed to be constant in each region. Inhomogeneous solutions can be obtained by making use of the fact that the first derivative of the field has to exhibit a discontinuity at $z=z_{\textrm{TMD}}$, so that the second derivative contains a delta distribution. We thus have the boundary conditions:
\begin{equation}
\begin{split}
&E_{\textrm{L}}(z_{\textrm{TMD}})=E_{\textrm{R}}(z_{\textrm{TMD}}), \\
&E'_{\textrm{L}}(z_{\textrm{TMD}})=E'_{\textrm{R}}(z_{\textrm{TMD}})+\mu_0\omega^2\chi(\omega)E(z_{\textrm{TMD}},\omega)\,.
\end{split}
\label{eq:E_T_wave_BC}
\end{equation}
These equations can be arranged in the form of a transfer matrix:
\begin{equation}
\begin{split}
&\left(
\begin{array}{c}
R^+\\
R^-\\
\end{array}
\right)
=
\boldsymbol{M}_{\textrm{TMD}}
\left(
\begin{array}{c}
L^+\\
L^-\\
\end{array}
\right)\,, \\
&\boldsymbol{M}_{\textrm{TMD}}=
\left(
\begin{array}{cc}
\frac{1}{2}\left(1+\frac{q_{\textrm{L}}}{q_{\textrm{R}}}\right)+i\xi(\omega) 
& \frac{1}{2}\left(1-\frac{q_{\textrm{L}}}{q_{\textrm{R}}}\right)+i\xi(\omega) \\
\frac{1}{2}\left(1-\frac{q_{\textrm{L}}}{q_{\textrm{R}}}\right)-i\xi(\omega) 
& \frac{1}{2}\left(1+\frac{q_{\textrm{L}}}{q_{\textrm{R}}}\right)-i\xi(\omega) \\
\end{array}
\right)
\end{split}
\label{eq:E_T_wave_TM}
\end{equation}
with 
\begin{equation}
\begin{split}
\xi(\omega)=\frac{\omega}{2\varepsilon_0 c n_{\textrm{R}}} \chi(\omega)
\end{split}
\label{eq:def_xi}
\end{equation}
and $z_{\textrm{TMD}}=0$. Here, $c=1/\sqrt{\varepsilon_0\mu_0}$ is the speed of light in vacuum. The transfer matrix $\boldsymbol{M}_{\textrm{TMD}}$ for the propagation of an electromagnetic wave through the active TMD layer, characterized by $\chi(\omega)$, is augmented by additional transfer matrices that describe the propagation through dielectric layers of the heterostructure encapsulating the TMD layer, see Fig.~\ref{fig:sketch_TM}. For every interface between dielectric slabs with refractive indices $n_i$ and $n_{i+1}$ (counting from left to right), we have the matrix
\begin{equation}
\begin{split}
&\boldsymbol{M}_i=
\frac{1}{2}\left(
\begin{array}{cc}
1+\frac{n_{i}}{n_{i+1}} 
& 1-\frac{n_{i}}{n_{i+1}} \\
1-\frac{n_{i}}{n_{i+1}} 
& 1+\frac{n_{i}}{n_{i+1}} \\
\end{array}
\right)\,,
\end{split}
\label{eq:TM_interface}
\end{equation}
while the free propagation within a slab is described by the matrix
\begin{equation}
\begin{split}
&\boldsymbol{N}_i=
\left(
\begin{array}{cc}
e^{i q_i d_i}
& 0 \\
0
& e^{-i q_i d_i} \\
\end{array}
\right)\,,
\end{split}
\label{eq:TM_free}
\end{equation}
with the wave number $q_i=\omega n_{i}/c$ and thickness $d_i$. Then, propagation through the whole heterostructure is given by the matrix product
\begin{equation}
\begin{split}
&\boldsymbol{M}_{\textrm{tot}}=
\left(\prod^{i_{\textrm{TMD}}+1}_{i=N-1}
\boldsymbol{M}_i\boldsymbol{N}_{i}\right)
\boldsymbol{M}_{\textrm{TMD}}
\left(\prod^{1}_{i=i_{\textrm{TMD}}-1}
\boldsymbol{N}_{i+1}\boldsymbol{M}_i\right)
\,,
\end{split}
\label{eq:TM_product}
\end{equation}
where the TMD represents the $i_{\textrm{TMD}}$-th layer and $N$ is the total number of layers. Note that the outermost layers ($i=1$ and $i=N$) are modeled as semi-infinite, with the $N$-th layer playing the role of a dielectric substrate while the first layer is usually given by air or vacuum. Assuming that a plane wave with unity amplitude impinges on the heterostructure from the left, we have so solve the $2\times 2$-matrix problem
\begin{equation}
\begin{split}
&\left(
\begin{array}{c}
t\\
0\\
\end{array}
\right)
=
\boldsymbol{M}_{\textrm{tot}}
\left(
\begin{array}{c}
1\\
r\\
\end{array}
\right)
\end{split}
\label{eq:E_T_wave_TM_tot}
\end{equation}
to obtain the reflection coefficient $r$ and the transmission coefficient $t$, which yields:
\begin{equation}
\begin{split}
&t= M^{11}_{\textrm{tot}}- \frac{M^{21}_{\textrm{tot}}}{M^{22}_{\textrm{tot}}} M^{12}_{\textrm{tot}}\,, \\
&r=-\frac{M^{21}_{\textrm{tot}}}{M^{22}_{\textrm{tot}}}
\,.
\end{split}
\label{eq:E_T_wave_TM_t_and_r}
\end{equation}
Reflection and transmission are then given by $R=|r|^2$ and $T=|t|^2$, respectively. Finally, linear absorption can be calculated as $A=1-R-\frac{n_{N}}{n_1}T$.

\subsubsection{Nonlinear optical response}\label{sec:nonlinear_optics}

\begin{figure}
\centering
\includegraphics[width=\columnwidth]{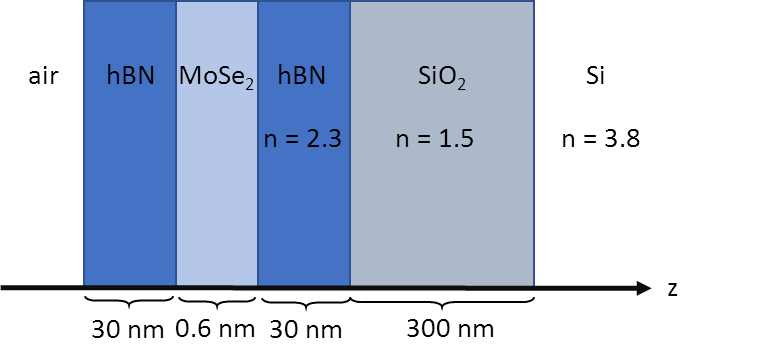}
\caption{Sketch of the dielectric heterostructure used in this work, where the refractive index $n$ and thickness of each layer are given.}
\label{fig:sketch_HS}
\end{figure}

In the case of nonlinear excitation, we solve the wave equation (\ref{eq:E_T_wave_FT_class}) numerically using the Crank-Nicolson method \cite{crank_practical_1947}. 
To this end, we decompose the electric field and the polarization into their circularly polarized components as shown in Eq.~(\ref{eq:def_pol_2d}),
\begin{equation}
\begin{split}
\boldsymbol{E}(z,t)&=\sum_{\sigma=\pm}\boldsymbol{e}_{\sigma}\big( \boldsymbol{E}(z,t) \cdot (\boldsymbol{e}_{\sigma})^* \big)=\sum_{\sigma=\pm}\boldsymbol{e}_{\sigma}E_{\sigma}(z,t)\,, \\
\boldsymbol{P}^{\textrm{2d}}(t)&=\sum_{\sigma=\pm}\boldsymbol{e}_{\sigma}\big( \boldsymbol{P}^{\textrm{2d}}(t) \cdot (\boldsymbol{e}_{\sigma})^* \big)=\sum_{\sigma=\pm}\boldsymbol{e}_{\sigma}P^{\textrm{2d}}_{\sigma}(t)
\,,
\end{split}
\label{eq:decompose_E_P}
\end{equation}
and introduce the (scaled) time derivative of the electric field as a new variable:
\begin{equation}
\begin{split}
\widetilde{\boldsymbol{E}}(z,t)&=\frac{n(z)}{c}\frac{\partial}{\partial t}\boldsymbol{E}(z,t)\,, \\
\frac{\partial^2}{\partial z^2}\boldsymbol{E}(z,t)& -\frac{n(z)}{c} \frac{\partial}{\partial t}\widetilde{\boldsymbol{E}}(z,t)
=\mu_0 \frac{\partial^2}{\partial t^2} \boldsymbol{P}^{\textrm{2d}}(t) |\xi(z)|^2
\,.
\end{split}
\label{eq:ddt_E}
\end{equation}
Next, we discretize the fields $\boldsymbol{E}(z,t)$ and $\widetilde{\boldsymbol{E}}(z,t)$ in space and time coordinates, with $\boldsymbol{E}^{j}_{i}$ denoting the electric field at position $z_i$ and time $t_j$. 
\begin{figure}
\centering
\includegraphics[width=\columnwidth]{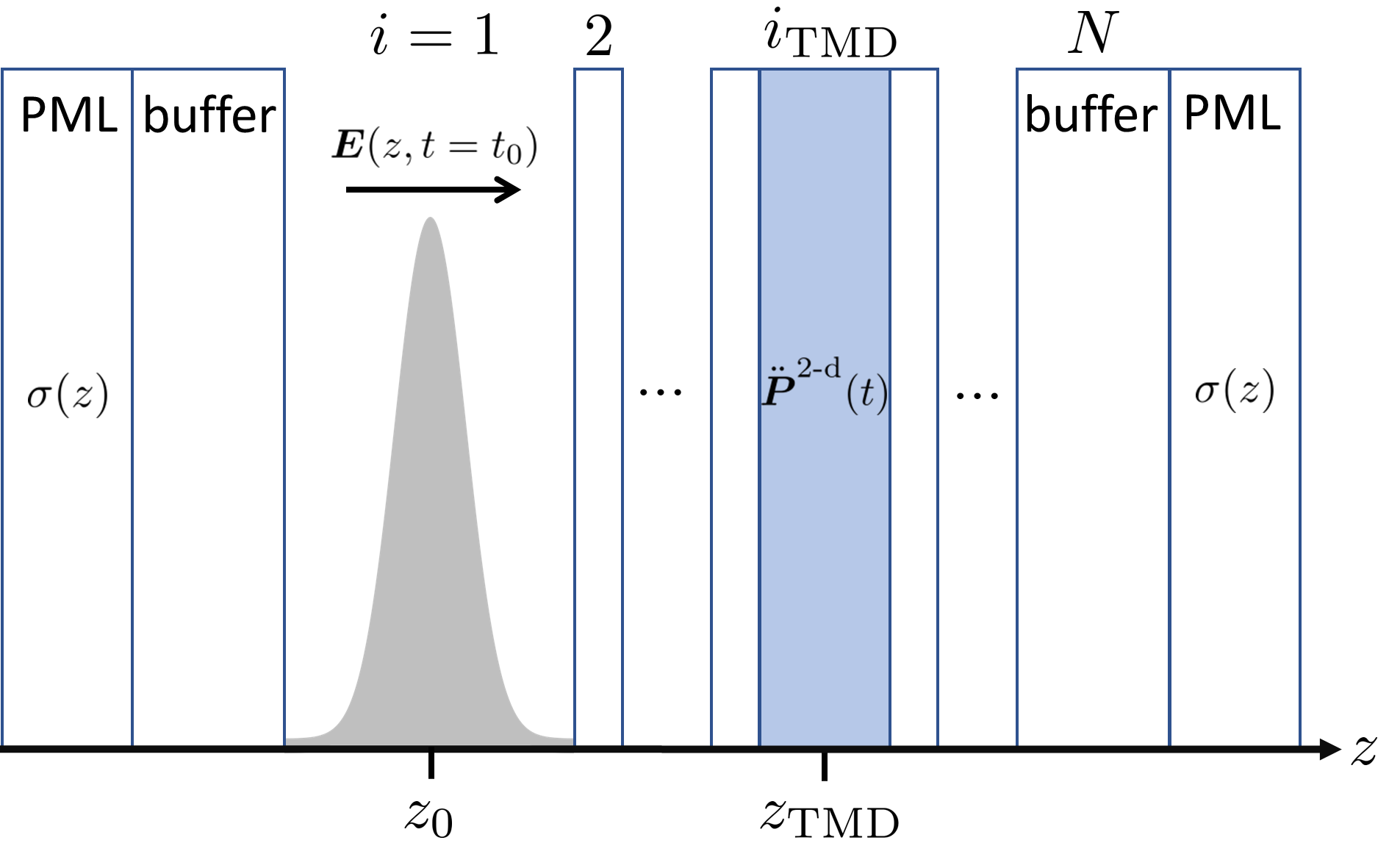}
\caption{Sketch of the setup for pulse propagation through a dielectric heterostructure encapsulating the active TMD layer at position $z=z_{\textrm{TMD}}$, which provides an inhomogeneity $\ddot{\boldsymbol{P}}^{\textrm{2d}}(t)$. The initial electric field envelope, hosted by (vacuum) layer $1$, is concentrated around the position $z=z_0$. Propagation takes place from left to right towards the heterostructure. The so-called perfectly matched layers (PML) on the outside feature a phenomenological conductivity $\sigma(z)$, which attenuates the electric field outside of the heterostructure. The PML are separated from the inner part of the setup by buffer layers. The left buffer layer is assumed to be vacuum, while the right layer constitutes a dielectric material (typically Si). Compare the layer indices of the transfer matrix setup in Fig.~\ref{fig:sketch_TM}.}
\label{fig:sketch_Crank_Nicolson}
\end{figure}
At a given time, the second spatial derivative on a non-uniform grid can be obtained by locally approximating the solution as a second-order polynomial:
\begin{equation}
\begin{split}
\frac{\partial^2}{\partial z^2}\boldsymbol{E}(z,t)\Big|_{z=z_i}&=\frac{2\boldsymbol{E}_{i+1}(t)}{h_i(h_{i-1}+h_i)}-\frac{2\boldsymbol{E}_{i}(t)}{h_{i-1}h_i}\\&+\frac{2\boldsymbol{E}_{i-1}(t)}{h_{i-1}(h_{i-1}+h_i)}
\end{split}
\label{eq:dz2_E}
\end{equation}
with $h_i=z_{i+1}-z_i$. The corresponding first-order derivative is given by:
\begin{equation}
\begin{split}
\frac{\partial}{\partial z}\boldsymbol{E}(z,t)\Big|_{z=z_i}&=\frac{h_{i-1}\boldsymbol{E}_{i+1}(t)}{h_i(h_{i-1}+h_i)}+\frac{(h_i-h_{i-1})\boldsymbol{E}_{i}(t)}{h_{i-1}h_i}\\&-\frac{h_{i}\boldsymbol{E}_{i-1}(t)}{h_{i-1}(h_{i-1}+h_i)}\,.
\end{split}
\label{eq:dz_E}
\end{equation}
The Crank-Nicolson scheme is the average of an explicit and an implicit Euler method, which means that the fields at given position are calculated as 
\begin{equation}
\begin{split}
\boldsymbol{E}_i(t_j)=\frac{\boldsymbol{E}^{j+1}_{i}+\boldsymbol{E}^{j}_{i}}{2}\,,
\end{split}
\label{eq:E_Crank_nicolson}
\end{equation}
while the time derivatives have the usual form,
\begin{equation}
\begin{split}
\frac{\partial}{\partial t}\boldsymbol{E}(z,t)\Big|_{t=t_j}=\frac{\boldsymbol{E}^{j+1}(z)-\boldsymbol{E}^{j}(z)}{\Delta t}
\end{split}
\label{eq:dtE}
\end{equation}
with the time step size $\Delta t$.
The resulting discretized equations are:
\begin{equation}
\begin{split}
\frac{1}{2}\Bigg[
&\frac{2\boldsymbol{E}^{j+1}_{i+1}}{h_i(h_{i-1}+h_i)}-\frac{2\boldsymbol{E}^{j+1}_{i}}{h_{i-1}h_i}+\frac{2\boldsymbol{E}^{j+1}_{i-1}}{h_{i-1}(h_{i-1}+h_i)}\\
+&\frac{2\boldsymbol{E}^{j}_{i+1}}{h_i(h_{i-1}+h_i)}-\frac{2\boldsymbol{E}^{j}_{i}}{h_{i-1}h_i}+\frac{2\boldsymbol{E}^{j}_{i-1}}{h_{i-1}(h_{i-1}+h_i)}
\Bigg]\\
-&\frac{n(z_i)}{c} \frac{\widetilde{\boldsymbol{E}}^{j+1}_i-\widetilde{\boldsymbol{E}}^{j}_i}{\Delta t}
=\mu_0 \frac{\partial^2}{\partial t^2} \boldsymbol{P}^{\textrm{2d}}(t)\Big|_{t=t_j} |\xi(z_i)|^2\,, \\
&\frac{n(z_i)}{c}\frac{\boldsymbol{E}^{j+1}_i-\boldsymbol{E}^{j}_i}{\Delta t}=\frac{\widetilde{\boldsymbol{E}}^{j+1}_i+\widetilde{\boldsymbol{E}}^{j}_i}{2}\,.
\end{split}
\label{eq:wave_eq_discret}
\end{equation}
One might introduce a rotating frame for the dynamical variables according to 
\begin{equation}
\begin{split}
\boldsymbol{E}(z,t)&=\hat{\boldsymbol{E}}(z,t)e^{-i\omega_{\textrm{rot}}(t-\frac{n_{\textrm{rot}}}{c} z)}\,, \\
\boldsymbol{P}^{\textrm{2d}}(t)&=\hat{\boldsymbol{P}}^{\textrm{2d}}(t)e^{-i\omega_{\textrm{rot}}t}
\end{split}
\label{eq:rot_frame}
\end{equation}
with an appropriately chosen frequency $\omega_{\textrm{rot}}$ and refractive index $n_{\textrm{rot}}$. It has to be noted that while this transformation can remove the (trivial) oscillation of a forward-propagating wave, it introduces an additional oscillation to a corresponding backward-propagating wave. It therefore depends on the chosen geometry and refractive-index profile whether the rotating frame simplifies the numerical treatment or not. In this work, we apply a rotating frame with $\omega_{\textrm{rot}}$ chosen as the carrier frequency of the pump field and $n_{\textrm{rot}}=1$.
It follows that
\begin{equation}
\begin{split}
\hat{\widetilde{\boldsymbol{E}}}(z,t)&=\frac{n(z)}{c}\Big[
\frac{\partial}{\partial t}\hat{\boldsymbol{E}}(z,t)-i\omega_{\textrm{rot}}\hat{\boldsymbol{E}}(z,t)
\Big]\,.
\end{split}
\label{eq:rot_frame2}
\end{equation}
Moreover, calculating the second-order spatial derivative of the electric field from Eq.~(\ref{eq:rot_frame}) introduces additional terms involving the first spatial derivative of $\hat{\boldsymbol{E}}(z,t)$ as well as the rotating field itself. The discretized wave equation (\ref{eq:wave_eq_discret}) can thus be transformed to the rotating frame in a straightforward way. 
The second time derivative of the macroscopic polarization in Eq.~(\ref{eq:wave_eq_discret}) is computed by interpolating the interband polarizations $\psi^{vc}_{\bk}$ as well as their first and second derivatives in time and using the definition (\ref{eq:def_pol_2d}). Interpolation of dynamical variables is naturally enabled by the ABM-method used to propagate the cluster equations.
\\
\\The refractive-index profile $n(z)$ is modeled by smearing out the index steps around layer interfaces at $z=z_{\textrm{int}}$ using Fermi functions:
\begin{equation}
\begin{split}
n(z)=
n_{\textrm{L}}+\Big(1-\frac{1}{1+e^{(z-z_{\textrm{int}})/\Gamma_n}}\Big)(n_{\textrm{R}}-n_{\textrm{L}})
\end{split}
\label{eq:n_Fermi}
\end{equation}
with $n_{\textrm{L}}$ and $n_{\textrm{R}}$ the refractive indices of the left and right layer materials, respectively, and the smearing parameter $\Gamma_n=1$ nm. For the $z$-confinement, we use a Gaussian model:
\begin{equation}
\begin{split}
|\xi(z)|^2=\frac{2\sqrt{\textrm{ln}(2)}}{d_{\xi}\sqrt{\pi}}
e^{-\left(\frac{z-z_{\textrm{TMD}}}{d_{\xi}}\right)^2 4\textrm{ln}(2)}
\end{split}
\label{eq:z_confinement}
\end{equation}
with the FWHM $d_{\xi}$ nm as an effective TMD layer width. We assume that the TMD layer has no influence on the refractive index as $d_{\xi}$ is much smaller than optical wavelengths. Hence the TMD layer is described by a dielectric slab of thickness $d_{\xi}$, where the left and right halves have the same refractive indices as the adjacent layers, respectively.
\\
\\The initial electric field and its derivative are modeled as
\begin{equation}
\begin{split}
\boldsymbol{E}(z,t_0)&=\boldsymbol{e}_{\textrm{p}}E_0\left(t-n_0\frac{z-z_0}{c}\right)e^{-i\omega_0\left(t-n_0\frac{z-z_0}{c}\right)}\Big|_{t=t_0}\,, \\
\widetilde{\boldsymbol{E}}(z,t_0)&=\boldsymbol{e}_{\textrm{p}}\frac{\partial}{\partial t}\Big[E_0\left(t-n_0\frac{z-z_0}{c}\right)e^{-i\omega_0\left(t-n_0\frac{z-z_0}{c}\right)}\Big]\Big|_{t=t_0}\,.
\end{split}
\label{eq:E_initial}
\end{equation}
Here, $E_0(t)$ is the envelope of the pump pulse, $z_0$ is the initial position of the pulse center, $\omega_0$ is the carrier frequency and $n_0$ is the refractive index in the region around $z_0$. The polarization vector is given by $\boldsymbol{e}_{\textrm{p}}$. The initial fields are chosen such that the wave initially propagates in positive $z$-direction (from left to right). Hence we place $z_0$ sufficiently far to the left, such that the envelope is located outside of the heterostructure, i.e. the magnitude of $E_0$ has decreased to $\eta E_0(t_0-n_0\frac{z-z_0}{c}=0)$ at the heterostructure surface. We choose $\eta=10^{-4}$.
The fluence of the pump pulse propagating in vacuum corresponds to the transmitted electromagnetic energy per area given by
\begin{equation}
 \begin{split}
  F=\int dt \left| \mathbf{S}(t) \right| = \varepsilon_0 c \int dt \left| \mathbf{E}(t) \right|^2\,, \end{split}
\label{eq:fluence}
\end{equation}
with $\textbf{S}$ denoting the Poynting vector. Here, $\mathbf{E}(t)=\boldsymbol{e}_{\textrm{P}}E_0(t)e^{-i\omega_0 t}$ is the pump laser field outside of the heterostructure, with the envelope $E_0(t)$, the carrier frequency $\omega_0$ and the polarization vector $\boldsymbol{e}_{\textrm{P}}$, see Eq.~(\ref{eq:E_initial}). 
\\
\\To ensure proper boundary conditions, i.e. outwardly propagating waves being attenuated outside of the heterostructure, we add so-called perfectly matched layers (PML) \cite{sjogreen_perfectly_2005}, which contain phenomenological absorbing terms. This is realized by replacing the spatial derivative in Eq.~(\ref{eq:ddt_E}) according to
\begin{equation}
\begin{split}
\frac{\partial}{\partial z}\rightarrow\frac{1}{1+i\frac{\sigma(z)}{\omega}}\frac{\partial}{\partial z}\
\end{split}
\label{eq:PML_replacement}
\end{equation}
with the phenomenological conductivity $\sigma(z)$. We transform the resulting equation to the frequency domain by assuming a harmonic time dependence of the fields ($\frac{\partial}{\partial t}\rightarrow -i\omega $), neglecting the spatial derivative of $\sigma(z)$ and making use of the fact that the $\ddot{P}$-inhomogeneity does not overlap spatially with the PML region. After cancelling all $1/\omega$-terms, the equation is transformed back to the time domain, which leads to the replacement
\begin{equation}
\begin{split}
&\frac{\partial^2}{\partial z^2}\boldsymbol{E}(z,t)\rightarrow \\
&\frac{\partial^2}{\partial z^2}\boldsymbol{E}(z,t)
-2\sigma(z)\frac{n(z)}{c}\widetilde{\boldsymbol{E}}(z,t)-\sigma^2(z)\frac{n^2(z)}{c^2}\boldsymbol{E}(z,t)\,.
\end{split}
\label{eq:PML_replacement_final}
\end{equation}
The new terms can be transformed to the rotating frame and straightforwardly included in the discretized wave equations.
The profile $\sigma(z)$ is chosen such that the conductivity is quadratically switched on inside the PML with the coefficient $\frac{1000\textrm{ps}^{-1}}{(10\mu\textrm{m})^2 n_{\textrm{PML}}}$, where $n_{\textrm{PML}}$ is the refractive index of the PML material. This gradual switching justifies the assumption of a negligible z-derivative of $\sigma(z)$ and ensures that no artificial reflection occurs at the PML interface. The width of the PML is $10$ $\mu$m. Finally, the PML are separated from the inner part of the setup by buffer layers of width $5$ $\mu$m. The left buffer layer is assumed to be the same material as the initial pulse region (typically vacuum), while the right buffer layer simulates a practically semi-infinite dielectric substrate material (typically Si). As refractive index of the PML, we use the indices of the adjacent buffer layers, respectively.
The complete setup consisting of heterostructure, initial pump pulse and auxiliary layers is schematically shown in Fig.~\ref{fig:sketch_Crank_Nicolson}. 
\\
\\ The spatial coordinate is discretized by using equidistant grids with spacing $h=0.25$ nm in regions of $\pm 10$ nm around the refractive index steps, $h=0.1$ nm in a region of $\pm 1$ nm around the TMD layer and $h=20$ nm far away from the heterostructure. The latter is chosen to properly resolve the spatial oscillations, whose wavelengths depend on the local refractive index, in all parts of the supercell. Between the different grids, we gradually vary the spacing $h$ to avoid numerical instabilities. In particular, to switch from the fine grids around the outer boundaries of the heterostructure to the coarse grids outside, we increase $h$ by a factor $1.01$ in each step.
\\
\\The discretized wave equations (\ref{eq:wave_eq_discret}) in the rotating frame, with additional absorbing terms, can be written in matrix form with respect to the spatial coordinates:
\begin{equation}
\begin{split}
\boldsymbol{A}
\left(
\begin{array}{c}
\hat{E}_{+}\\
\hat{E}_{-}\\
\hat{\widetilde{E}}_{+}\\
\hat{\widetilde{E}}_{-}\\
\end{array}
\right)^{j+1}
=
\boldsymbol{B}
\left(
\begin{array}{c}
\hat{E}_{+}\\
\hat{E}_{-}\\
\hat{\widetilde{E}}_{+}\\
\hat{\widetilde{E}}_{-}\\
\end{array}
\right)^{j}
+\boldsymbol{C}^j
\,.
\end{split}
\label{eq:wave_eq_matrix}
\end{equation}
Hence, to propagate the solution from time $t_j$ to time $t_{j+1}$, one has to solve the above matrix problem. 
The solution is performed by means of a modified Thomas algorithm (tridiagonal matrix algorithm), which explicitly takes into account the tridiagonal form of the $\hat{E}$-part of the matrix $\boldsymbol{A}$. First of all, we note that the matrix problem can be solved for each polarization direction separately. Referring to $\boldsymbol{x}$ as the solution vector (containing the fields at time $t_{j+1}$) and $\boldsymbol{r}$ as the RHS vector for a given polarization and assuming that we have a total number of $N$ spatial grid points, the $\hat{E}$-equations take the form
\begin{equation}
\begin{split}
&a_i x_{i-1}+b_i x_i+c_i x_{i+1}+d_i x_{i+N}=r_i\,, \\
&a_1=0,c_N=0,i=1..N
\,,
\end{split}
\label{eq:wave_eq_matrix_E}
\end{equation}
while the $\hat{\widetilde{E}}$-equations are
\begin{equation}
\begin{split}
e_i x_i+f_i x_{i+N}=r_{i+N},i=1..N
\,.
\end{split}
\label{eq:wave_eq_matrix_Etilde}
\end{equation}
We use Eq.~(\ref{eq:wave_eq_matrix_Etilde}) to express $x_{i+N}$ in terms of $x_i$,
\begin{equation}
\begin{split}
x_{i+N}=S_i+T_i x_i\,,
\,.
\end{split}
\label{eq:wave_eq_matrix_Etilde2}
\end{equation}
and insert it into Eq.~(\ref{eq:wave_eq_matrix_E}). Then we use the ansatz
\begin{equation}
\begin{split}
&x_i=P_i x_{i+1}+Q_i\, \\
&x_{i-1}=P_{i-1} x_{i}+Q
\end{split}
\label{eq:wave_eq_matrix_ansatz}
\end{equation}
to derive recursive relations for the coefficients $P_i$ and $Q_i$. Since $a_1=0$, the recursion starts with $P_1=\frac{-c_1}{b_1+d_1 T_1}$ and $Q_1=\frac{r_1-d_1 S_1}{b_1+d_1 T_1}$. The solution $x_i$ for $i=1..N$ is obtained by using $c_N=0$, i.e. $x_N=Q_N$, and using Eq.~(\ref{eq:wave_eq_matrix_ansatz}) recursively. Finally, $x_{i+N}$ can be directly obtained via Eq.~(\ref{eq:wave_eq_matrix_Etilde2}).
We find that an (equidistant) time step size $\Delta t=50$ as is sufficient (corresponding to a free-space propagation of $15$ nm). To feed the electric field back into the EOM for correlation functions at any required time $t$, we solve Eq.~(\ref{eq:wave_eq_matrix}) for a time $t_j$ on the equidistant grid, where the fields are already known, such that $\Delta t = t - t_j$. The field $\boldsymbol{E}^{\textrm{2d}}$ at the TMD position is then calculated via
\begin{equation}
\begin{split}
\boldsymbol{E}^{\textrm{2d}}(t)=\int dz |\xi(z)|^2 \boldsymbol{E}(z,t)
\end{split}
\,,
\label{eq:E_field_average}
\end{equation}
see Eqs.~(\ref{eq:FT_av_E}) and (\ref{eq:light_matter_fact}).
\\
\\The dielectric heterostructure modelled in this work is schematically shown in Fig.~\ref{fig:sketch_HS}. It is inspired by the experimental setup used in Ref.~\cite{wang_optical_2019}. As TMD layer thickness we assume $0.65$ nm \cite{kylanpaa_binding_2015}. The refractive indices at optical frequencies are $n=2.3$ for hBN \cite{artus_natural_2018}, $n=1.5$ for SiO$_2$ \cite{malitson_interspecimen_1965}, and $n=3.8$ for Si \cite{aspnes_spectroscopic_1980}.

\subsection{Perturbation theory}\label{sec:perturbation}

In this section, we derive perturbative expressions for the dephasing of excitonic coherences induced by carrier-phonon interaction. To this end, we first introduce
two-particle operators
\begin{equation}
\begin{split}
X^{\dagger}_{\alpha,\bq}&=\sum_{\bk,v,c}\phi^{vc}_{\alpha,\bq}(\bk)a_{\bk-\bq,\textrm{c}}^{\dagger} a_{\bk,\textrm{v}}^{\phantom\dagger}, \\
a_{\bk-\bq,\textrm{c}}^{\dagger} a_{\bk,\textrm{v}}^{\phantom\dagger}&=\sum_{\alpha}(\phi^{vc}_{\alpha,\bq}(\bk))^*X^{\dagger}_{\alpha,\bq} \,.
\end{split}
\label{eq:X_op}
\end{equation}
The wave functions $\phi^{vc}_{\alpha,\bq}(\bk)$ are solutions of the Bethe-Salpeter equation (BSE) in the absence of photoexcited carriers 
\begin{equation}
 \begin{split}
&(\varepsilon^{\textrm{c}}_{\bk-\bq}-\varepsilon^{\textrm{v}}_{\bk}-E_{\alpha,\bq})\phi^{vc}_{\alpha,\bq}(\bk) \\
-&\sum_{\bk',v',c'}
V^{c,v',v,c'}_{\bk-\bq,\bk',\bk,\bk'-\bq}\phi^{v'c'}_{\alpha,\bq}(\bk')=0\,,
\end{split}
\label{eq:BSE}
\end{equation}
with two-particle eigenenergies $E_{\alpha,\bq}$. The total momentum is denoted by $\bq$, 
while $\alpha$ is the quantum number that belongs to the relative motion of electron and hole. 
The wave functions fulfill
orthonormality and completeness relations:
\begin{equation}
\begin{split}
\sum_{\bk,v,c}(\wfc{v}{c}{\alpha}{\bq}{\bk})^*\wfc{v}{c}{\alpha'}{\bq}{\bk}&=\delta_{\alpha,\alpha'}, \\
\sum_{\alpha}\wfc{v}{c}{\alpha}{\bq}{\bk}(\wfc{v'}{c'}{\alpha}{\bq}{\bk'})^*&=\delta_{\bk,\bk'}\delta_{v,v'}\delta_{c,c'}\,.
\end{split}
\label{eq:ortho_exciton}
\end{equation}
With the definition (\ref{eq:X_op}) we can introduce microscopic interband polarizations in the exciton representation:
\begin{equation}
\begin{split}
\psi_{\alpha,\bn}=\ev{X^{\phantom\dagger}_{\alpha,\bn}}&=\sum_{\bk,v,c}(\phi^{vc}_{\alpha,\bn}(\bk))^*\psi^{v,c}_{\bk}\,, \\
\psi^{v,c}_{\bk}&=\sum_{\alpha}\phi^{vc}_{\alpha,\bn}(\bk)\psi_{\alpha,\bn}\,.
\end{split}
\label{eq:psi_X}
\end{equation}
Similarly, phonon-assisted polarizations (\ref{eq:def_Xi}) are transformed:
\begin{equation}
\begin{split}
\Xi_{\alpha,\bq,j}=\Delta\ev{D^{\dagger}_{\bq,j}X^{\phantom\dagger}_{\alpha,\bq}}&=\sum_{\bk,v,c}(\phi^{vc}_{\alpha,\bq}(\bk))^*
\Delta\ev{D^{\dagger}_{\bq,j}a_{\bk,\textrm{v}}^{\dagger} a_{\bk-\bq,\textrm{c}}^{\phantom\dagger}} \\
&=\sum_{\bk,v,c}(\phi^{vc}_{\alpha,\bq}(\bk+\bq))^*
\Xi^{v,c}_{\bk,-\bq,j} \,, \\
\Xi^{v,c}_{\bk,\bq,j}&=\sum_{\alpha}\phi^{vc}_{\alpha,-\bq}(\bk-\bq)\Xi_{\alpha,-\bq,j}\,.
\end{split}
\label{eq:Xi_X}
\end{equation}
We explicitly introduce analogous correlations with flipped band indices:
\begin{equation}
\begin{split}
\widetilde{\Xi}_{\alpha,\bq,j}=\Delta\ev{D^{\dagger}_{\bq,j}X^{\dagger}_{\alpha,\bq}}
&=\sum_{\bk,v,c}\phi^{vc}_{\alpha,\bq}(\bk)
\Xi^{c,v}_{\bk,\bq,j} \,, \\
\Xi^{c,v}_{\bk,\bq,j}&=\sum_{\alpha}(\phi^{vc}_{\alpha,\bq}(\bk))^*\widetilde{\Xi}_{\alpha,\bq,j}\,.
\end{split}
\label{eq:Xi_X2}
\end{equation}
Starting from the EOM of interband polarizations (\ref{eq:EOM_psi_k_corr}) in the absence of photoexited carriers, i.e., without the coherent-phonon and carrier-carrier-doublet contributions, 
we can formulate EOM for $\psi_{\alpha,\bn}$ using (\ref{eq:psi_X}). We then expand all correlation functions on the RHS in terms of excitonic quantities, discard inter-band carrier-phonon processes and apply the orthonormality relation (\ref{eq:ortho_exciton}). The resulting EOM is:
\begin{equation}
\begin{split}
i\hbar\frac{d}{dt}\psi_{\alpha,\bn}&=E_{\alpha,\bq}\psi_{\alpha,\bn}-\boldsymbol{d}_{\alpha}\cdot \boldsymbol{E}^{\textrm{2d}} \\
&+\sum_{\beta,\bq,j}G^{\alpha,\beta}_{\bq,j}\Big[
\Xi_{\beta,\bq,j}+(\widetilde{\Xi}_{\beta,\bq,j})^*
\Big]
\end{split}
\label{eq:EOM_psi_X}
\end{equation}
with effective matrix elements
\begin{equation}
\begin{split}
\boldsymbol{d}_{\alpha}=\sum_{\bk,v,c}(\phi^{vc}_{\alpha,\bn}(\bk))^*\boldsymbol{d}^{c,v}_{\bk}
\end{split}
\label{eq:eff_dipole_ME}
\end{equation}
and
\begin{equation}
\begin{split}
G^{\alpha,\beta}_{\bq,j}=\sum_{\bk,v,c}(\phi^{vc}_{\alpha,\bn}(\bk))^*\Big(
&g^{c,c}_{\bk,\bq,j}\phi^{vc}_{\beta,\bq}(\bk)\\-&g^{v,v}_{\bk+\bq,\bq,j}\phi^{vc}_{\beta,\bq}(\bk+\bq)
\Big)
\end{split}
\label{eq:eff_elph_ME}
\end{equation}
in the exciton picture. Starting from Eq.~(\ref{eq:EOM_Xi}), we repeat the above procedure for the phonon-assisted polarizations 
$\Xi^{v,c}_{\bk,\bq,j}$ and $\Xi^{c,v}_{\bk,\bq,j}$ to obtain the EOM 
\begin{equation}
\begin{split}
i\hbar\frac{d}{dt}\Xi_{\alpha,\bq,j}&=(E_{\alpha,\bq}-\hbar\omega_{\bq,j}-i\gamma_{\textrm{phen}})\Xi_{\alpha,\bq,j}
\\
&+n_{\bq,j}\sum_{\beta}\psi_{\beta,\bn}\big(G^{\beta,\alpha}_{\bq,j}\big)^*
\end{split}
\label{eq:EOM_Xi_X}
\end{equation}
and 
\begin{equation}
\begin{split}
i\hbar\frac{d}{dt}\widetilde{\Xi}_{\alpha,\bq,j}&=(-E_{\alpha,\bq}-\hbar\omega_{\bq,j}-i\gamma_{\textrm{phen}})\widetilde{\Xi}_{\alpha,\bq,j}
\\
&-(1+n_{\bq,j})\sum_{\beta}(\psi_{\beta,\bn})^*G^{\beta,\alpha}_{\bq,j}\,,
\end{split}
\label{eq:EOM_Xi_tilde_X}
\end{equation}
where we describe dephasing due to coupling to higher-order correlations by a phenomenological constant $\gamma_{\textrm{phen}}$. 
To adiabatically eliminate the phonon-assisted polarizations, we apply a Markov-approximation to Eqs.~(\ref{eq:EOM_Xi_X}) and (\ref{eq:EOM_Xi_tilde_X}), extracting trivial oscillations $e^{-iE_{\alpha,\bn}/\hbar t}$ from the excitonic polarizations. The resulting EOM for the excitonic polarizations in Markov approximation is:
\begin{equation}
\begin{split}
i\hbar\frac{d}{dt}\psi_{\alpha,\bn}&=E_{\alpha,\bq}\psi_{\alpha,\bn}-\boldsymbol{d}_{\alpha}\cdot \boldsymbol{E}^{\textrm{2d}} \\
&+\sum_{\beta}\psi_{\beta,\bn}\Gamma_{\alpha,\beta}
\end{split}
\label{eq:EOM_psi_X_Born_Markov}
\end{equation}
with the coefficient
\begin{equation}
\begin{split}
\Gamma_{\alpha,\beta}&=\sum_{\nu,\bq,j}G^{\alpha,\nu}_{\bq,j}\big(G^{\beta,\nu}_{\bq,j}\big)^*\times\\
\Big[
&\frac{1+n_{\bq,j}}{E_{\beta,\bn}-E_{\nu,\bq}-\hbar\omega_{\bq,j}+i\gamma_{\textrm{phen}}}\\
+&\frac{n_{\bq,j}}{E_{\beta,\bn}-E_{\nu,\bq}+\hbar\omega_{\bq,j}+i\gamma_{\textrm{phen}}}
\Big]\,.
\end{split}
\label{eq:EOM_psi_X_Born_Markov_coeff}
\end{equation}
Similar expressions have been derived in Refs.~\cite{selig_dark_2018,chan_exciton_2023}.
Assuming that phonon-induced coupling between different relative-motion states $\alpha,\beta$ is weak, Eq.~(\ref{eq:EOM_psi_X_Born_Markov}) describes a driven excitonic oscillator with its frequency renormalized by the polaron shift $\textrm{Re} \left\{\Gamma_{\alpha,\alpha}\right\}$ and a damping $-\textrm{Im} \left\{\Gamma_{\alpha,\alpha}\right\}$ corresponding to an inverse lifetime of the coherence.
\\
\\ By calculating the Fourier transform of Eq.~(\ref{eq:EOM_psi_X_Born_Markov}) and neglecting off-diagonal coefficients, we obtain an algebraic expression for the excitonic polarization in the frequency domain. Making use of Eqs.~(\ref{eq:psi_X}), (\ref{eq:def_pol_2d}), and (\ref{eq:def_chi}), we arrive at the following explicit form of the optical susceptibility:
\begin{equation}
\begin{split}
\chi(\omega)&=\frac{1}{\varepsilon_0}\sum_{\alpha}\frac{|\boldsymbol{d}_{\alpha}\cdot\boldsymbol{e}_{\textrm{p}}|^2}{E_{\alpha,\bn}+\Gamma_{\alpha,\alpha}-\hbar\omega}
\,.
\end{split}
\label{eq:chi_PT}
\end{equation}
Via the transfer matrix approach described in Appendix~\ref{sec:linear_optics}, it can be used to compute the linear optical response including exciton-phonon coupling in perturbation theory.

\subsection{Brillouin zone sampling}\label{sec:numerical}

To numerically evaluate the cluster equations for a set of Bloch states from the first Brillouin zone (BZ), we perform the thermodynamic limit ($N\rightarrow\infty$, $\mathcal{A}\rightarrow\infty$, $N/\mathcal{A}=\textrm{const.}$) by replacing momentum sums according to
\begin{equation}
\begin{split}
\sum_{\bk}\rightarrow\frac{\mathcal{A}}{(2\pi)^2}\int d^2\bk= \frac{N A_{\textrm{UC}}}{(2\pi)^2}\int d^2\bk\approx 
\frac{N A_{\textrm{UC}}}{(2\pi)^2}\sum_{\bk_i}\Delta
\end{split}
\label{eq:thermo_limit}
\end{equation}
with the crystal unit cell area $A_{\textrm{UC}}$ and the k-point weight $\Delta$. 
The resulting prefactor $N A_{\textrm{UC}}$ is partly compensated by a rescaling of interaction matrix elements. The Coulomb matrix elements (\ref{eq:Coul_ME}) are rescaled by $N A_{\textrm{UC}}$, which yields matrix elements proportional to $A_{\textrm{UC}}$. Similarly, the carrier-phonon matrix elements (\ref{eq:carr_phon_ME_bloch}) are rescaled by $\sqrt{N A_{\textrm{UC}}}$. Any remaining factors $\mathcal{A}$ or $\sqrt{\mathcal{A}}$ in the EOM can be absorbed by an appropriate rescaling of correlation functions. 
For example, we replace $c_{\lambda_1,\lambda_2,\lambda_3\lambda_4}^{\bq,\bk',\bk}$ by $c_{\lambda_1,\lambda_2,\lambda_3\lambda_4}^{\bq,\bk',\bk}\mathcal{A}$, $\ev{D^{\dagger}_{\bn,j}}$ by $\frac{1}{\sqrt{\mathcal{A}}}\ev{D^{\dagger}_{\bn,j}}$ and $\Xi^{\nu,\nu'}_{\bk,\bq,j}$ by $\Xi^{\nu,\nu'}_{\bk,\bq,j}\sqrt{\mathcal{A}}$. We also replace the exciton wave functions $\phi^{vc}_{\alpha,\bq}(\bk)$ by $\phi^{vc}_{\alpha,\bq}(\bk)\sqrt{\mathcal{A}}$, so that the rescaled functions have a well-defined normalization in the thermodynamic limit.
\\
\\The first BZ and corresponding integrals according to Eq.~(\ref{eq:thermo_limit}) are discretized using a Monkhorst-Pack mesh. 
Moreover, we reduce phase space by considering momentum states in limited regions around the high-symmetry points $K$, $-K$, $\Lambda$ and $-\Lambda$ (the latter being halfway between $\Gamma$ and $K/-K$, respectively).
To compute the linear absorption spectra in Fig.~\ref{fig:absorption}, we use a $36\times 36\times 1$-mesh with momentum states limited to circles with radius $2$ nm$^{-1}$ around the high-symmetry points. For all calculations on nonlinear excitation dynamics, we use a $30\times 30\times 1$-mesh and radius $1.7$ nm$^{-1}$. 
To obtain the results shown in Fig.~\ref{fig:absorption}, we take into account only one spin subsystem to reduce the numerical effort for the three-phonon correlations.
The resulting numbers of EOM for the absorption calculations are $2.8\times 10^9$ for two spin subsystems and $1.4\times 10^9$ for one spin subsystem, respectively. For the nonequilibrium calculation, we have to solve $2.1\times 10^9$ EOM. We repeat the calculation of absorption spectra for $T=300$ K on the coarse mesh to obtain the correct exciton energies for resonant pumping.
\\
\\Care has to be taken to properly treat the long-wavelength limit ($\bq=\bn$) of Coulomb interaction. To this end, we write the (screened) Coulomb matrix elements in the Wannier basis as $V_{\alpha\beta}(\bq)=\frac{1}{|\bq|}\widetilde{V}_{\alpha\beta}(\bq)$. Assuming that any other function $f(\bq)$ that is integrated over with a Coulomb matrix element has a weak momentum dependence around $\bq=\bn$, we explicitly integrate the $1/q$ behavior in polar coordinates:
\begin{equation}
\begin{split}
&\sum_{\bq_i} V_{\alpha\beta}(\bq_i) f(\bq_i)\Delta\\\approx
&\int\limits_{\bq\approx\bn} d^2\bq  V_{\alpha\beta}(\bq)f(\bq)+\sum_{\bq_i\neq\bn} V_{\alpha\beta}(\bq_i) f(\bq_i)\Delta\\
\approx& \widetilde{V}_{\alpha\beta}(\bn)f(\bn)  \int\limits_{\bq\approx\bn} d\phi dq q \frac{1}{q}+\sum_{\bq_i\neq\bn} V_{\alpha\beta}(\bq_i) f(\bq_i)\Delta  \,,
\end{split}
\label{eq:coul_pole}
\end{equation}
where the $\bq\approx\bn$-integration is performed over a Voronoi cell of the Monkhorst-Pack mesh.

\subsection{Absorption of higher excitons}\label{sec:abs_B}

\begin{figure}
\centering
\includegraphics[width=1.\columnwidth]{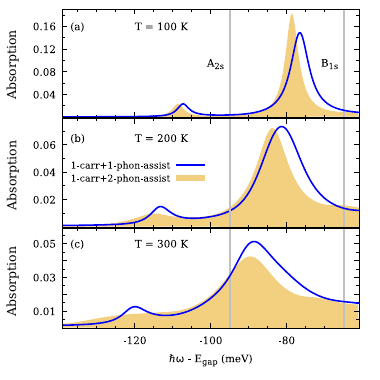}
\caption{A$_{\textrm{2s}}$- and B$_{\textrm{1s}}$-exciton absorption of hBN-encapsulated monolayer MoSe$_2$ at different temperatures. Different levels of approximation including polarizations assisted by one or two phonons are denoted by \textit{1-carr+n-phon-assist}, see the scheme in Fig.~\ref{fig:sketch_cluster}. The vertical line indicates the exciton energies without carrier-phonon interaction. Optical transition energies $\hbar\omega$ are shown relative to the band gap $E_{\textrm{gap}}$.}
\label{fig:absorption_B}
\end{figure}

For completeness, we show numerical results for the absorption of higher-lying exciton states in Fig.~\ref{fig:absorption_B} involving one- and two-phonon-correlations. Calculations are performed similar to those for Fig.~\ref{fig:absorption}, with the difference that both spin subsystems are taken into account here. Compared to A$_{\textrm{1s}}$-excitons, an increased number of exciton-phonon scattering channels into lower states leads to significantly stronger line broadening and polaron shifts. However, the trend of overestimating the linewidth at small temperatures and underestimating it at elevated temperatures in the \textit{1-phon-assist} approximation is also observed for higher excitons.

\end{document}